# DENSE AND LONG-TERM MONITORING OF EARTH SURFACE PROCESSES WITH PASSIVE RFID — A REVIEW

Mathieu Le Breton [1,2], Frédéric Liébault [3], Laurent Baillet [2], Arthur Charléty [2], Éric Larose [2], Smail Tedjini [4]

[1] Géolithe Innov, Crolles, Frances
[2] Université Grenoble Alpes, ISTerre, Grenoble, France
[3] Université Grenoble Alpes, INRAE, ETNA, Grenoble, France
[4] Université Grenoble Alpes, Grenoble-INP/LCIS, Valence, France

**Abstract**—Billions of Radio-Frequency Identification (RFID) passive tags are produced yearly to identify goods remotely. New research and business applications are continuously arising, including recently localization and sensing to monitor earth surface processes. Indeed, passive tags can cost 10 to 100 times less than wireless sensors networks and require little maintenance, facilitating years-long monitoring with ten's to thousands of tags. This study reviews the existing and potential applications of RFID in geosciences. The most mature application today is the study of coarse sediment transport in rivers or coastal environments, using tags placed into pebbles. More recently, tag localization was used to monitor landslide displacement, with a centimetric accuracy. Sensing tags were used to detect a displacement threshold on unstable rocks, to monitor the soil moisture or temperature, and to monitor the snowpack temperature and snow water equivalent. RFID sensors, available today, could monitor other parameters, such as the vibration of structures, the tilt of unstable boulders, the strain of a material, or the salinity of water. Key challenges for using RFID monitoring more broadly in geosciences include the use of ground and aerial vehicles to collect data or localize tags, the increase in reading range and duration, the ability to use tags placed under ground, snow, water or vegetation, and the optimization of economical and environmental cost. As a pattern, passive RFID could fill a gap between wireless sensor networks and manual measurements, to collect data efficiently over large areas, during several years, at high spatial density and moderate cost.

Table of content

# 1    Why passive RFID in geosciences?

## 1.1    *What is RFID: Back to basics*

Radio Frequency Identification (RFID) is a set of technologies that enables to detect and identify a target using a wireless radiofrequency communication channel. Active RFID uses a conventional radiofrequency communication, in which a device with a battery emits its own radiofrequency wave to an interrogator (many protocols exist such as Bluetooth, LoRA or LTE) in order to communicate its identifier. In this review, we focus instead on passive RFID, where only the interrogator emits a radiofrequency wave. A passive target, called "tag," receives this wave, modulates it to embed information, and reflects the modulated wave towards the interrogator. Backscatter communication is a method in which a passive RFID tag receives a radiofrequency excitation from an interrogator, and backscatters a modulated signal to send back a response to the interrogator.

Historically, backscattered communication was first patented as a preliminary concept to communicate between an active and a passive radiotelegraphy station (Brard, 1930). Later, Stockman (1948) explained the physics behind the communication by reflected powers. He concluded that "considerable research and development work has to be done before the remaining basic problems in reflected-power communication are solved, and before the field of useful applications is explored." At the same time, Theremin secretly released in 1945 a fully passive device for wireless microphone spying (see Nikitin, 2012), called the "Thing." It consists of a simple electromagnetic cavity, with one of its walls replaced by a vibrating membrane sensitive to sound waves. Under the illumination by a radiofrequency signal at its resonant frequency, the "Thing" backscattered a signal modulated in amplitude by the ambient sound waves. It is probably the very first passive wireless sensor and is arguably the ancestor of the RFID tag. On the other hand, the radar systems which began to develop also during the 1940s are among the most relevant and useful techniques exploiting backscatter signals. It gave rise to the development of the Identify Friend or Foe System very widely used by the allies during the Second World War (see Obe, 2003). With the development of electronic systems and the advent of semiconductor materials and integration technologies in the 1950s, it was possible to develop integrated and compact systems like the modern labels. The best-known device inherited from the 1960s is the Electronic Article Surveillance method, which enabled the first commercial application of RFID (Minasy, 1970). The method is effective for preventing shoplifting from stores or item removal from buildings.

Today, industrial tags mostly use microelectronic chips. Such tags are essentially an antenna or a coil, connected to an application-specific integrated circuit (the RFID chip). The chip can switch its input impedance between two states, thus alternating the radiofrequency power reflected by the tag between two levels (see Fig. 1). This alternation will be interpreted by the reader as a digital signal, containing data such as the identifier of the tag stored in the chip memory. Physically, today's tags send data back to the reader either by magnetic coupling through a coil (e.g., Low-Frequency tags at 125 kHz and High Frequency tags at 13.56 MHz) or by backscattering through an antenna (e.g., ultra-high frequency tags at 866–960 MHz) (see Fig. 2). Besides, many examples in this review use battery-assisted passive (*BAP*) tags (also called semi-passive). These are essentially passive tags, whose chip receive extra power from an external battery, to increase the chip capabilities and its reading distance. While passive tags (either batteryless or battery-assisted) use the backscattering communication, active tags use an RF front-end architecture for communication. Research also investigates and develop chipless tags, to reduce the tags cost by encoding information physically on the tag instead of using a microelectronic chip. While promising, chipless RFID is not yet widespread in the industry; therefore it is still too preliminary to be usable in real operations for geoscience. Overall, the different technologies produced by the RFID industry today offer new opportunities for sensing the earth surface, discussed in the next section.

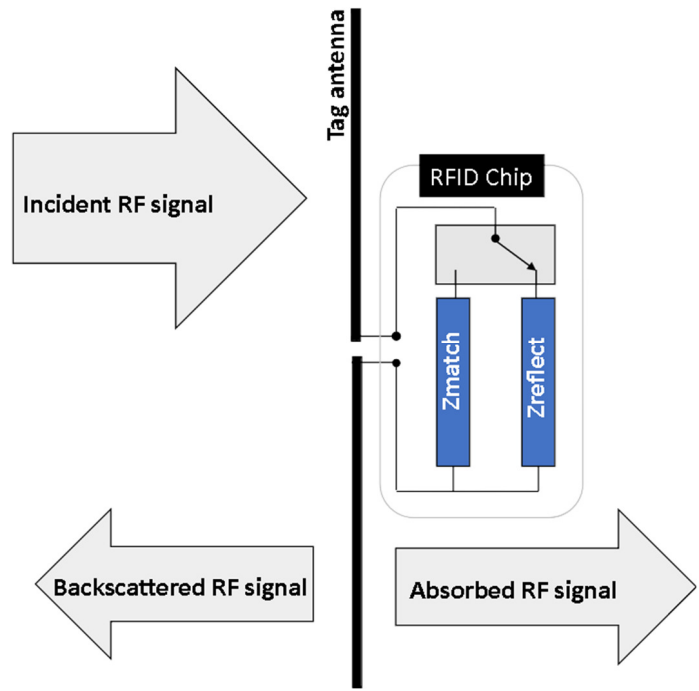

**Fig. 1: When receiving an incident radiofrequency signal, Ultra-High Frequency (UHF) tags send information to the reader by switching the impedance of the RFID chip between two states: Zmatch which maximizes the absorbed power (necessary to power up the chip) and Zreflect which maximizes the reflected power (In practice, both states reflect and absorb something). Resulting alternance of backscattered power level is interpreted by the reader as a digital signal. The performance of the RFID chip depends both on its ability to power-up using the absorbed power, and on the reader's ability to distinguish between the reflected signals of the two states.**

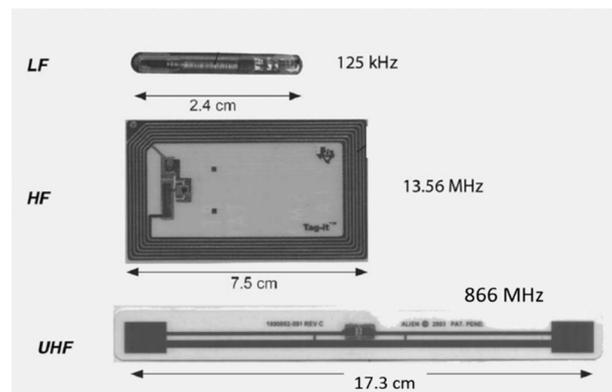

**Fig. 2: Main types of passive RFID tags, working by magnetic coupling at low-frequency (LF) and high frequency (HF), or by backscattered wave propagation (UHF). Turns refer to the number of loops of the conductor for the antenna. From (Dobkin, 2008)**

## 1.2 The interest of passive RFID in geosciences

Passive RFID is widely used to identify goods for logistics, transportation and retail. Today, the implementation of ambitious concepts such as the Internet of things (Xu et al., 2014) is opening the technology to completely new applications (Duroc and Tedjini, 2018). Indeed, passive RFID tags can now offer, besides the identification function, new functions of sensing and of centimetric localization (Zannas et al., 2020; Li et al., 2019), whose research is continuously increasing since 2003 (Fig. 3). The addition of sensing an external parameter to tag identification capabilities enables the concept of ubiquitous sensing (Want, 2004) in many sectors, such as food transport (Bibi et al., 2017), health care (Bianco et al., 2021), industry (Occhiuzzi et al., 2019), agriculture (Maroli et al., 2021), smart cities (Fahmy et al., 2019) civil engineering (Zhang et al., 2017; Duan and Cao, 2020), and geoscience (covered by the present review).

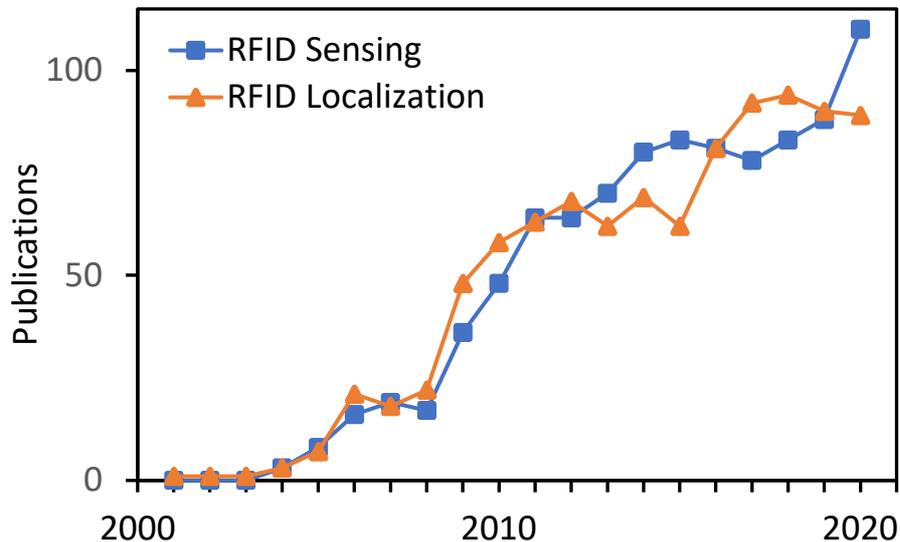

**Fig. 3: Number of publications listed in Web of Science for RFID sensing and RFID localization. The search uses the Expanded and Emerging Citation Index, with the words [RFID+Radio frequency identification] combined with [Sensing+Sensor] or [Localization+Location+Positionning] in the title and abstract. Inspired from (Zhang et al., 2017)**

In geosciences, the augmented functions of passive RFID tags offer new ways to deploy and operate sensors. The data from RFID tags is collected either with fixed or mobile readers (Fig. 4, e-f). Fixed readers can continuously interrogate the tags that are within its reading area. Readers can read tags passing near a portal (e.g., placed at a river section), or tags staying within a larger zone (e.g., on a landslide). However, using fixed readers requires to install and maintain a permanent infrastructure, and one reader can cover only a limited area. On the contrary, mobile readers can be moved around to scan a large area, either manually or with a vehicle, a robot or an unmanned aerial vehicle (UAV). Therefore, RFID can complement wired sensors, wireless sensor networks (WSN), remote sensing and manual operation of sensors (Fig. 4, a-d).

WSN specifically, in which sensing nodes communicate wirelessly one to another up to a central node, play a major role in the Internet of things. WSNs have become a mature technology in earth science both for operational and research monitoring (e.g., global positionning systems (GPS), micro-seismicity, tilting, soil moisture, strain). WSNs are also evolving towards lower cost of materials and lower energy consumption (Wixted et al., 2017): they reach material costs of typically 200-2000€ per node with battery lifetime of 1 – 10 years. Compared to WSNs, RFID sensors offer a longer lifetime and a lower cost per sensor, but has smaller read distance (see 6.3), fewer functionalities, and less standardization regarding sensing techniques (Atzori et al., 2010). As an example, WSNs deployed and maintained on geotechnical sites consists typically of 5 – 50 nodes. Yet, operating a large-scale WSN with hundreds of nodes during decades remains expensive due to the cost of material and maintenance. Passive RFID appears more competitive under these conditions due to its low-cost both in materials and maintenance. Besides, conventional RFID is well standardized and obeys several regulations and norms (sometime specific to geographic regions). This favors the interoperability of RFID components from different providers and the rapid implementation of RFID in real applications worldwide. The application that are very sensitive to the cost per sensor (material and long-term maintenance) should particularly benefit from passive RFID over WSNs.

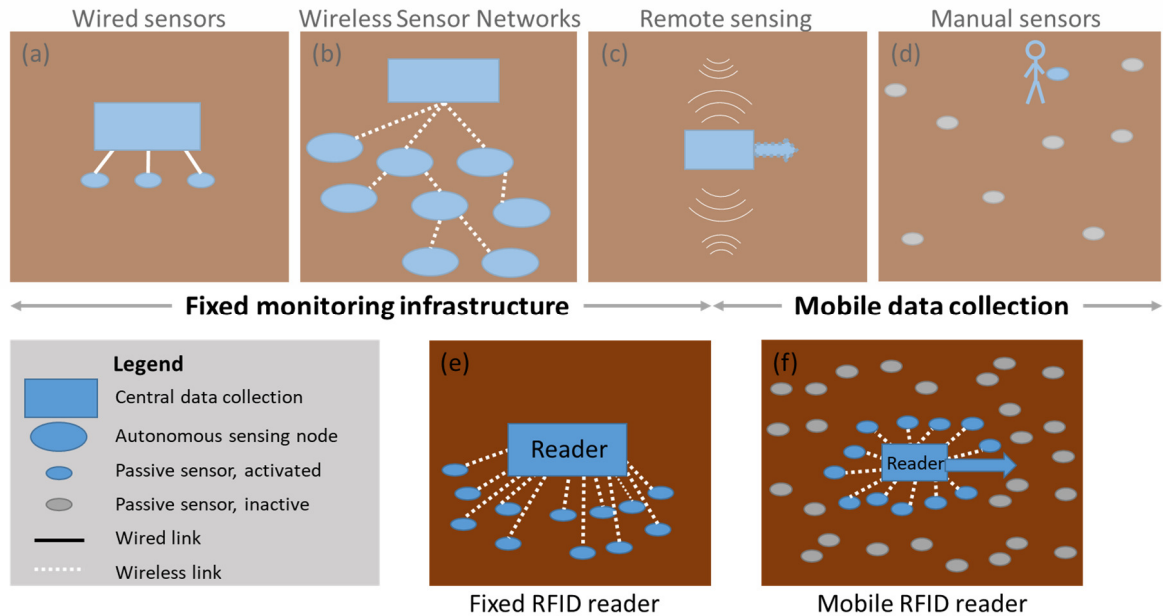

**Fig. 4: (a-d) Schematics ways to collect data from sensors in geosciences. Fixed monitoring infrastructure allows for continuous data, whereas mobile data collection allows for lightweight infrastructure. The blue square represents an autonomous acquisition system that stores the data or communicates it to a server over remote network (3G, Sigfox…). The sensing units are either dependent to the central acquisition device (small blue round) or autonomous in terms of data acquisition, communication, and powering (larger blue rounds). The communication is either wired or wireless communication link. The main deployment used for monitoring of earth surface processes today consists in (a) wired sensors connected to an acquisition system; (b) wireless sensor networks of autonomous nodes which communicate wirelessly one to another locally up the main node; (c) remote sensing units, either mobile or fixed, that acquire data from the environment reflections or from passive reflectors; (d) manual collection of data measured with non-communicating sensors (installed permanently or carried by an operator) that need a human operator. RFID systems allow for new deployment possibilities, with dense arrays of low-cost wireless tags and rapid data collection. RFID adds two new topologies: (e) a fixed reader which reads passive tags continuously within its read range, or (f) a mobile reader which collects data over a larger area with lightweight infrastructure that consist only in the passive sensors.**

Passive sensors in general—including RFID—allow affordable deployment of many sensors for long duration, up to decades. Sensors operated manually or with remote sensing (e.g., laser reflectors, radar reflectors, and simple visual sensors) also present low-cost, zero-energy consumption, and reduced maintenance. However, their data collection requires either a manual reading or complex remote sensing instruments. Compared to other passive sensors, RFID tags exhibit several advantages. First, they are produced industrially at a large scale, allowing for reliable, interoperable and low-cost tags. Second, their standard communication protocols ease the data collection of hundreds of tags almost simultaneously. Third, unique identification number of each tag eases the data inventory of large-scale networks. Fourth, RFID tags offer the capability to switch easily from manual data collection to automatic continuous monitoring (see Fig. 4); in geotechnics for example, it is common to run manual surveys over a large area when the risk is low, and to install locally a continuous monitoring if the risk increases (Burland et al., 2012). Finally, RFID also allows for quick data collection by an operator, with a result that is independent of his skills.

Geoscientists have used RFID progressively, following the technologies available on the market (see Fig. 5). First, they inserted low-frequency tags (LF: 125 kHz) into pebbles, to trace their mobility during flow events in rivers. The tags are later detected with either manual surveys or permanent detection portals at short distance from the interrogator (<1 m) (Lamarre et al., 2005; Schneider et al., 2010), to measure their motion between surveys. More recently, RFID tags have been used to measure the displacement of slow-moving landslides (0.01–10 m/year) during several months, with a centimetric accuracy (Le Breton et al., 2019). The tags worked at ultra-high frequency (UHF, 868 MHz) by means of backscattering waves, at a long distance from the interrogator, up to 60 m in 2018 with commercial off-the-shelf devices, and currently growing (see 6.3). Augmented tags can, in addition, provide information on their environment (Zannas et al., 2020). They were used to monitor soil temperature (Luvisi et al., 2016), soil moisture (Pichorim et al., 2018; J. Wang et al., 2020; Deng et al., 2020), displacement of rock structures (Le Breton et al., 2021), and snow water equivalent of dry snow (Le Breton, 2019). Also, potential applications on rock slopes could use sensors developed for measuring crack opening (Caizzone and DiGiampaolo, 2015), vibrations (Jayawardana et al., 2016, 2019) or tilt (Vena et al., 2019). Finally, RFID tags can be deployed in wider areas thanks to an increasing reading range (Durgin, 2016; Amato et al., 2018), the interrogation by an unmanned aerial vehicle (Ma et al., 2017; Buffi et al., 2019; Wang et al., 2015), and the deployment of tags in harsh areas such as buried under the ground (Abdelnour et al., 2018), or below snow or vegetation (Le Breton, 2019). This article presents how RFID technologies were applied for geoscientific and geotechnical monitoring, in the laboratory or in the field outdoors. Then it discusses the key technical challenges that may increase the value of the technique in regards of geoscientific applications.

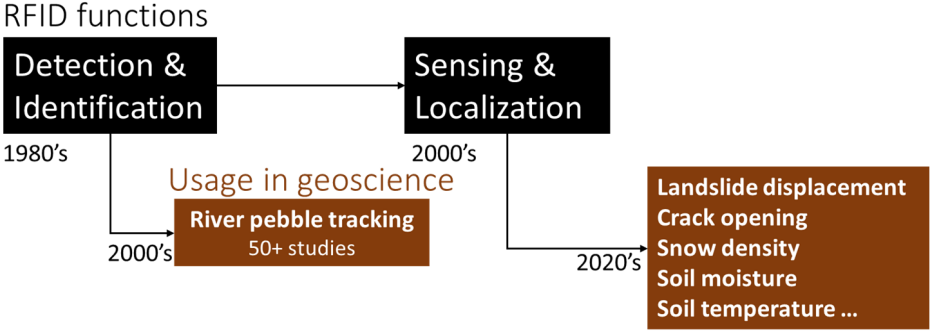

**Fig. 5: Overview of the RFID functions and their application to monitor earth surface dynamics, with the period of appearance in research.**

### 1.3 Sensing tags

Numerous applications have turned tags into wireless sensors (Marrocco, 2010; Costa et al., 2021) by augmenting its capabilities (Tedjini et al., 2016). The information provided can go from simple 1-bit threshold sensing to more quantitative measurements. In our opinion, the techniques of sensing tags fall into three categories: dedicated sensors, antenna-based sensing, and recently propagation-based sensing (see Table 1).

**Table 1: Categories of RFID sensing approaches, and their characteristics**

| | Dedicated sensor | Antenna-based sensing | Propagation-based sensing |
|---|---|---|---|
| Sensing examples | —Temperature<br>—Displacement threshold<br>—Tilt<br>... many more examples | —Moisture<br>—Chemical<br>—Proximity<br>—Temperature | —Human activity<br>—Object vibration<br>—Snow density<br>—Vegetation moisture |
| Data used | Value in chip memory | —Received Signal strength<br>—Phase<br>—Self-tuning code | —Received Signal strength<br>—Phase |
| Physical principle | —Analog sensor (embedded sensor, embedded analog-to-digital converter, external microcontroller)<br>—Digital sensor (serial bus, contact pins) | —Material near the tag which position or permittivity is changed.<br>—Tag antenna alteration (shape, permittivity of substrate, actuator) | —Attenuation in medium<br>—Phase delay<br>—Multipath interferences<br>—Depolarization, scattering |
| Measured zone | Sensor vicinity | Tag vicinity (0 – 0.1 m) | Far field (0.1 – 100 m) |
| Specific Constraints | Few chips can use sensors; The sensor needs extra power. | The tag antenna often needs adaptation. | The location of tags and reader must be adequate. |
| Comments | Often the most accurate. More expensive tags that consume more energy. | Often less accurate, can provide qualitative insights.<br>More suitable for threshold values | Remote measurement over large areas.<br>Can use any commercial tag. |

Connecting a dedicated sensor to the tag is the first way to transform RFID tags into sensors. In this approach, the chip powers the sensor, acquires the data, and sends the data numerically over the RFID radio channel. This approach is the most conventional and often the most accurate. Indeed, the sensor is designed to be sensitive to the measured parameter, and should note be influenced by the quality of the communication channel (however, in practice, the received power may influence the measured value on batteryless tags). We group the dedicated sensors into three categories: the sensors built into the RFID chip, the sensors connected to an input on the RFID chip, and the sensors connected to an intermediate microcontroller (see Table 2). The sensors directly built into chips are the simplest to use but offer few options for measurements (today's chips can measure only temperature, strain or hall effect). In contrast, the sensors connected to an RFID chip input allow many more possibilities: switch and encoders can connect on anti-tampering and general-purpose input/output (GPIO) contact inputs (e.g., Zhu et al., 2021); resistive and capacitive sensors can connect on analog inputs (e.g., Pichorim et al., 2018; A. Abdelnour et al., 2019); digital sensors can connect as slaves on digital communication buses such as SPI (e.g., Vena et al., 2019). Finally, the sensors can be connected to an external microcontroller, which acquires the sensed data and writes it into the RFID chip memory through a digital communication bus. A microcontroller adds new sensing capabilities such as multi-sensors, better time sampling, better sensitivity or pre-processing (e.g., De Donno et al., 2014; Jayawardana et al., 2016). But it also increases the complexity, cost and power consumption of the tag. Overall, dedicated sensors attached on RFID tags allow measuring many physical parameters, such as temperature, light, acceleration, dielectric constant, humidity, conductivity, strain, pressure, and much more. The counterpart is that they require to design a dedicated tag for each sensed parameter (Węglarski and Jankowski-Mihułowicz, 2019). Their exploitation to monitor natural earth processes is covered in more details in the next sections.

Table 2: Selection of UHF RFID chips with extra sensing functionalities

| RFID Chip (launch date) | Read Sensitivity (dBm) Passive | / BAP | Built-in Sensors and data logging | Input sensing interface Via microcontrol | Digital | Contact | Analog | Self-tuning |
|---|---|---|---|---|---|---|---|---|
| Impinj MonzaX (2012) | −19 | | | I²C | | | | |
| NXP Ucode I2C (2011) | −18 | −23 | | I²C | | | | |
| SERMA PE3001 (2009) | −6 | | | SPI | | | | |
| Fujitsu MB97R8110 (2019) | −12 | | Logger 28 kbit | SPI | SPI | 3 pin | | |
| Farsens Rocky100 (2017) | −14 | −35 | Temperature | SPI | SPI | 5 pin | | |
| EM-Micro EM4325 (2012) | −7 | −31 | Temperature | SPI | SPI* | 4 pin | | |
| Axxon AZN50x (2022) | -14 | -22 | Temperature; Logger 4k values | | | 2 pin | | |
| NXP Ucode GiM+ & GiL+ (2010-2011) | −18 | −27 | | | | 1 pin | | |
| Asygn 3213 (2020) | −12 −13 | -16 | Temperature; strain; Hall effect. | | | | 12 bit | |
| AMS SL900A (2013) | −7 | −15 | Temperature; battery level; Logger 1 kbit. | | | | 10 bits (×2) | |
| Axxon Magnus S3 (2015) | −16 | | Temperature; Tag RSSI | | | | | 9 bit |
| Impinj Monza R6 (2012) | −22 | | | | | | | 3 bit |
| NXP Ucode 8 | -23 | | | | | | | 2 bit |

Antenna-based sensing is the second way to transform RFID tags into wireless sensors. Indeed, any change in the shape, substrate properties, or immediate vicinity (near-field region) of the tag antenna, may also change the antenna impedance. As the impedance matching between the tag antenna and the RFID chip governs the performance of RFID tags, the modification of the antenna impedance will alter the parameter of the tag-to-reader communication. Such alteration can be measured on different communication parameters, in particular the maximum read range, the received signal strength indicator (RSSI), the resonance frequency, the phase shift (Caccami et al., 2015), the group delay and the power of activation. This concept of antenna sensitivity, which can provide extremely low-cost sensing tags, has been exploited in several publications. Bhattacharyya et al. (2009) exploited the perturbation of a tag antenna by a metal plate to measure the inflection of a bridge. Other tags exploited the deformation of their antenna caused by external strain (Occhiuzzi et al., 2011a) or temperature (Bhattacharyya et al., 2010b), for instance. Occhiuzzi et al. (2011b) loaded the antenna with carbon nanotubes which are sensitive to gas concentration. Manzari et al. (2014a) added a thermistor on the antenna. Nguyen et al. (2013) measured the dielectric permittivity of meat in the food industry. Antenna-based RFID sensors were reviewed, for general applications (Occhiuzzi et al., 2013) and also more specifically for structural health monitoring (Zhang et al., 2017). We emphasize that most of the published papers on antenna-based RFID sensors focus on the concept of the device and on the validation of the proposed structure. The effective variation of the backscatter signal with the sensed parameters is usually demonstrated with experimental laboratory measurements. Yet, using the sensor in real applications would request to mitigate several influences. For example, the signal amplitude strongly depends on the tag orientation and on multipath interferences. Besides, the phase is sensitive to the tag position (Nikitin et al., 2010) and to the environment (Le Breton et al., 2017). In practice, the use of antenna-based sensing would often require calibration, obtained for example by associating some reference devices to correct the RFID measurements (Marrocco, 2010). Additionally, several methods demonstrated in the laboratory, such as frequency sweeping, will also be highly restricted by regulation (e.g., ETSI-302-208 from the European Telecommunication Standards Institute). Therefore we deduce that antenna-based sensing should be limited to the use as a threshold indicator, to distinguish between two states (or three sometimes) of a system.

More recently, self-tuning RFID chips were exploited to directly measure the changes in the antenna impedance. These chips have the capability to stay tuned to the antenna impedance, and transmit to the reader a value indicating the exact amount of impedance correction to be made by the chip. We consider this recent case as antenna-based sensing because the tag antenna is the sensor, yet it also has some advantage of dedicated sensing, because the measurement is made by the chip. Since the antenna impedance is influenced by materials at its vicinity, this correction value can thus be used to estimate parameters, such as sensing moisture (Swedberg, 2015), chemical composition (Caccami and Marrocco, 2018), temperature (Zannas et al., 2018), or electric potential (Nanni et al., 2022). Self-tuning measurements are made directly by the tag, and therefore are more independent to unwanted influences (distance, multipath, orientation) than phase and RSSI sensing indicators, given that they receive the adequate signal power (Caccami and Marrocco, 2018).

Propagation-based sensing is a third way of RFID sensing. It consists in measuring the influence of the propagation environment on the RFID signal, to estimate changes of properties in the propagation channel. These changes occur through phenomena of direct wave transmission, depolarization, or multipath interferences. The first phenomenon concerns the direct wave transmission: a change in the medium placed between a tag and the interrogator antenna will modify the phase or RSSI measured by the interrogator. Such change was exploited to detect a human presence (Hussain et al., 2020), estimate the quantity of fresh dry snow (see 5.3) (Le Breton et al., in preparation), or estimate the water content of a vegetal layer (Le Breton, 2019). The second phenomenon is the depolarization caused by scattering. For example, He et al. (2020) proposes to cross-polarize a tag and reader antenna at a perpendicular orientation. When no scattering occurs, most of the signal is lost due to the cross polarization, which cancels the communication. However, in the presence of a scattering body nearby, the linearly polarized wave partly depolarizes into a perpendicular component, reestablishing the communication. This can detect the presence of a scattering body, such as a human body. The last phenomenon, multipath interferences, can cause changes in the signal phase and strength from only tiny changes in the surroundings of a tag. It was used to detect the vibration of an object near a tag and measure the vibration frequency (Yang et al., 2020). Overall, propagation-based sensing seems of particular interest for geoscience, because it can estimate average properties of a complex medium over large distances and do not require customized tags. Propagation-based sensing principles are also very similar to geophysical methods such as ground-penetrating radar, polarimetric radar or GPS interferometry (Bradford et al., 2009; Kim and Zyl, 2009; Larson, 2016; C. Lin et al., 2016); the main difference is that the reflectors (the tags) are easily identified but need to be placed in advance. Propagation-based sensing has provided a decent measurement accuracy for snow sensing and for vibration sensing. However, the tags and reader must be placed adequately, for example to allow communication across the snow, or reflections on the vibrating object.

The accuracy of RFID sensing is often a challenge compared to traditional sensing systems. Obtaining accurate measurements may require special care. For example, that often requires to mitigate the influences due to environmental fluctuations, such as temperature, rain, snow or icing (Le Breton et al., 2017; Le Breton, 2019). Another care can consist in controlling the signal power delivered by the interrogating antenna (e.g., Caccami and Marrocco, 2018; Camera and Marrocco, 2021). In regard to accuracy, antenna-based sensing seems the most difficult approach, because it is sensitive to many influence parameters. Therefore, in our opinion, it should be used only to determine the qualitative state of a system, for instance a binary state: on/off system, open/closed circuit or below/above a threshold. Dedicated sensing and propagation-based sensing seem the most appropriate for geoscience observations. In addition, combining different sensing approaches can increase the accuracy. For example, the measurement of phase by the reader can be calibrated against the temperature measured by the tag (Le Breton et al., 2017). In addition, data provided by sensors can be calibrated against the on-chip received signal strength (Camera and Marrocco, 2021).

To conclude this section, the technologies available today allow for the design of effective RFID devices that enable innovative solutions for environmental sensing. To add new application of RFID sensors and exploit the very competitive cost of RFID tags, it has been proposed early to map large spaces with numerous tags (Capdevila et al., 2010). Moreover, industrial companies are planning the future potential of RFID sensing, for example with the definition of the RAIN RFID standards "a wireless technology aimed at connecting billions of everyday items to the Internet, enabling businesses and consumers to identify, locate, authenticate, and engage each item" (*Rain RFID eBook*, 2020).

### 1.4   *Localization of tags*

In addition to sensing, tags can also be located in space, using their signal amplitude or the variations of phase delay. The simplest method is to detect a tag and conclude that it is nearby, within the detection range of the interrogator. The method has been used extensively to track the displacement of pebbles in rivers (see section 2) using low-frequency tags with sub-metric detection range. This technique has, nevertheless, a spatial resolution that depends on the detection range, of several decimeters to meters. The localization of tags at longer distances uses UHF tags (Miesen et al., 2011), and exploits the received signal strength (RSSI) or the phase of arrival.

Signal-strength-based methods have been initially introduced for tag localization (Griffin and Durgin, 2009; Ni et al., 2003), followed by phase-based methods which usually offer the best accuracy, reaching one centimeter or less (Scherhäufl et al., 2015; Wang et al., 2016; Zhou and Griffin, 2012). In practice the phase-based methods show the best potential for observing ground displacements (Le Breton et al., 2019). In this review we will present phase-based methods only, which we consider more adapted among other localization methods (Balaji et al., 2020). Backscattering communication used in passive RFID can easily measure the phase difference arrival because the reader transmits and receives the same carrier wave, compared to active RFID which uses two distinct RF front-ends on the tag and reader.

The phase of arrival in free space $\varphi$ relates to the delay of propagation, and depends on the speed of light in the air c, the distance between the tag and the reader r , the carrier frequency f, and an offset $\varphi_0$ caused by the devices:

$$\varphi = -\frac{4\pi f}{c} r + \varphi_0 \qquad (1)$$

In practice, the phase of arrival measured by the reader is ambiguous, wrapped within $0-\pi$ or $0-2\pi$ (Miesen et al., 2013b), reducing its intrinsic value as a single measurement. However, the Phase Difference of Arrival (Vossiek and Gulden, 2008; Nikitin et al., 2010) $\delta\varphi$ measured between two different phase readings $\varphi_n$ is exploitable for localization. It relates to the change in tag-reader radial distance $\delta r$ that occurred between the two reading positions :

$$\delta\varphi = \varphi_2 - \varphi_1 = -\frac{4\pi f}{c} \delta r \qquad (2)$$

Where represents the measured phases at different positions related. Note that this equation holds only for phase shifts smaller than the reader measurement ambiguity ($\pi$ or $2\pi$).

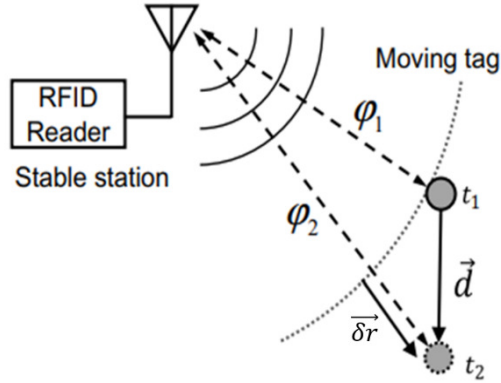

**Fig. 6: Schematic description of the Phase-Difference of Arrival measurement principle. A displacement $\vec{d}$ between two positions $t_1$ and $t_2$ implies a variation in the phase measurement from $\phi_1$ to $\phi_2$ linked to the radial displacement $\delta r$. From (Nikitin et al., 2010; Le Breton et al., 2019)**

Simplest usage of this approach estimate 1D distance or displacement between a tag and a station antenna (see Fig. 6). Optionally it can locate the tag in 2D or 3D using multiple station antennas and trilateration (Scherhäufl et al., 2015). The two main phase-based localization schemes exploit either the variation of the phase in time which is a relative localization technique, or the variation of phase with signal frequency which allows for absolute ranging. The latter shows less accuracy and seems not suited to track slow displacements (Le Breton et al., 2019).

Using arrays of tags allows estimating the tilt of the array relatively to the station antenna (by comparing the phase differences between tags) (Fig. 7). For reciprocity reasons, this approach also holds for arrays of antenna facing fixed tags (Scherhäufl et al., 2015). The approaches mentioned previously allow for a centimeter-scale error for punctual measurements as well as month-long monitoring in outdoor conditions (Charlety et al., in preparation).

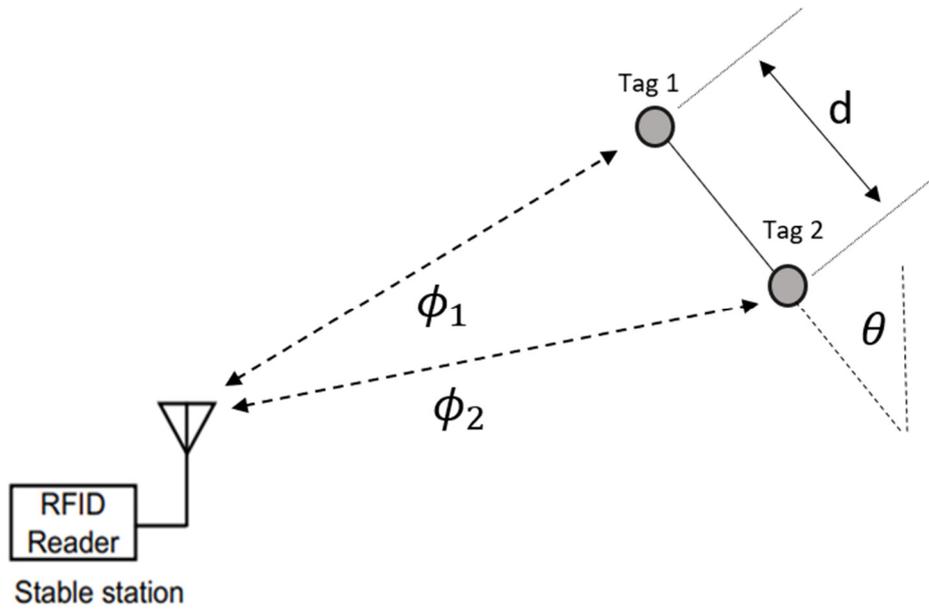

**Fig. 7: Schematic of a uniform linear array consisting of 2 tags separated by a known distance $d$. The difference between the measured phases $\varphi_1$ and $\varphi_2$ is related to the array angular orientation $\theta$, also called tilt. This approach is used for localization and tilt measurements. Inspired from (Nikitin et al., 2010)**

The synthetic aperture radar (SAR) is another use of phase measurements, used typically for satellite or ground-based radars. Applied to RFID, it implies moving one or more reader antennas along known trajectories (Fig. 8) (Buffi et al., 2019; Wu et al., 2019), often carried by a mobile robot, a UAV (see Section 6.4). Most SAR use a similar theoretical approach, which consists in the minimization of a cost function comparing real phase measurements and simulated measurements computed from the antenna trajectory and a probable tag position (Motroni et al., 2018). Optimization algorithms like particle swarms or Kalman filters are often used to decrease the computational cost of such calculations (Bernardini et al., 2020; Gareis et al., 2020). SAR localization has proven to reach a centimeter-scale precision. However, the shape of the reader trajectory has a strong impact on the localization accuracy, and has to be chosen carefully to locate the tag in all investigated dimensions (Bernardini et al., 2021).

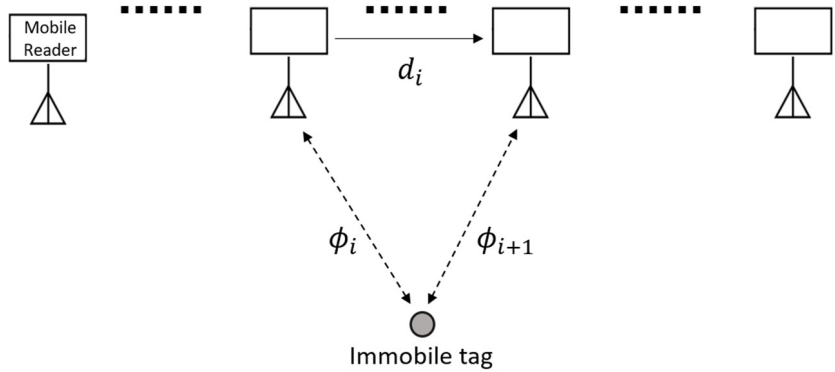

**Fig. 8: Principle of the Synthetic Aperture Radar localization approach. A series of phase measurements $\phi_i$ is performed along a given reader trajectory, with known displacements $d_i$ between consecutive measurements. This approach assumes a constant tag position. Inspired from (Wu et al., 2019)**

One of the main limitations of RFID phase localization is the phase wrapping ambiguity and the influence of multipathing (Faseth et al., 2011; Ma et al., 2018; DiGiampaolo and Martinelli, 2020). A way to reduce these effects is the Time of Flight method (Arnitz et al., 2010; Arthaber et al., 2015), which avoids ambiguity as the full-time shift is measured. It can also reduce the influence of multipath when the time of first arrival is measured, and when a wide-band pulse is used. However it requires custom devices that are not on the market nor standardized. Nonetheless, RFID localization using standard commercial equipment has proven doable and is currently a flourishing research subject, with a centimetric to decimetric spatial resolution.

To conclude, RFID localization is nowadays investigated in various fields of application and through several approaches. It shows great capability and adaptability both in indoor and outdoor environments, with multiple ways to bypass the measurement limitations. More developments in this domain are surely to come.

## 2    Mobility of coarse sediments in rivers and coasts

In earth sciences, RFID tags (also called PIT tags, for Passive Integrated Transponders) were first used to track the mobility of coarse sediments in rivers during flow events for the study of bedload transport (Nichols, 2004; Lamarre et al., 2005). Then they were rapidly applied to study the transport of coarse sediments in coasts (e.g., Allan et al., 2006; Curtiss et al., 2009; Osborne et al., 2011). Other applications were reported to track the transport of woody debris in rivers (MacVicar et al., 2009; Ravazzolo et al., 2015) and boulders in debris flow (McCoy et al., 2011; Graff et al., 2018). But these applications have stayed relatively limited compared to fluvial bedload tracing and coastal sediment tracing, for which around 60 papers and 15 have been published between 2004 and 2021, respectively.

### 2.1    Design and implementation of RFID tracing experiment

The very first field experiments with RFID tags in rivers showed high recovery rates (>80%) for relatively large populations of deployed tracers (>100) (Nichols, 2004; Lamarre et al., 2005). Soon after these early promising results, RFID tags rapidly became the reference technology for bedload tracing in rivers, and they progressively replaced the most commonly used techniques, such as painted or magnetic stones (Hassan and Ergenzinger, 2003). We can estimate from the literature that during the last 20 years, several tens of thousands RFID tracers have been injected in rivers all around the world. This wide adoption of passive RFID tags in sediment transport studies is based on several decisive advantages: they provide low-cost (<5 €), long-lived (>10 yr), and small size (<35 mm) tracers, that can be remotely identified under several decimeters of sediments. Although buried magnetic stones can also be detected with recovery rates comparable to RFID tags, their identification is only possible after time-consuming and intrusive surveying of alluvial deposits. The only alternative solution for remote identification of tracer stones today are the battery-powered radio transmitters (e.g., Habersack, 2001). However, the need for a battery reduces the duration of the experiments, and their higher cost reduces the number of tracers deployed.

The most commonly used passive RFID tags in sediment transport studies are glass-encapsulated waterproof low-frequency (134.2 kHz) tags of 23 mm or 32 mm length. Transponders are generally inserted into rocks by drilling or notching natural stones (Fig. 9a), but artificial stones can also be used (e.g., Schneider et al., 2014; Olinde and Johnson, 2015). The range of investigated grain sizes is constrained by the size of the transponder, and the smallest tracers are generally comprised between 20 mm and 40 mm in intermediate axis (*b*-axis). Once equipped with a tag, RFID tracers are deployed in river channels, and may be transported by one or multiple flow events. Their displacement is surveyed after a period by the manual detection and mapping of displaced tracers using a mobile antenna and a positioning system (e.g., differential GPS, total station) (Fig. 9b). The limited detection range of low-frequency RFID tags (<1 m) implies time-consuming field surveys. Another strategy is to use a stationary antenna at a cross-section to automatically detect RFID tracers crossing the monitoring section (Schneider et al., 2010; Mao et al., 2017; Stähly et al., 2020; Casserly et al., 2021). RFID tracing experiments in rivers are typically based on hundreds of tracers surveyed at seasonal or yearly time intervals. The quality of data from tracing experiments strongly depends on the recovery rate, which is generally above 70% for small streams, but which strongly decline for large rivers (see compilations of recovery rates in Chapuis et al. (2014, 2015).

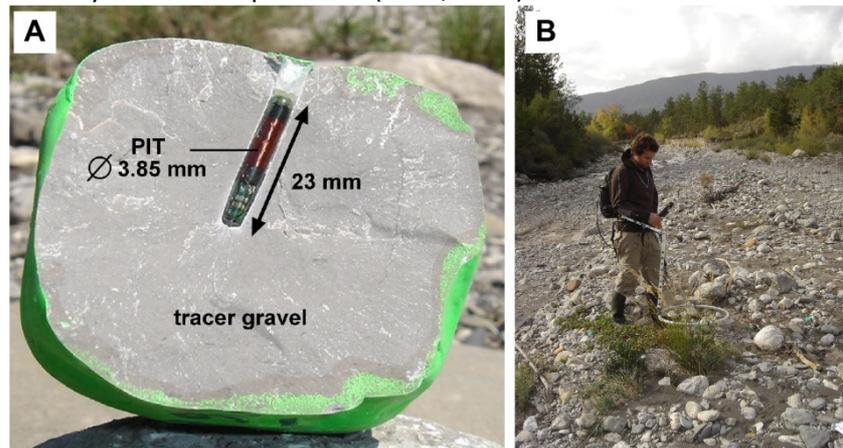

**Fig. 9: (a) Tracer gravel equipped with a glass-encapsulated low-frequency passive RFID tag and (b) mobile antenna used for the remote detection of RFID tracers in river channels. Adapted from (Liébault et al., 2012).**

### 2.2   Sediment transport studies using RFID tags.

One of the most popular applications of RFID tags in fluvial sediment transport is the study of grain dispersion by flow events along river reaches. This grain-scale approach of bedload transport has been conceptualized by Einstein in the late 1930s, who considered bedload as a random process of individual particle displacements, resulting from a succession of step lengths and rest periods, exponentially distributed (Einstein, 1937). RFID tracers have therefore been widely used to constrain the distribution of cumulative bedload transport distances in a variety of fluvial environments, including small and medium boulder-bed streams (Lamarre and Roy, 2008a, 2008b; MacVicar and Roy, 2011; Phillips et al., 2013; Schneider et al., 2014; Olinde and Johnson, 2015), plane-bed (Imhoff and Wilcox, 2016; Papangelakis et al., 2019a), wandering or braided gravel-bed streams (Bradley and Tucker, 2012; Liébault et al., 2012), and large rivers (Arnaud et al., 2017) (Fig. 10).

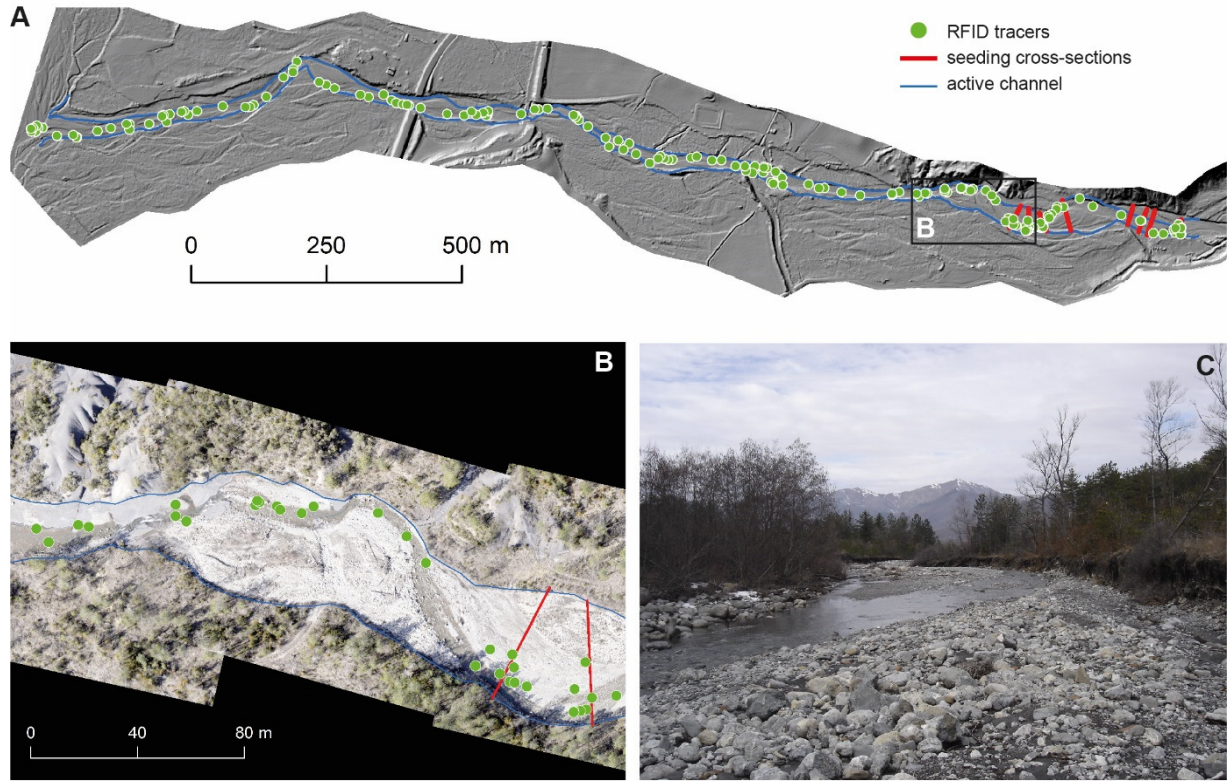

Fig. 10: Example of RFID tracer dispersion observed in a wandering gravel-bed river during a 3-month period after which frontrunners have been located at more than 2 km from their seeding location: (A) dispersion map on hillshade view of a LIDAR digital terrain model; (B) zoom on the position of recovered RFID tags in the active channel, showing preferential deposition of tracers close to the wet channel which can be viewed as the active bedload transport zone during the investigated period; (C) picture showing the morphological configuration of the active channel (adapted from Liébault et al., 2012)

Controlling factors of bedload dispersion have been investigated in several studies. The most commonly used hydraulic predictor for the mean displacement of the tracer population is the excess specific stream power (Lamarre and Roy, 2008b; Schneider et al., 2014; Houbrechts et al., 2015; Arnaud et al., 2017; Papangelakis et al., 2019a; Gilet et al., 2020). It is computed by considering the difference between the peak stream power during the survey period and the critical stream power for incipient motion of bedload transport. Some time-integrated flow predictors have been also proposed, like the impulse framework, a dimensionless impulse integrating the cumulative excess shear velocity normalized by grain size (Phillips et al., 2013; Phillips and Jerolmack, 2014; Imhoff and Wilcox, 2016; Gilet et al., 2020), or the cumulative stream energy (Schneider et al., 2014) and the effective runoff (Olinde and Johnson, 2015). Scaling laws of transport distances can be successfully fitted with flow predictors; however they remain site-specific because bedload dispersion in river channels is a complex phenomenon driven not only by the flow, but also by interactions between channel morphology and sediment transport.

RFID tags have been widely used to investigate grain-size effects on bedload dispersion in different fluvial environments, by comparing displacement lengths and/or virtual velocity of tracers of different sizes (MacVicar and Roy, 2011; Liébault et al., 2012; Schneider et al., 2014; Dell'Agnese et al., 2015; Rainato et al., 2018; Mao et al., 2020). This kind of analysis is particularly interesting for characterizing full and partial mobility regimes of bedload transport using data from tracer populations with wide grain-size distributions (Wilcock, 1997; Vázquez-Tarrío et al., 2018). RFID tags with wide grain-size distributions also offer the possibility to test in the field physically-based formulations of incipient motion conditions for bedload transport (Phillips and Jerolmack, 2014; Houbrechts et al., 2015; Petit et al., 2015; Ivanov et al., 2020; Roberts et al., 2020).

Morphological controls on bedload dispersion is another key topic addressed with RFID tags. A strong difference of mobility between tracers deployed in gravel bars and low-flow channels has been observed in a wandering gravel-bed river during flow conditions of moderate intensity. That suggests that for regular flow events, the contribution of gravel bars to bedload transport is limited (Liébault et al., 2012). Morphological controls on bedload dispersion can also be inferred by comparing transport distances with the mean longitudinal spacing between morphological units of the channel. It has been for example shown in small boulder-bed streams that the average distance between steps and pools or between steps controls displacement lengths (Lamarre and Roy, 2008b; Mao et al., 2020). Another approach is to combine tracer surveys with topographic monitoring and hydraulic modeling of the river channel to make inferences about links between morphodynamics and tracer entrainment and deposition (MacVicar and Roy, 2011; Milan, 2013; Chapuis et al., 2015).

Another RFID application in sediment transport concerns the tracking of debris flows in headwater channels. Only few attempts have been undertaken up to now, because the processes of channel scouring and filling during debris flows are much more intense than those related to bedload transport; RFID tracers may therefore become buried under sediment deposits much thicker than the detection range of passive RFID tags. Debris flows are also known as fast-moving mass movements typically propagating over long distances along rough, steep, and inaccessible channels, making surveying with mobile antennas particularly difficult. However, a recent tracing experiment in the French Alps with RFID tags revealed a recovery rate of 31% after a debris flow, with maximum travel distances of about 720 m (Graff et al., 2018) (Fig. 11). Another example of RFID tracking in a very active debris-flow torrent of the Southern French Alps revealed that medium-size boulders (70 cm diameter) can be transported over distances of 1.5 km during small debris flow events (Bel, 2017).

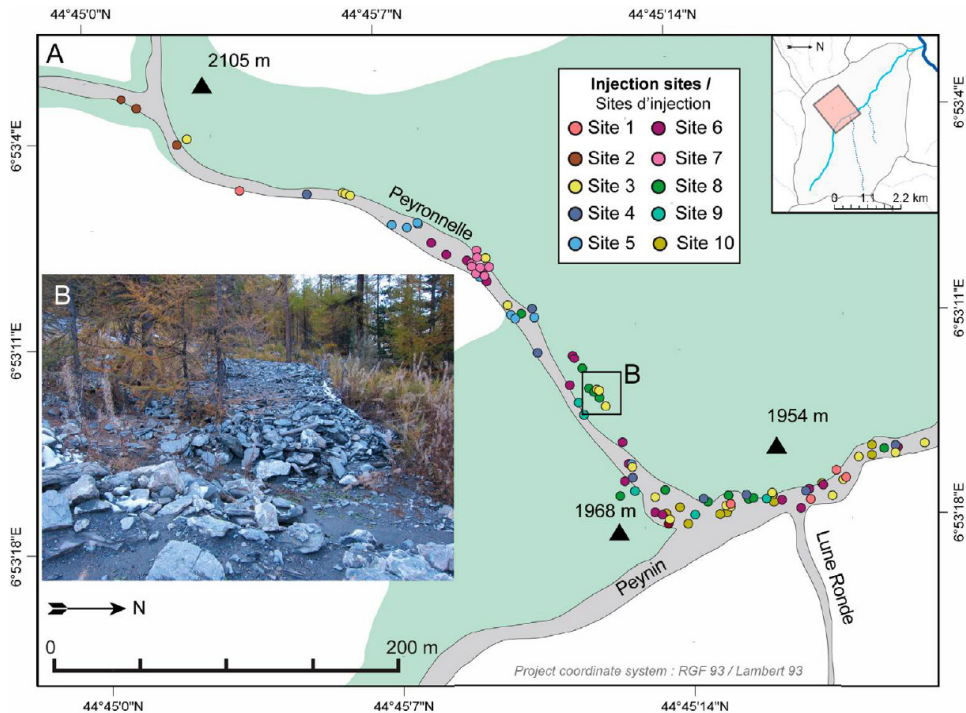

**Fig. 11: An example of passive RFID tracer dispersion induced by a debris flow in a small steep-slope torrent of the French Alps (Peyronelle Torrent); the debris fan appears in green, and active channels in gray; the picture in panel B shows a debris-flow lobe deposited in a secondary channel, in which several RFID tracers were detected. From (Graff et al., 2018)**

### 2.3   RFID tags applications in river management

Although RFID tags have been early used to assess the physical effects of hydraulic works dedicated to aquatic habitat restoration in river channels (Carré et al., 2007; Biron et al., 2012), their diffusion in the science of river management has stayed limited until very recently. The last four years have seen the emergence of numerous RFID tracing experiments dedicated to the monitoring of river restoration projects. Most of them are addressing the question of the propagation of gravels artificially introduced in river channels (sediment replenishment) to mitigate the morphological impact of dams (Arnaud et al., 2017; Brousse et al., 2020a; Stähly et al., 2020). But other tracing experiments were dedicated to the effect of weirs (Casserly et al., 2020; Peeters et al., 2020; Magilligan et al., 2021) or check-dams (Galia et al., 2021) on bedload transport, or to the comparison of restored and unrestored riffle-pool sequences in a small urbanized gravel-bed stream (Papangelakis and MacVicar, 2020). RFID tags have also been recently used in a restored braided river to assess the sediment continuity between the restored reach and its main upstream sediment source (Brousse et al., 2020b). All these recent studies reveal that the range of applications of bedload tracing with RFID tags is very large.

### 2.4   Coarse sediment tracking in coastal environments

Passive RFID tags have been early used to study coarse sediment transport in coastal environments (Allan et al., 2006; Bertoni et al., 2010; Curtiss et al., 2009). Although the number of reported field experiments (around 15) is much less than those of fluvial environments, coastal sediment transport represents an important application field of passive RFID in geosciences. PIT tags have been successfully employed to investigate longshore and cross-shore transport of cobbles along mixed sand-gravel beaches, in relation to marine currents directions, wave climates, and submerged artificial structures for beach protection (Allan et al., 2006; Curtiss et al., 2009; Bertoni et al., 2010; Miller et al., 2011; Bertoni et al., 2012; Dolphin et al., 2016). Seasonal patterns of sediment transport have been documented, by comparing tracer movements induced by different forcing mechanisms, such as wind waves, tides, and wakes (Curtiss et al., 2009; Osborne et al., 2011), or by detecting uni- or multi-directional movements as a function of incident wave angles (Miller et al., 2011, p. 20). Coastal RFID tracers also provide information on size-sorting effects and abrasion rates (Allan et al., 2006; Dickson et al., 2011), and on morphometric indicators of sediment flux, such as beach active width and mobile layer depth (Miller et al., 2011; Miller and Warrick, 2012). The tracing experiment of the Elwha River delta reported by Miller et al. (2011) was notably used to reconstruct volumetric sediment fluxes at two different sites with contrasting angles of breaking waves, showing more intense sediment transport for the site with less oblique angle, a field observation in contradiction with predictions from longshore sediment transport models. More recently, RFID tags have been used to study intertidal boulder transport during storm events, with direct implications for the morphological evolution of rocky coasts in the UK (Hastewell et al., 2019, 2020) and north-west Spain (Gómez-Pazo et al., 2021). Field studies reported in the UK notably revealed that very large boulders (more than 10 t) can be transported during contemporary storms of moderate intensity. Passive RFID in coastal environments was also employed to document pebble movements related to swash processes on low-energy beaches (Bertoni et al., 2013), and to typhoon conditions on a gravel beach (Han et al., 2016). The movement of detrital coral fragments has been also tracked along a fringing reef with passive RFID (Ford, 2014). Mobile RFID antennas commonly used in wadable river channels have been adapted for applications to coastal environments requiring underwater surveying. A waterproof RFID reader has notably been developed and included in a toolbox specifically designed for beaches (Benelli et al., 2012).

Table 3: Technical indicators on studies that tracked coastal sediments using RFID tags.

| Reference | Tags | Area (m²) | Recovery rate (%) | Duration of the survey | Survey repetitions | Transport max (m) | Comments |
|---|---|---|---|---|---|---|---|
| (Gómez-Pazo et al., 2021) | 80 | 2 500 | 50-75 | 39 mo | 6 | 20 | Large boulders |
| (Hastewell et al., 2019, 2020) | 104 | 30 000 | 91 | 19 mo | 15 | 21 | Large boulders |
| (Dolphin et al., 2016) | 940 | 81 387 | > 70 | 36 mo | 35 | 1300 | Motorized survey |
| (Han et al., 2016) | 200 | 2 500 | 37 | 4 mo | 2 | 27 | Uses UHF tags |
| (Ford, 2014) | 382 | 20 000 | 4 | 6 mo | 1 | 312 | |
| (Bertoni et al., 2013) | 145 | 2 000 | 90 | 24 h | 2 | 19 | |
| | 78 | 1 500 | 97 | 24 h | 2 | 5 | |
| (Bertoni et al., 2012; Benelli et al., 2012) | 102 | 1 500 | 52 | 2 mo | 1 | 82 | Underwater |
| (Dickson et al., 2011) | 180 | 250 000 | 22 | 7 mo | 8 | 2 500 | |
| (Miller et al., 2011) | 128 | 1 500 | 83 | 12h / 12 mo | 9 | 70 | |
| (Bertoni et al., 2010; Benelli et al., 2012) | 96 | 2 500 | 77 | 2 mo | 1 | 50 | Underwater |
| (Curtiss et al., 2009; Osborne et al., 2011) | 96 | 3 000 | ~ 80 | 14 mo | ~30 | > 50 | |
| (Allan et al., 2006) | 400 | 5 000 | 18–66 | 20 mo | 8 | 57 | |

# 3   Displacement of unstable terrains

## 3.1   *Landslide and soil displacements*

The displacement of landslides—typically a few centimeters to a few meters per year—is often spatially heterogeneous and irregular in time. Landslides sometimes display catastrophic acceleration phases that need to be monitored to obtain early warnings and reduce their risk to the population. Landslide monitoring greatly benefit from dense data in space and time (e.g., Intrieri et al., 2019; Lacroix et al., 2020). Several monitoring methods already exist (Angeli et al., 2000), such as GPS (Benoit et al., 2015; Gili et al., 2000), laser total stations, photogrammetry (Travelletti et al., 2012), LIDAR (Jaboyedoff et al., 2012), Radar interferometry (Herrera et al., 2009), or local radiofrequency networks (Kenney et al., 2009; Intrieri et al., 2018). Among them, radiofrequency techniques (GPS, radar, wireless sensor networks) are the most reliable over the year since they can operate during rainfall, snowfall, heavy fog, or in the presence of vegetation cover. Reducing the cost of reliable monitoring techniques is a challenge, emphasized in recent reviews related to landslide monitoring (Intrieri et al., 2019; Lacroix et al., 2020). For example, low-cost GPS solutions were developed (Squarzoni et al., 2005; Buchli et al., 2012; Benoit et al., 2015), but they still cost around a few k€ per measured point, which is too costly to monitor hundreds of points.

Recently, Le Breton et al. (2019) have monitored a landslide using localization methods of UHF RFID tags. The radial displacement (1D) of nineteen tags was measured relatively to a base station installed on a stable ground, continuously for 5 months, using the phase difference of arrival method (Nikitin et al., 2010). The measurements were validated against a long-range wire extensometer, and surveys using a total station. The RFID technique appears little sensitive to most environmental effects, particularly when using appropriate devices and calibration (Le Breton et al., 2017), as compared to the existing long-range wire extensometer. The accuracy of the RFID technique was 1 cm during normal weather and always less than 8 cm during the hardest conditions like heavy snow events, for which the long-range wire extensometer was no longer reliable. This work showed that RFID localization techniques can properly measure the ground displacements of a few centimeters up to several meters per year.

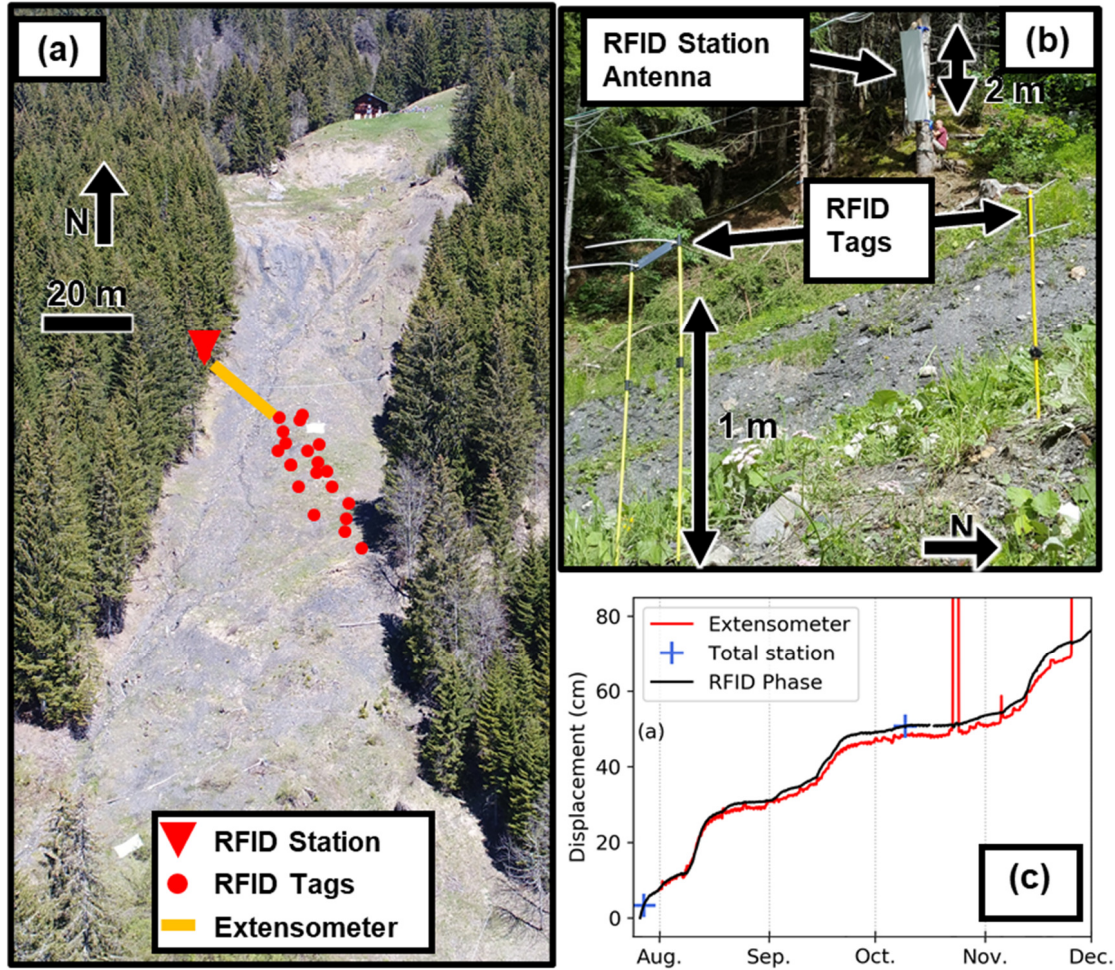

Fig. 12. Instrumentation of Pont-Bourquin landslide using RFID tags. (a) aerial view of the landslide, (b) picture of the instruments on the field, and (c) results of RFID measurements for one tag, validated by a wire extensometer and laser positioning. From Le Breton et al. (2019).

Since its validation on the Pont-Bourquin landslide (Fig. 12), some of the authors then deployed the method on several additional landslides, in order to test and validate various operational configurations (Charléty et al., 2022a). Another configuration was tested on the slope above the Rieu Benoit River, next to the Valloire ski resort, French Alps. In this case the goal was to monitor different sets of tags, in different directions, from one single RFID reader. We thus deployed two antennas in two different directions, connected to the same RFID reader device.

On the Harmalière landslide, a clayey slow moving landslide displaying a regressive character, a set of 4 antennas were deployed around the same RFID reader (see Fig. 13 for field configuration) with a total opening of 5 m, to track the tag positions in 2D or 3D (Charléty et al., 2022b). In this specific configuration, the group of antennas was deployed on a stable ground, parallel to the movement of the landslide, addressing the same set of about 20 tags. The poor radial sensitivity of a single antenna for phase change detection was there compensated with the use of additional antennas installed at different positions.

The Villa Itxas Gaianas landslide (Delbreil, south of France) is also instrumented since January 2022. Continuous RFID measurements and manual GPS surveys measured a motion of to 2 meters in six months. This setup confirmed the usage RFID displacement monitoring in coastal place, subject to strong wind and corrosive sea spray.

In these use cases, the read range (the maximum tag-station distance) appears critical to monitor large landslides. Owing to cumulative development, the read range has gradually increased from 60 m (2018) to 120 m (2020), and further on (see details in 6.3).

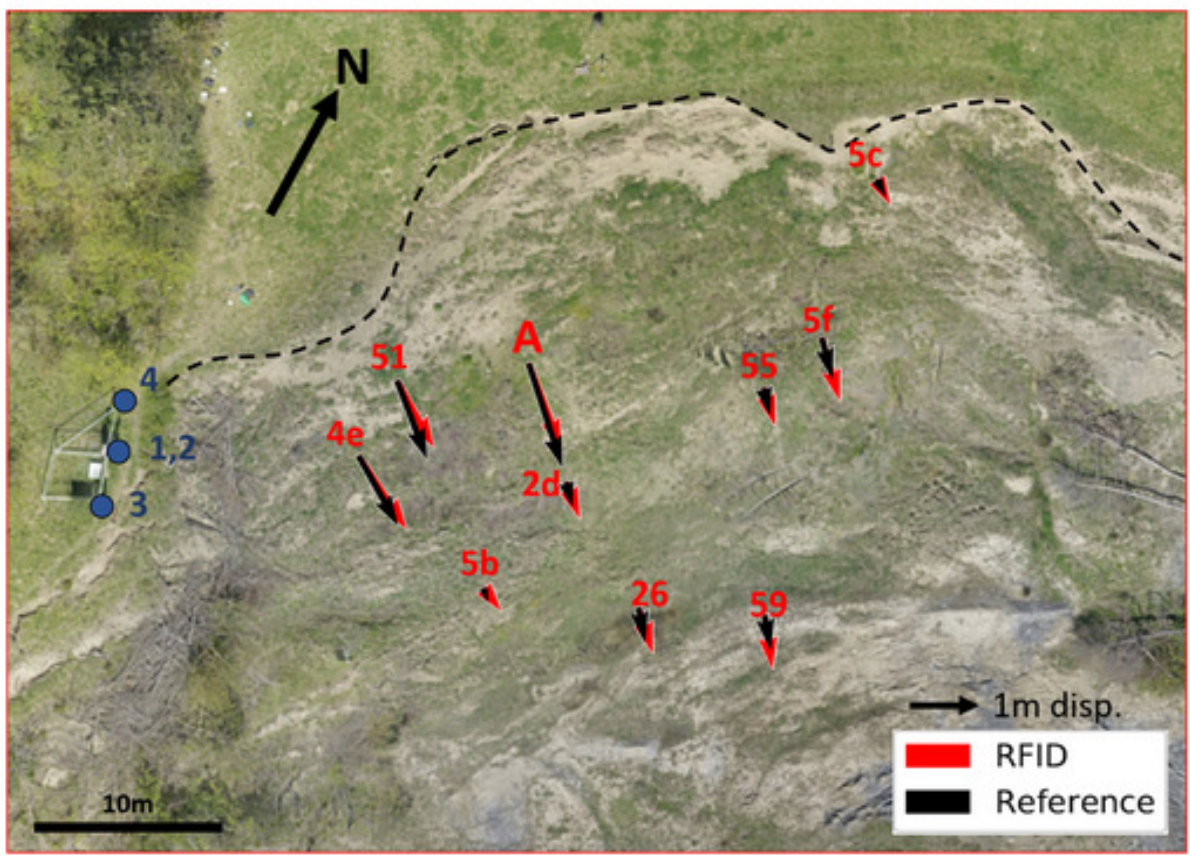

**Fig. 13: Installation of RFID tags for measuring displacements of 2D on Harmalière landslide, with the cumulated displacement measured over 6 months indicated with an arrow, using a GPS as a reference. From (Charléty et al., 2022b).**

## 3.2   Tilt of unstable boulders and slopes

Tilt monitoring can be valuable to provide early warning (Intrieri et al., 2012) before failure of rotational landslides (Hungr et al., 2014) and other unstable elements. The ability to deploy low-cost and low-power sensors appears critical when the zones to monitor are highly heterogeneous and need monitoring during several years. Low-cost WSNs have been proposed in pebbles (Gronz et al., 2016), on rotational landslides, on shallow landslides (Uchimura et al., 2010) and on unstable boulders (Dini et al., 2020) (Fig. 14). The latter study highlights the need for numerous sensors, to monitor each prone-to-fall boulder independently. Yet, WSN nodes are more expensive than RFID tags. And they consume much more power than RFID tags, which is operationally challenging for continuous monitoring of a large number of nodes.

Again, RFID seems like an appealing alternative, for example to track the tilt of many boulders independently. Tilt of RFID tags can be measured either with dedicated sensors or with localization methods. Localization methods can measure a tilt from the relative displacement between several tags fixed on the same object, or from the optimal polarization of the reader antenna, in line with the tag (Gupta et al., 2014; Lai et al., 2018). Localization has the advantage to work with any tag. But in practice the accuracy is easily deteriorated by environmental influences such as multipathing, and the measurement quality depends on the system geometry. In practice, dedicated sensors are likely to provide more reliable results with simpler installations, than propagation-based approaches.

Dedicated RFID tilt sensors use either MEMS accelerometers and switch sensors. MEMS accelerometers usually measure 3-axis acceleration, providing complete orientation of the object relatively to the vertical. They can be interfaced either directly on tags inputs (Vena et al., 2019) or through a microcontroller (Farsens, n.d.; Jayawardana et al., 2016). Unfortunately, these studies did not characterize the tilt accuracy. But general studies on ultra-low power accelerometers indicate that the accuracy should remain <1° (Łuczak et al., 2017). That is sufficient for boulder and landslide applications.

Alternatively, switch sensors consist in a continuity loop that is open or closed depending on the tag orientation. Such sensors can be connected to the tag antenna, in order to detune it when its tilt reaches a threshold (Philipose et al., 2005; Shi et al., 2017; Ziai and Batchelor, 2017). On use cases that require permanent detection even when the tag is not interrogated, irreversible latching sensors can be used (W. Wang et al., 2020a). However, antenna-based sensing can be challenging in the field, due to its difficulty to differentiate a change of signal due to the sensor from other effects such as the tag deterioration by environmental influences. Connecting these switch sensors to tags equipped with digital input pins (see Table 2) could in practice increase the measurement reliability.

To conclude, boulders and landslides may be monitored with a tilt switch for threshold alert, or with MEMS accelerometers to quantify their rotation along time. Propagation-based and antenna-based sensing allows for simple implementations, but are less reliable and accurate. In practice the usage of chips with dedicated sensor inputs seems preferable, particularly in an early-warning system that is deployed to secure the population.

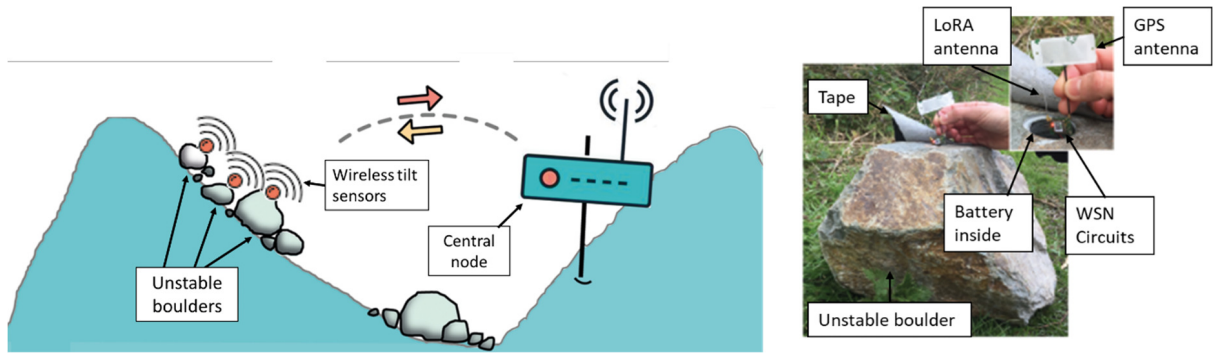

**Fig. 14: Example of a wireless sensor network, to monitor the tilt of boulders on a slope. Use of passive RFID tags can be envisioned on a similar use case. Modified from (Dini et al., 2020)**

### 3.3    Millimetric displacement of fissures

Rocks falling from cliffs present a risk to the population, which can be investigated and mitigated by monitoring rock compartments. The current monitoring techniques, reviewed by Lambert (2011), include insitu sensors such as extensometer, crackmeters, laser distance-meters or strain gauges. These reach subcentimetric to submillimetric accuracy but end up expensive to monitor large areas. The remote interferometric radar technique is well adapted to monitor large areas, but it measures mostly line of sight displacements, and its high cost dedicates it for high-risk zones only. When the risk is lower or not yet identified, passive visual targets can be installed on fissures with manual data collection, which is adequate on the ground but more difficult on a cliff. RFID sensors could allow deploying numerous passive targets, but interrogated wirelessly. They also gives the ability to switch seamlessly from manual wireless data collection over large areas (when the risk is low) to automatic continuous monitoring on a specific area (when an activity is detected). This strategy is indeed suggested for geotechnical monitoring of limited risk zones (Dunnicliff et al., 2012, p. 94).

Several tags have been designed and tested in the laboratory for measuring strain, displacement and crack formation. Antenna-based strain sensors were first developed to measure large strain up to 50% deformations (Merilampi et al., 2011; Occhiuzzi et al., 2011a; Hasani et al., 2013). Then, much smaller strains were measured with an accuracy of the order of 20 µm/m in the laboratory (Yi et al., 2013, 2015) (Fig. 15c), for structural health monitoring and crack detection (Zhang et al., 2017) (Fig. 15a). In practice, however, searching for the antenna resonant frequency (Fig. 15b) is impractical out of the laboratory, particularly in regions under the European ETSI-302-208 regulation that limits the available bandwidth. Displacement sensors were also developed, to measure the distance between the tag and a metallic object at its vicinity (Bhattacharyya et al., 2009) or between two tags (Caizzone and DiGiampaolo, 2015). To improve the accuracy, it is possible to connect dedicated strain gauge sensors on the tag (DiGiampaolo et al., 2017; Jayawardana et al., 2019). Some RFID chips that sense their own deformation could also be used (ASYGN AS321X), given a special care on the mechanical properties of the tag, its long-term behavior, and its fixation.

Threshold displacement detection with RFID tags appear as a simple but a reliable approach. It was proposed to detect defects appearing on a surface (Nappi and Marrocco, 2018), contact with a screw or a magnet (Fischbacher et al., 2020; Inserra et al., 2020) or rupture of a part of an antenna (W. Wang et al., 2020b). Threshold displacement tags were recently deployed on a cliff to detect potential displacement of rock compartments (Le Breton et al., 2021) (Fig. 16). Each tag is connected to a simple actuator placed on the two sides of a fissure, to provide 1-bit information when displacement exceeds a threshold. Laboratory experiments demonstrated the ability to raise an alarm just after reaching a displacement threshold of 1 mm.

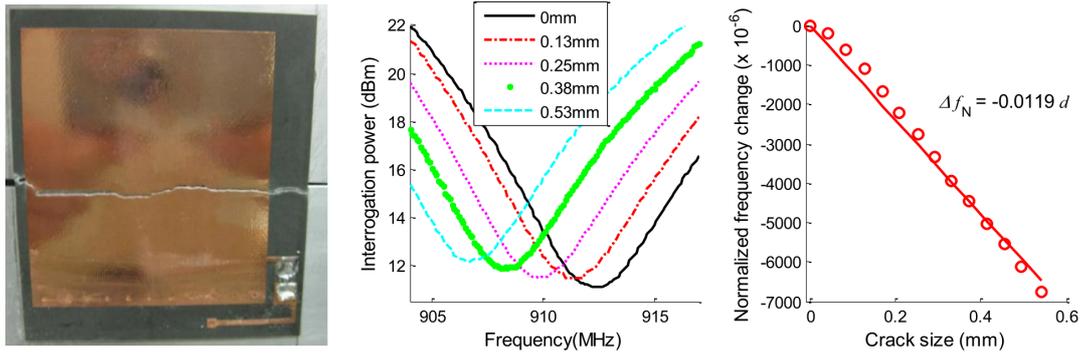

**Fig. 15: (a) Propagation of a crack on a tag antenna. (b) The resonant frequency of the antenna, centered on UHF RFID frequencies, moves while the crack opens. (c) As a result, the normalized resonant frequency versus the crack size shows a potential accuracy <0.1 mm. From (Yi et al., 2013, 2015)**

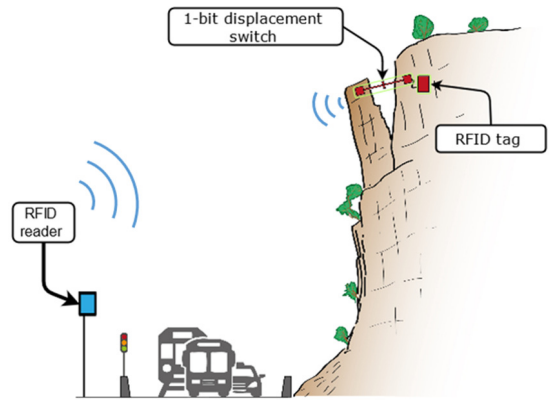

**Fig. 16: Principle of a 1-bit sensing tag deployed on a cliff to detect ruptures. From (Le Breton et al., 2021) with the courtesy of Géolithe and R. Leroux-Mallouf.**

# 4   Ground vibrations

Vibration measurements can be used to monitor the resonant frequency of structures (civil or geological), or to detect transient vibration events that exceed a threshold (construction work, blast, earthquake). Measuring resonant frequencies exploits weak ambient vibrations and requires high-accuracy seismic-grade accelerometers providing noise levels below µg (Scudero et al., 2018) (C. Wang et al., 2020). Detecting transient vibrations requires permanent monitoring, presenting a challenge of power consumption. We evaluate their feasibility for a practical usage, and the challenges to be resolved.

## 4.1   Resonant frequency of structures

Ambient noise seismology exploits vibration from non-controlled and diffuse sources, to characterize the subsurface and detect changes over time. In particular, arrays of seismic sensors placed at short distances (over tens to hundreds of meters) for long duration can monitor dynamic parameters of near-surface geological structures (Larose et al., 2015) such as their resonant frequency. This frequency provides insights on the structure's mechanical behavior (Del Gaudio et al., 2014;

Larose et al., 2015; Kleinbrod et al., 2019; Colombero et al., 2021) such as the stability of a rock column (Lévy et al., 2011; Bottelin et al., 2013, 2017). This section evaluates whether RFID vibration sensors could be used to monitor resonant frequency of a geological structure or not.

RFID tags were recently developed to measure the vibration modes of civil structures (Jayawardana et al., 2016, 2019) (Fig. 17). A tag connected to an ultra-low-power MEMS accelerometer was used to retrieve the fundamental vibration frequency of a metallic beam. This system could measure the beam's fundamental frequency within 1 − 40 Hz, with a precision of ±0.01 Hz (at 6 Hz). That worked for values of acceleration within 30 − 300 mg, equivalent to displacements of 0.02−0.2 mm. In 2019, the authors improved the results by adding a strain gauge to measure oscillating deformation of the beam of the order of 100-10 με. DiGiampaolo et al. (2017) proposed a similar experiment, using only a strain gauge to measure the vibrations of a beam.

Simpler implementations were also proposed, using an acceleration actuator attached on the tag antenna (Rahmadya et al., 2020). Standard tags were also used, by measuring the oscillation of the phase caused either by the displacement of the tag itself (Yang et al., 2017) (Li et al., 2021), or by the displacement of an object near the tag that modifies the multipath interferences (Yang et al., 2020). However those simpler approaches could measure at most displacements of 2 mm, which is much less accurate than dedicated sensors.

The mentioned dedicated sensors can measure the typical 0.1-100 Hz vibration frequencies observed on rock columns (Moore et al., 2018). Yet, in practice the sensor's noise levels of 90 − 250 μg/√Hz (see Table 4) is far greater than the ambient vibrations levels of 1 ng to 10 μg at 1 − 10 Hz (Peterson, 1993). Therefore, today, RFID vibration sensors may offer applications of modal analysis only for measuring strong vibrations. This occurs at the vicinity of strong vibration sources (transportation, factories, explosives) or on sites that present high resonance factors (e.g., buildings, large bridges). As we expect future improvements in the sensor's sensitivity, we envision that RFID accelerometers will be applicable to more and more applications.

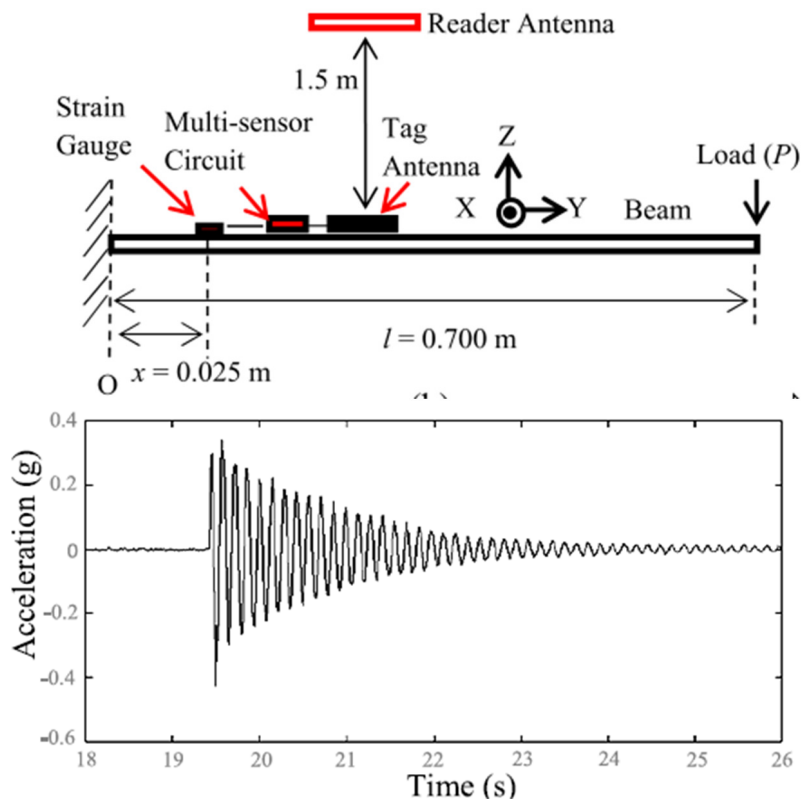

**Fig. 17: (a) Experiment on a tag with an accelerometer, to measure the vibrations of a cantilever beam. (b) Transient acceleration measured after removing the load at t=19.5s, showing a resonance at 6 Hz. From (Jayawardana et al., 2016, 2019)**

**Table 4: Examples of ultra-low power commercial MEMS accelerometers.**

| Accelerometer Sensor | Resolution | Noise level horizontal (μg/√Hz) * | Power consumption | Used in |
|---|---|---|---|---|
| ADLX362 | 1 mg ±2 g | n.a. (wakeup) 550 @2g 175 @2g | 0.5 μW @6 Hz 3.5 μW @100Hz 45 μW @100Hz | (Jayawardana et al., 2016, 2019; Konstantakos et al., 2019; Vena et al., 2019) |
| Bosh BMA400 | 1 mg ±2 g | 2600 eq. (11mg) @4g 180 @4g | 1.5 μW @25Hz 26 μW continuous | |
| ST IIS2DLPC | 1 mg ±2 g | (5.5 mg) @2g | 6.3 μW @50Hz | |

*Noise levels are given on X and Y-axis. They are often higher (x1.5 to x3) on Z-axis.

### 4.2 Detection of transient vibrations and shocks

Soil vibrations are important to quantify because they can damage buildings, depending on the vibration amplitude and frequency (

 Table 5). That can occur near construction works or explosive blasts (Segarra et al., 2015). Monitoring the vibrations ensures that buildings are not damaged following international standardization rules (ISO 4866, 2010) and can also apply to fragile or unstable geological structures (Bottelin et al., 2020). Ultra-low power MEMS can detect events with very little power, by providing a triggering signal only when a vibration exceeds a threshold (Evans et al., 2014). In a low-power seismic acquisition station, Konstantakos et al. (2019) monitored vibrations using the ADLX362 sensor (Table 4) which consumed only 3 μA current (and can go down to 0.3 μA in wake-up mode). When vibration amplitude exceeds a threshold, the sensor wakes-up a complete system to properly record the event.

Detecting transient vibrations is challenging with RFID, because it requires enough power to enable permanent monitoring. RFID tags equipped with ultra-low-power MEMS accelerometers may do so, using battery assistance (Jayawardana et al., 2019), energy harvesting (e.g., small solar cells), or radiofrequency energy stored temporarily in a capacitor (e.g., Vena et al., 2019), as discussed in 6.3. In terms of sensitivity, the ADLX362 accelerometer offers 1 mg sensitivity and 0.55 mg/√Hz noise level. This should detect the vibration levels that can damage a mud-brick house, of 2–3 mm/s at 5–20 Hz, corresponding to 6–36 mg (Yan et al., 2017).

Shock detection may also be interesting, for example to detect if a part of infrastructure has suffered from a shock and needs maintenance verification. Such shocks imply strong vibrations, which should be detectable with 1-bit threshold vibration sensors attached on RFID tags. A tag equipped with a simple 2-state switch sensor could detect variations of acceleration (>0.5 g on the example) (Occhiuzzi et al., 2010) (Fig. 18). However, such tag would need a constant interrogation to record a transient vibration: it can record vibration only when it is interrogated and it has no memory. Alternatively, shock sensing may be useful not only in real time, but also to detect if an event has occurred in the past. This may be feasible by connecting the sensor to a tag that offers tampering detection input. Alternatively, the information can be stored mechanically, requiring zero sensing power to detect the event, by attaching a bistable accelerometer to a tag (Todd et al., 2009). Latching bistable sensors can change their mechanical state durably when acceleration exceeds a certain threshold. The threshold values go from just above 1 g (Mehner et al., 2015) to hundreds of g's (Hansen et al., 2007; Reddy et al., 2019) and can depend on the vibration frequency (

Fig. 19). If such a sensor were implemented on a field, the threshold value would have to be chosen in accordance with the strength of the expected events to monitor. Technical improvements of these latching sensors could improve their usability if they were implemented on RFID systems. Recent

developments allow latching sensors to record the amplitude of the highest vibration event (Mehner et al., 2015; Reddy et al., 2019), or to be reset electrically to their initial position (Mehner et al., 2015; Reddy et al., 2019).

In conclusion, RFID tags can detect strong shocks (several g's) using zero-power bistable sensors. Small shocks or motions (>0.5 g) can be detected with a one-bit reversible switch sensor. Soil vibrations that are strong enough to damage buildings (>5 mg), generated for example by human activity or earthquakes, should be detectable using ultra-low-power Micro Electro Mechanical Systems (MEMS) accelerometers. Proof of concept tags have been designed in the laboratory for sensing such vibrations, and remain to be adapted for field applications.

**Table 5: Range of structural response for various sources, that may damage structures. Adapted from (ISO 4866, 2010).**

| Vibration excitation | Frequency range (Hz) | Displacement range (μm) | Velocity range (mm/s) | Acceleration range (mg) | Continuous or Transient |
|---|---|---|---|---|---|
| **Traffic** road, rail, ground-borne | 1 – 100 | 1 – 200 | 0.2 – 50 | 2 – 100 | C/ T |
| **Blast vibration** ground-borne | 1 – 300 | 100 – 2 500 | 0.2 – 100 | 2 – 5 000 | T |
| **Air over pressure** | 1 – 40 | 1 – 30 | 0.2 – 3 | 2 – 50 | T |
| **Pile driving** ground-borne | 1 – 100 | 10 – 50 | 0.2 – 100 | 2 – 200 | T |
| **Machinery outside** Ground-borne | 1 – 100 | 10 – 1 000 | 0.2 – 100 | 2 – 100 | C/ T |
| **Machinery inside** | 1 – 300 | 1 – 100 | 0.2 – 30 | 2 – 100 | C/ T |
| **Human activity inside** | 0.1 – 30 | 5 – 500 | 0.2 – 20 | 2 – 20 | T |
| **Earthquakes** | 0.1 – 30 | $10 - 10^5$ | 0.2 – 400 | 2 – 2 000 | T |

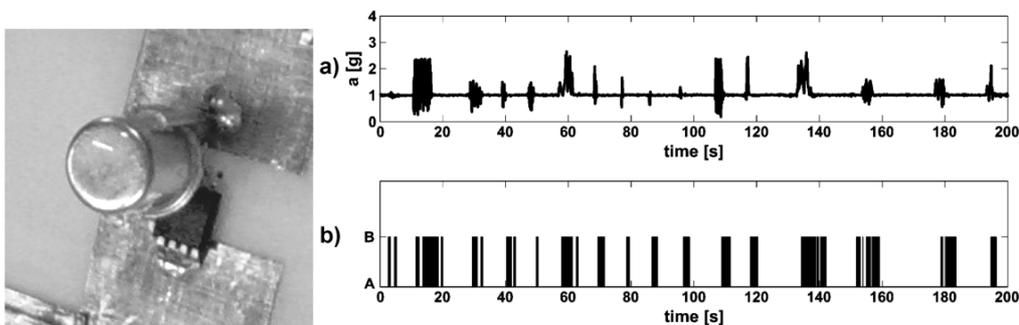

**Fig. 18: (left) 1-bit motion detector attached to a tag and (top right) records by a dedicated vertical accelerometer and (lower right) by the 1-bit RFID sensor. From (Occhiuzzi et al., 2010)**

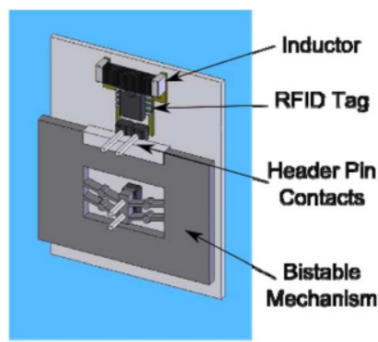

# 5 Chemico-physical properties of geomaterials

## 5.1 Soil moisture

Soil moisture determination is important in the fields of agricultural, geotechnical, hydrological and environmental engineering (Lekshmi et al., 2014). Three electrical methods are used today to estimate soil moisture (Schmugge et al., 1980; Lekshmi et al., 2014; Dane and Topp, 2020, p. 4). They measure either the electrical conductivity between electrodes, the change of capacitance of a sensor, or the time delay of a wave transmitting in the soil (transmissiometry/reflectometry). Many other non-electrical methods allow such estimation, for example through the influence of soil moisture on underground air humidity or on daily temperature fluctuations.

Soil conductivity directly depends on moisture and water salinity, and to a lesser extent on the temperature and on the soil solid phase. Therefore, the conductivity method requires regular calibration on each location to mitigate the effect of soil and salinity. The method consists in measuring the electrical resistance between two probe electrodes when two additional electrodes inject a current. That is implemented today on commercial passive RFID tags (FARSENS HYDRO H402), and on multi-sensor tags (Cappelli et al., 2021).

Capacitive sensors (Fig. 20.a) are considered more accurate than conductivity sensors to estimate soil moisture, because they are sensitive mostly to the soil in-phase dielectric permittivity, which in turn depends mostly on water content for wet soils. The in-phase dielectric constant is one for air, 3 − 5 for dry soil and about 80 for water at 20 °C at 1 MHz to 1 GHz frequencies. Therefore water content holds the major influence factor on dielectric permittivity in wet soils (Peplinski et al., 1995). Capacitive sensors connected to analog sensing tags (SL900A) have provided quite accurate measurements in the laboratory (Fonseca et al., 2017; Pichorim et al., 2018) including during several days (Korošak et al., 2019).

We consider antenna-based sensing as a group of capacitive sensors. In the early stages, such tags were demonstrated to detect the presence of liquid water near the tag (Bhattacharyya et al., 2010a), because the water strongly reduces the received signal strength. Soil moisture was then estimated using received signal strength. However the RSSI indicator is sensitive to many unwanted parameters and has a limited range of measurement. To improve the accuracy, a reference tag can be placed above the ground (J. Wang et al., 2020). To maximize the dynamic range, the tag antenna can be tuned to the permittivity of dry soil (Alonso et al., 2017). Sweeping the carrier frequency—allowed mostly in laboratories—can instead determine the tag resonant frequency and its amplitude. The soil moisture control mostly the resonant frequency (Kim et al., 2014), and its salinity controls the amplitude of the resonance (Dey et al., 2019, 2016; Hasan et al., 2015, 2013). Very preliminary results also investigated the usage of self-tuning tags to estimate the antenna detuning (Stoddard et al., 2019). In terms of application, antenna-based moisture sensors (Zhou et al., 2016) monitored drying concrete and could retrieve the water content over a very large range of values, from very wet to dry. Aroca et al. (2016, 2018) buried tags at different depths below the ground, and observed a loss of RSSI, caused both by propagation attenuation—as they suggest—and by the coupling of soil with the tag antenna. J. Wang et al. (2020) presented for the first time a real-world application of soil moisture sensing (Fig. 20.b), using the antenna of commercial low-cost tags for monitoring the moisture of pots in a green plant. After calibration, they could determine the soil moisture with 5% accuracy on one pot, and in a large-scale experiment they could determine which pot required watering with 95% of success. They optimized the system by using several existing concepts. First, they observed that using the indicator of power of activation (i.e., the minimal power transmitted by the reader for which the tag is activated), instead of the RSSI, divides by two the influence of propagation. Then, they used a reference tag to make differential measurements independently from the pot location. Next, they averaged the measurements over long periods to eliminate rapid fluctuations caused by human activity. Finally, they calibrated the measurement against a standard measurement, for a pot on which the soil is reproducible.

The reflectometry (or transmissiometry) method measures the change in time delay of a propagated wave between two electrodes. It is among the most accurate method for measuring soil moisture (Lekshmi et al., 2014), because it measures the propagation over a volume of soil. Therefore it depends much less on the mechanical coupling of the sensor with the soil than with capacitive sensors. Fonseca et al. (2018) have implemented the transmissiometry method with a UHF RFID tag (Fig. 20.c): the wave propagates along a line connected to the tag, which includes a 10 MHz oscillator for measurements. The results were linear within 10–32% of moisture and up to 0.15 S/m conductivity.

Alternative indicators can be related to soil moisture. For example, an air humidity sensor buried in a cavity would measure either a low or saturated humidity in dry and wet soil, respectively (Cappelli et al., 2021). In complement, tags were also proposed to measure not only soil moisture, but temperature, air humidity, soil conductivity and luminous intensity (Korošak et al., 2019; Deng et al., 2020; Cappelli et al., 2021).

In conclusion, many approaches have been tested to measure soil moisture. Among them, transmissiometry is the most accurate because it distinguishes moisture from salinity. Antenna-based sensing has also proven efficient in an industrial green house, using a reference tag and a calibration on a reproducible soil. Most studies installed the tag antennas above ground and the sensor within the ground. One design also buried the tag, which could be promising for long-term monitoring (i.e., for agriculture), but strongly reduces the reading range (see 6.2.1).

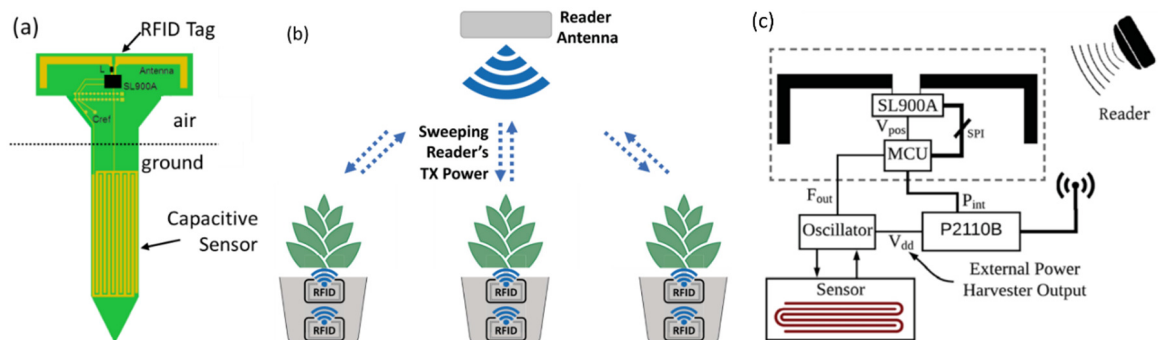

**Fig. 20: Examples of soil moisture sensor tags. (a) Dedicated capacitive sensor (Pichorim et al., 2018); (b) antenna-based sensing based on activation power using reference tags (J. Wang et al., 2020); and (c) time-domain transmissiometry using a microcontroller and an energy harvesting module (Fonseca et al., 2018)**

### 5.2 Soil temperature

The daily and yearly temperature variations significantly influence many near-surface processes. Temperature is a very important parameter to monitor over several decades, particularly in regions sensitive to global warming. For example, in permafrost and mountainous areas, a small temperature change near 0 °C may have important consequences (e.g., Ravanel et al., 2013). The temperature of soils, rocks or snow can be very heterogenous in space, particularly along depth. Monitoring temperature over long duration at low cost can be of interest, particularly with buried passive wireless systems that don't require ground wiring that can easily be damaged (agriculture, theft, animals, grass mowing…). Arrays of batteryless RFID tags could monitor the temperature at different depths and locations in soil or rock materials, during years, with little cost and maintenance, as already demonstrated in logistics (Jedermann et al., 2009). Temperature data can also rectify the thermal influences of another measurement to increase its accuracy (e.g., Le Breton et al., 2017).

Today, RFID temperature sensors are embedded in RFID chips and produced industrially (see Table 2). They offer 1–2° C accuracy, depending on the calibration procedure, the temperature range (Occhiuzzi et al., 2018), and the powering level on batteryless tags (Camera and Marrocco, 2021). Researchers have also developed ultra-low consumption temperature sensors for batteryless tags (Opasjumruskit et al., 2006; Yin et al., 2010; Vaz et al., 2010; Wang et al., 2014), which could reach 0.4 °C accuracy (Tan et al., 2019). Many other original temperature sensors have been designed. They were based on antenna sensing (Qiao et al., 2013; Wang et al., 2019; Zannas et al., 2018), irreversible threshold detectors (Bhattacharyya et al., 2011; Babar et al., 2012), or electromechanical encoders (Zhu et al., 2021). Among all these sensors, the embedded temperature sensors should be preferred for geoscience applications because of their advanced maturity.

Soil temperature sensing using RFID tags was first suggested by Hamrita and Hoffacker (2005), and implemented much later on the field, to compare the soil and air temperature (Luvisi et al., 2016). The experiment buried a semi-passive tag with temperature data logging (Easy2Log RT0005, Caen RFID) buried 10 cm into the ground, read from a handheld RFID reader. The aim was to monitor the efficiency of a soil treatment for pathogens and weed, called solarization, that consists in covering the soil with a film to increase its temperature. The solarization increased the soil temperature relatively to the air temperature. Once the solarization film had broken, the resulting decrease of relative soil temperature could easily be measured thanks to the sensing tag (Fig. 21). Another implementation of buried soil temperature sensor has been also proposed by Deng et al. (2020), as a proof of concept for underground communication using a robot.

In conclusion, RFID temperature sensing is a mature and easy to use integrated sensors technology, offering a 1–2°C accuracy, and a low thermal signature. It could be particularly useful to monitor materials that are sensitive to temperature fluctuations such as snow or permafrost.

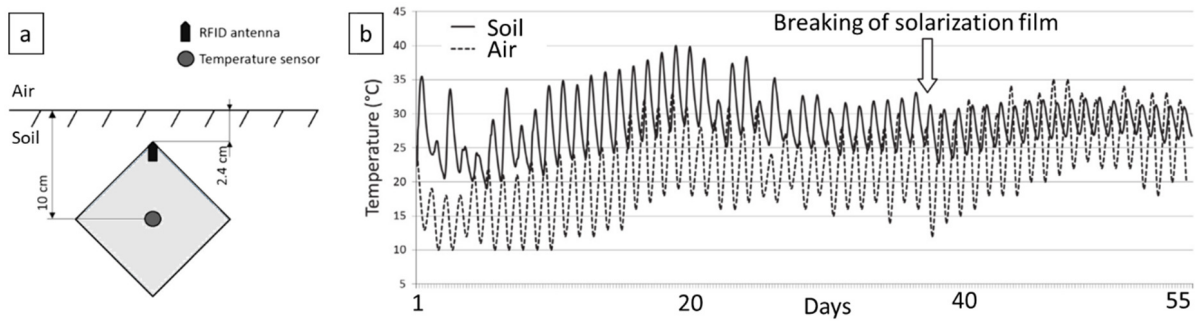

**Fig. 21: (a) Buried temperature data logging RFID tag, allowing (b) a comparison of air and soil temperature. In this example, the data allows detecting the rupture of a thermal solarization film that was keeping the soil hotter than the air. Modified from (Luvisi et al., 2016).**

### 5.3  Snow and ice

Snowpack monitoring is used for example to predict the quantity of snowmelt water to come in rivers for hydroelectric management or to estimate risks of an avalanche near ski resorts. The properties of the snowpack—the snow accumulated on the ground—strongly varies in time and space. Its measurement often requires the combination of multiple techniques (Kinar and Pomeroy, 2015). Some macroscopic properties of a snowpack such as thickness, liquid water content, density, and total water equivalent content, control the snow dielectric permittivity and can be measured with microwaves(Tedesco, 2015). In particular, the influences of antenna coupling, transmission through snow and multipath interference are used today for snow monitoring (Denoth, 1994; Bradford et al., 2009; Larson, 2016; Koch et al., 2019). These influences similarly affect the RFID signal (Le Breton, 2019), opening the way for sensing applications.

The snow water equivalent (SWE) of a snowpack can be monitored from the variations of phase delay over time, due to changes in the wave velocity propagation (Le Breton et al., 2022). The experiment consisted in placing a tag on a support very close to the ground, and adding dry snow between the tag and antenna which increases the phase delay and allows estimating the increase of snow water equivalent. The method was first validated in the laboratory on dry snow of different densities, then outdoors in a mountainous environment during several snow event, and finally during an entire winter on Col de Porte, the national snow test site of Meteo France (see Fig. 22).

The snowpack temperature can also be monitored with RFID tags embedding a temperature sensor, like used in soil (see previous section). The concept was demonstrated at Col de Porte, simultaneously as the SWE experiment. A vertical array of 23 tags installed before the snowfall provided a continuous vertical profile of temperature data, with 1 °C accuracy, during the whole winter (Le Breton et al., 2022). It is an example of monitor simultaneously two parameters with an RFID system—the SWE and snowpack temperature—combining both propagation-based sensing and a dedicated sensor.

Frost deposition and melting on a tag also detunes its antenna, leading to variations of the phase and RSSI (Le Breton, 2019). This influence has been recently exploited to estimate the quantity of ice deposited on a tag, and to detect its melting (Wagih and Shi, 2021a, 2021b).

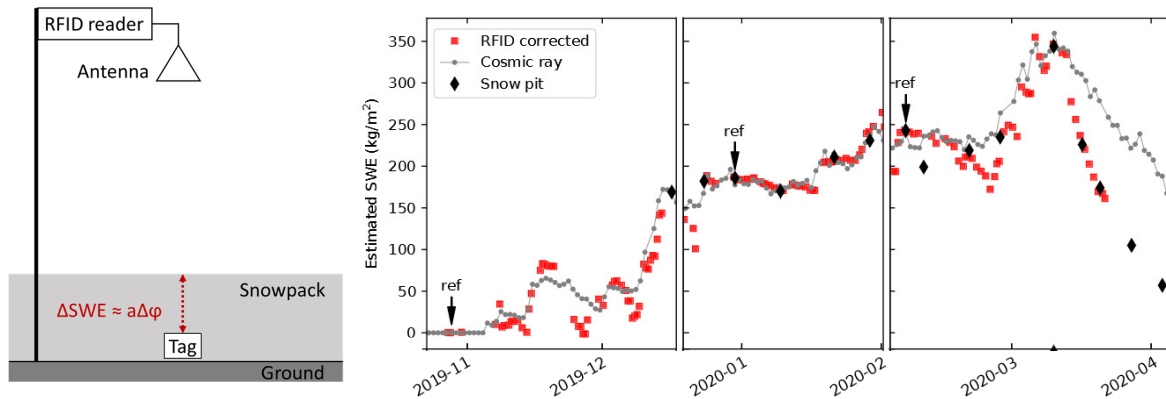

**Fig. 22: (a) Installation of a permanent RFID reader outdoors, interrogating a tag placed just above ground and covered by snow (b) result of the SWE estimated with RFID over a full winter on Col de Porte instrument testing, compared with snow pits and cosmic ray measurements of the SWE. From Le Breton et al., 2022**

### 5.4 Potential for mechanical, gas, luminosity and water sensing

Other RFID sensors that might be used to monitor near-surface processes are cited in this section.

Mechanical sensors of pressure (Salmerón et al., 2014) or weight (Catarinucci et al., 2020) (e.g., FARSENS EVAL01-Zygos-RM), might be of interest for example for geotechnical monitoring (Dunnicliff et al., 2012).

Air humidity sensing has been proposed on several research tag designs. They exploit antenna-based sensing (Siden et al., 2007; Virtanen et al., 2011), dedicated sensing (Pursula et al., 2013; Yu et al., 2015), and sometimes multi-sensing platforms (Cappelli et al., 2021; Korošak et al., 2019). Commercial tags are also available, greatly easing the ability for air humidity sensing (ASYGN AS321X Humidity; EVAL02-Hygro-Fenix-RM). In addition to humidity sensing, tags were also designed to detect other gases, such as ammonia (Karuppuswami et al., 2020; Occhiuzzi et al., 2011b; Yang et al., 2009), octane, ethanol (Manzari et al., 2014b), and methane (Zhang et al., 2018). These gas sensors might find applications for example in underground mines or on geothermal and volcanic areas.

Luminous intensity sensors have been developed for research (De Donno et al., 2014; Falco et al., 2017; Salmerón et al., 2014) and commercial tags (FARSENS, AS321X-EVAL-Light). Multispectral sensing tag may even quantify the luminosity in different bandwidth, within the visible spectrum only (Boada et al., 2020) or also in the infrared and ultraviolet spectra (Escobedo et al., 2016). This might be useful for example to estimate the effect of solar radiation energy on soils, rocks, or on the snowpack.

Water can also be monitored using RFID, with tags that measure the level of water (Bhattacharyya et al., 2010a; Capdevila et al., 2011; Melia-Segui and Vilajosana, 2019), the pH (Kassal et al., 2013; Nanni et al., 2022) and the concentration of electrolytes (Rose et al., 2015; Occhiuzzi et al., 2015; Kantareddy et al., 2018; Hillier et al., 2019). On the field, an array of water sensing tags might serve to evaluate where groundwater may emerge at the surface, for example on a landslide, or to track pollution.

# 6 Key challenges

RFID tags may offer advantages in harsh environments: tags do not require batteries, and can be interrogated across opaque media. Their affordability allows getting them destroyed.

## 6.1 Bedload tracing

The short detection range of passive low-frequency RFID tags (134 kHz) typically used in bedload transport studies is one of the most limiting technical drawbacks to consider. It implies the loss of deeply buried tracers, and time expensive field surveys when tracers spread over wide or long channel reaches. Standard operating procedures for bedload tracing with RFID tags have been proposed to optimize the detection range (Chapuis et al., 2014). This study investigated parameters such as the 3D shape of antenna detection zones for different RFID tags, different orientations, different types of antennas, and different media of tag deposition (open-air, dry and saturated sediment, water). It showed a maximum detection range of 0.7 m for 23-mm-long tags that are oriented perpendicular to the loop of the antenna, and a significant decrease of recovery rate for burial depths above 0.5 m. It has been therefore recommended to insert RFID tags parallel to the short axis (c-axis) of stones to favor a perpendicular orientation of tags with respect to the channel surface for maximizing the detection range. The comparison of detection ranges in different media showed a small effect of dry or saturated granular media on the detection range, as compared with open-air conditions. However, a significant reduction of the detection range (40 to 50% decrease) has been obtained for water submergence as compared to open-air conditions. Similar experiments of PIT tag detection were implemented directly in the field using 23 mm and 32 mm transponders and different types of antenna. The experiments confirmed that a granular medium (burial in the sediment deposit) does not have a strong effect on the detection range, with a decrease of less than 10% (Arnaud et al., 2015). These field experiments also showed a small effect of water submergence on tag detection, contrary to what has been observed in the sandbox experiments of Chapuis et al. (2014).

These experimental studies provide useful recommendations for the design of RFID protocols maximizing the detection range of tags. However, whatever the size and type of passive low-frequency tags and antennas, no disruptive technological step was found to increase the detection range of PIT tags. The only solution to date is to use active UHF RFID tags (e.g., COIN-ID 433.92 MHz from ELA innovation). These tags can be detected at distances up to 80 m in open-air, and 2.5 m below water or in a saturated granular medium (Cassel et al., 2017a). Another decisive advantage of these UHF tags is their anti-collision protocol (these are standard for passive UHF tags, but not LF tags) allowing the simultaneous detection of several tags close to each other. This means that active UHF RFID tags can be deployed in columns buried in the subsurface of river channels to constrain the active layer depth, as a kind of RFID scour chain. This has been recently tested in two large braided gravel-bed rivers (Brousse et al., 2018, 2020b), allowing the quantification of scour and fill depths which would have not been seen with traditional scour chains (Fig. 23). The first reported bedload tracing experiments with those tags revealed, in a large gravel-bed river, frontrunners that have traveled more than 3 km after a 5-year return period flood (Brousse et al., 2020a). The tracers were recovered at a high rate (71%). At the same site, the field-testing of several detection and positioning protocols built from ground-based and UAV surveys shows that high recovery rates can be obtained from short-time field surveys (Cassel et al., 2020). The main drawbacks of these tags compared to PIT tags are their higher cost (unit price around €40), larger size and volume (36 mm diameter and 10 mm thick). Additionally the lifespan, which depends on the chosen frequency of emission, is generally less than 5 yrs, which is relatively short. Their cylindrical shape does not facilitate their insertion in natural stones like for glass-encapsulated PIT tags, and necessitates the casting of artificial stones (Cassel et al., 2017a, 2017b). However, active UHF RFID tags might partially replace passive tags in the future, especially for bedload tracing in large rivers for short duration. Passive RFID tags may still have interest for long-term studies in small rivers.

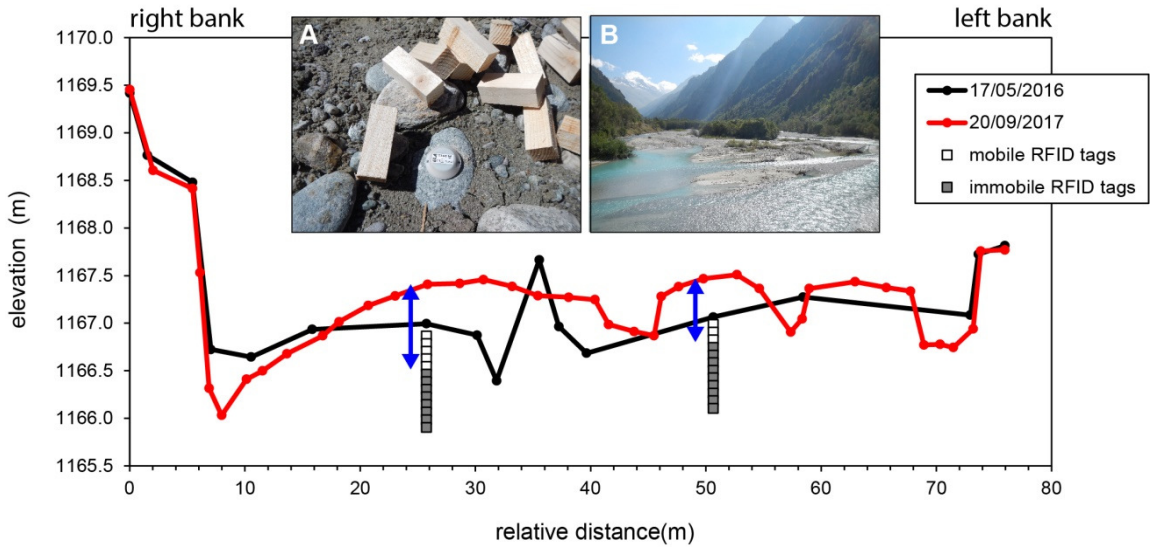

**Fig. 23: Example of using the anti-collision property of active UHF RFID tags to survey the active layer (blue arrows) of river channels during flow events, by inserting columns of RFID tags in the alluvial bed; RFID columns are used to detect the maximum bed scouring level, and the cross-section topographic resurvey is used to measure the thickness of sediment deposition; RFID columns have been made by superposing RFID tags and pieces of wood visible at the inset A; the May 2016 cross-section has been surveyed during the RFID column deployment; those data have been obtained in a very active braided river channel (Vénéon River, inset B), with an active layer thickness of around 1-m on the right bank during the investigated period (adapted from Brousse et al., 2018)**

A second major challenge of bedload tracing with RFID is to be able to know the burial depth of tracers. This information is crucial for the determination of the active layer of the bed during flow events. By combining the depth of the active layer with distances of transport, it becomes possible to estimate the time-averaged bedload flux (Haschenburger and Church, 1998; Liébault and Laronne, 2008; Houbrechts et al., 2012). The burial depth of low-frequency RFID tags can be constrained at fixed locations by considering the decay of the received input voltage by an antenna (Tsakiris et al., 2015). It has been shown that this decay scales with the cube of the distance between the tag and the reader, whatever the medium (air, water, gravel, sand). However, the orientation of the tag modulates the signal received by the reader, and this strongly limits the application of the decay law for tracers buried in alluvial deposits for which the orientation is variable and unknown. To fix this problem, it has been recently proposed to use a rotating weighted inner ball, which maintains the vertical orientation of a 12 mm PIT tags into stones. It is known as the Wobblestone device (Papangelakis et al., 2019b; Cain and MacVicar, 2020). During its first field-testing, this innovative device allowed to measure the burial depth of tracers with an accuracy of around 2 cm, down to a burial depth of 15 cm (Cain and MacVicar, 2020). That confirmed the great potential of this new RFID tracing device.

A third key challenge of RFID bedload tracing is to reduce the size of tags. This will allow surveying the displacement of the fine tails of bedload grain-size distribution (<15–20 mm), for which tracking solutions are still lacking today. The smallest tags that have been used up to date in bedload transport studies are the 12 mm glass-encapsulated PIT tags. However those tags have not been so much used because their detection range is strongly reduced (Chapuis et al., 2014). Any progress on the miniaturization of RFID tags will have therefore direct applications in bedload tracing. In this regards, tiny tags with a size of 1–4 mm² (e.g., models Murata LXMSJZNCMF-210, Hitachi PK2020B) are used today to track insects at 1–3 cm distance from the reader (Nunes-Silva et al., 2019). Research is investigating smaller tags with 0.03mm² size (Bhanushali et al., 2021), and even intracellular tags (Yang et al., 2021). Yet, smaller tags generally means smaller reading range, therefore smaller recovery rate. The strong challenge is that such smaller tags afford long-enough reading range.

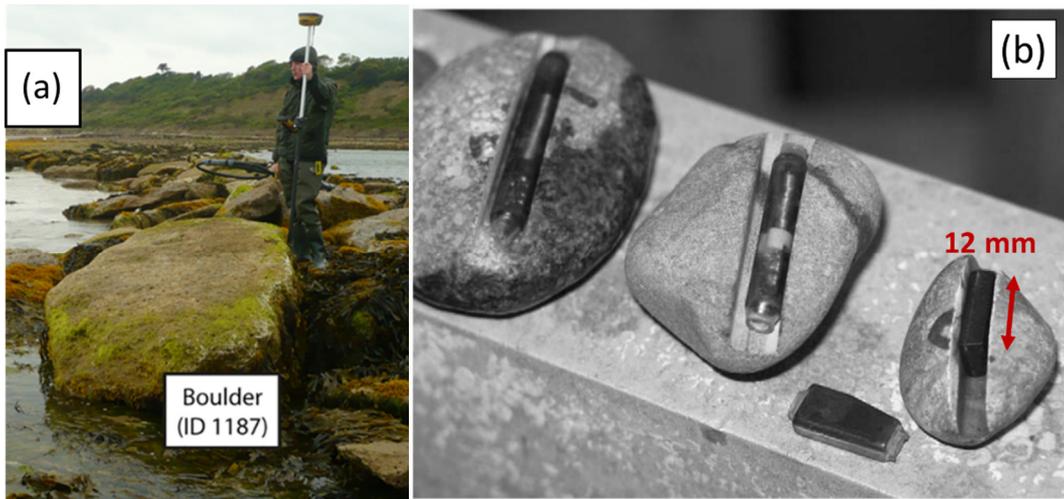

**Fig. 24: Tracking of sediments from (a) large boulders to (b) smallest gravels. Today, the tracking of smallest sediments is limited by the size of the tag. Modified from (Hastewell et al., 2020; Curtiss et al., 2009)**

The last key challenge is to reduce the time required for detecting the tags. Indeed, manual surveys of pebble and cobble tracers are time consuming, requiring a compromise between the area surveyed and the spatial density of the surveying directly linked to the recovery rate. Only one study could survey a large area (80 000 m²) every month, with very good recovery rate (70-80% after 3 years), using a thousand passive tags (Dolphin et al., 2016). To reach this performance, they used a quad with an array of detectors to scan a large area densely in a limited time. This rationalization of the surveying procedures can be a key enabler for the emergence of professional services of coarse sediments tracking.

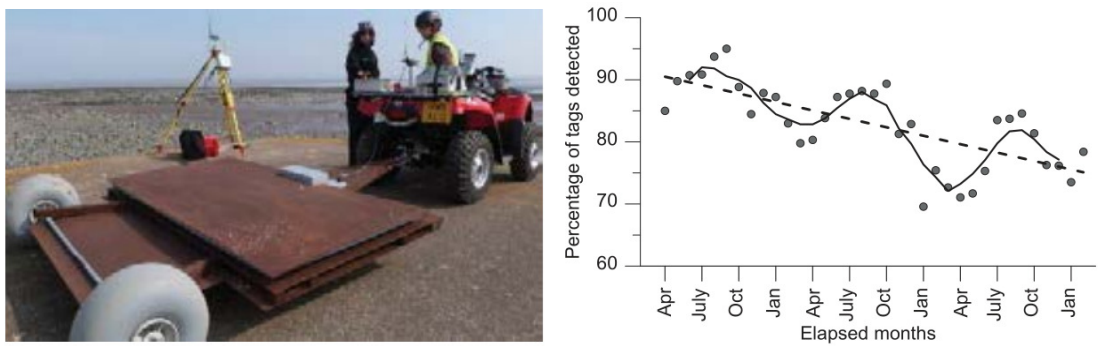

**(left) Array of four RFID readers carried by a quad which increases the velocity of surveying to 4m² / second. (right) The frequent surveying showed a recovery rate that oscillates yearly, with a decreasing trend. From (Dolphin et al., 2016)**

### 6.2   Tags under soil, water, snow or vegetation

### 6.2.1   General sources of signal loss

Tags placed outdoors may be covered by soil, water, snow or vegetation. These media can influence both antenna parameters and wave propagation, leading to a deterioration of the reading range or loss of localization accuracy (Fig. 25). Four main causes are identified: the attenuation of the wave within the medium, the partial reflection at the air-medium interface, the near-field coupling of the tag antenna with the medium (Fig. 26). Depending on the context, multipath effect may also have to be considered. These losses occur twice, on forward and backward communication.

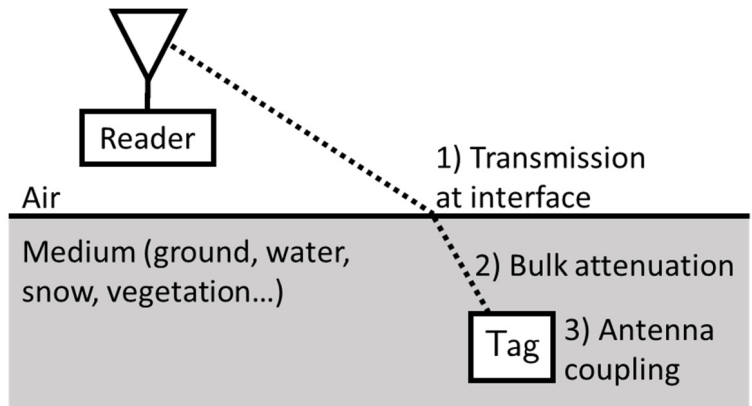

**Fig. 25: Major sources of signal loss identified when a tag is under a different medium than air**

For the attenuation, it is well established that the propagation of radiofrequency signals is very sensitive to the geometric and physical characteristics of the propagation medium. Its effect also depends strongly on operating frequency. The radiofrequency waves penetrating the ground undergo a very significant attenuation. As a rule of thumb, the penetration depth is inversely proportional to the square root of the product of the wave frequency and on the medium conductivity; the conductivity itself often depends on the liquid water content and salinity of the medium. Therefore the performances of RFID technologies will depend on the application environment and the RFID frequency (from 125 kHz to 2.4 GHz), as discussed in (Dziadak et al., 2009; Kumar and Sommerville, 2012). Numerous examples in the literature are dedicated to the implementation of RFID solutions in severe environment from an electromagnetic point of view.

Additionally, a wave will reflect a part of its energy when at the interface between the air and the medium. The resulting reflection coefficient $R$ depends on the dielectric permittivity of the medium, on the angle of incidence and on the wave polarization (see Balanis (2012), 5.3). The transmission coefficient ranges from 0.2 to ~1 (water to air) under the most favorable normal incidence, and decreases towards zero at parallel incidence. The reflection coefficient is minimized either for normal incidence, but also around the Brewster angle which occurs in general on vertical polarization (more exactly, on the component of the wave polarization that is perpendicular to the interface). Therefore, at least three strategies can be used to reduce the loss: normal incidence (i.e., interrogation from a UAV), Brewster's angle, or installing also the station antenna into the medium to avoid crossing the interface.

The coupling of the tag antenna at the proximity of the medium affects the antenna behavior in several ways. On the one hand, the medium will detune the antenna, leading to a shift of its resonant frequency, resulting in a power loss. On the other hand, the propagation loss in the medium will also decrease the amplitude of the antenna resonance. Finally, the radiation pattern and efficiency of the antenna may also be affected when placed at the proximity of an interface.

Also, multipath and scattering may influence the wave propagation, even for a tag placed above the medium. A change in the medium, for example due to vegetation growth, or to the formation of a snow layer on the ground, can then result in a modification of the phase or signal strength. That may lead to inaccuracy of localization, antenna-based sensing of propagation-based sensing. In the following sections, we study the influence of these media caused by the effects described hereinbefore.

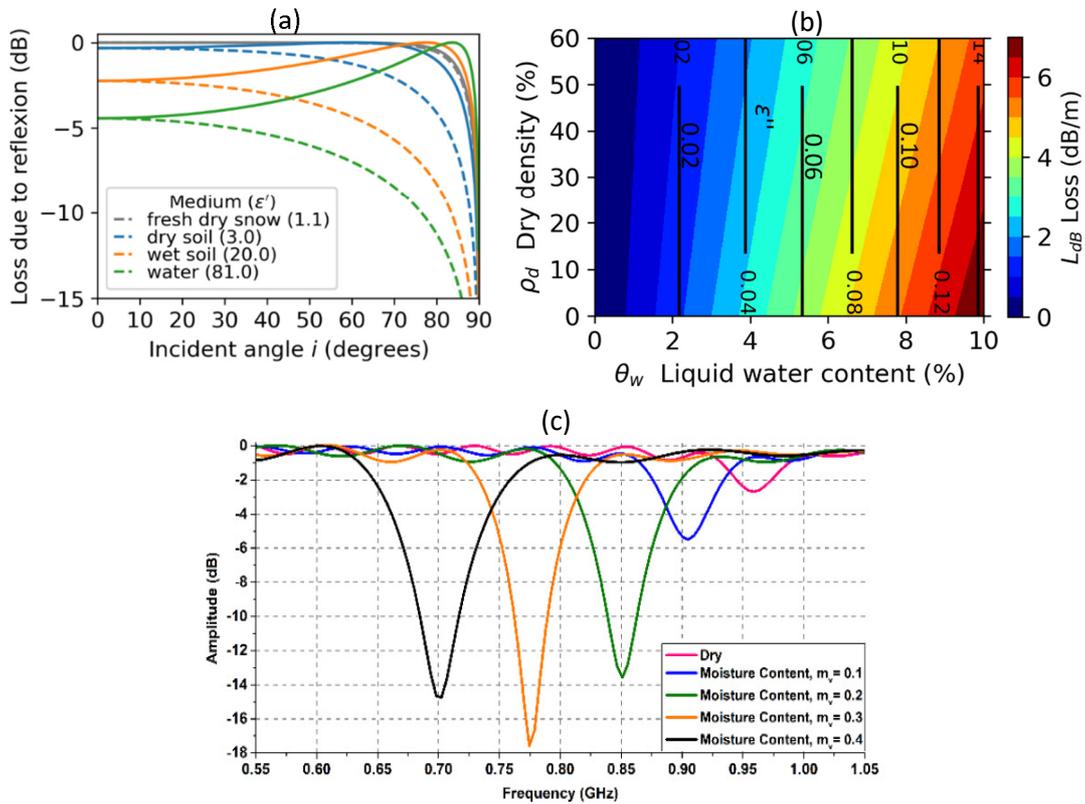

**Fig. 26: The main type of losses into a medium. (a) Loss due to reflection at an interface 1-R², that accounts twice for backward and forward propagation; (b) bulk attenuation in a snow medium depending on its density and liquid water content; (c) change of tag antenna resonance with changing humidity in soil. From (Le Breton, 2019; Dey et al., 2019)**

### 6.2.2 Buried tags

Buried wireless sensor networks (Akyildiz and Stuntebeck, 2006; Akyildiz et al., 2009) offer many potential applications for soil sensing, such as detecting pipes and leakages (Sun et al., 2011; Sadeghioon et al., 2014; Raza and Salam, 2020a). However it undergoes severe bottlenecks to maintain the sensors communication and power (Raza and Salam, 2020b; Salam and Raza, 2020). In particular, as a long lifetime is required for this application (more than a decade), passive and robust devices offer more efficient solutions compared to battery-powered devices. Besides, batteries can be a source of soil contamination due to the chemical substances that compose the battery. Therefore, remotely powered devices, such as passive RFID tags, are preferred. Buried RFID tags were applied for sensing soil temperature and moisture (Deng et al., 2020; Luvisi et al., 2016), and could be useful for monitoring soil in agriculture, for hydrological monitoring, or to detect leakages in buried networks.

However, the high permittivity (up to ε=20) and conductivity of the soil are challenging for reading tags. The permittivity contrast causes antenna detuning and wave reflection at the air interface, and the conductivity causes attenuation. Both greatly reduce the depth at which buried tags can be read, particularly at UHF. Examples of different passive tags and their performances are shown on Table 6.

The attenuation by the soil at UHF, observed both with buried transmitters (Bogena et al., 2009) and buried RFID tags (Elboushi et al., 2020), depends mainly on three physical parameters of the soil: its moisture, the type of porous media, and the conductivity of the pore water. In soils with 0–30% moisture, UHF tags could be interrogated down to depth of 0.6 m (Deng et al., 2020). Low-frequency tags allow potentially larger ground penetration depths, because the attenuation of the soil increases with the frequency. Low-frequency tags serve to track underground pebbles in river beds (see section 2), but also to mark buried construction assets (3M EMS) or metallic pipes (Vyas and Tye, 2019; Zarifi et al., 2017).

**Table 6: Examples of buried passive RFID tags**

| Reference | Depth (m) | Soil moisture | Where | Frequency | Comments |
|---|---|---|---|---|---|
| (3M EMS) | 1.8 | Not indicated | Outdoors | 169 kHz | Industrial device |
| (Vyas and Tye, 2019) | 1.25 | None | Laboratory | 105/15 kHz | Sensing tag on metal |
| (Luvisi et al., 2016) | 0.1 | 10–90% | Outdoors | 865 MHz | Temperature monitoring |
| (Li et al., 2018) | 1.2 | Dry | Laboratory | 915 MHz | Flood warning application |
| (Deng et al., 2020) | 0.6 | 30% | Outdoors | 915 MHz | Sensing soil moisture and temperature |
| (Khokhlova and Delevoye, 2019) | 0.1 | 25% | Laboratory | 40 MHz | The goal was to estimate tag depth. |
| (Elboushi et al., 2020) | 0.5 | Dry | Laboratory | 915 MHz | Characterize different soils |
| (Abdelnour et al., 2018) | 0.6 | 25% | Outdoor | 865 MHz | Exploits frequency doubling |

The loss due to antenna detuning is caused by a shift of the resonant frequency of the tag antenna, which can strongly reduce the performance of the tag. To mitigate this effect, it was proposed to pre-tune the tag antenna to the permittivity of a dry soil, instead of air (Alonso et al., 2017). However, the soil permittivity, and therefore the tag tuning, fluctuates considerably depending on the soil type and moisture. To adapt to the different soil properties, it is proposed to use wide-band buried antennas (Mondal et al., 2018b). However, using a wide-band antenna supposes nonconductive soil, which is very rare in practice. Another solution is to sweep the emitted RFID signal frequency to communicate at the tuning frequency of the tag (Salam et al., 2019). However regulation and commercial readers prevent this. Alternatively, improvement might be attained by deploying multiple tags each pre-tuned to a given permittivity, or self-tuning tags that adapts to strong changes of medium.

The contrast of permittivity at the air-soil interface strongly increases the reflection coefficient and decreases the transmission coefficient (Mondal et al., 2018a). It is important to note that the impact of the reflection coefficient counts twice: a first time for the interrogation signal from a reader and a second time for the backscatter signal from the tag. The strong reflection coefficient can pose challenges not only for the propagation within the soil but for the tag antenna and the radiofrequency electronics. As a rule of thumb, besides the degradation of reflection coefficients, the global transmission coefficient is harder to optimize and to monitor in general. Theoretically, the reflection coefficient can be minimized by interrogating with a vertical angle of incidence, or with an angle that has a total transmission (the Brewster angle, which occurs only at a vertical polarization). But in practice, the implementation of such a technique remains hazardous and very complex to exploit. Alternatively, the negative impact of reflection at the ground-air interface may be reduced by using harmonic tags. Harmonic tags generate a response signal at the double frequency of the incoming carrier signal, which the reader can more easily distinguish from the ground reflection (Abdelnour et al., 2018) (Fig. 27).

Additionally, any technique that increases the read range in the air (see 6.3), such as tags with high-gain antennas (Chang et al., 2009; Hautcoeur et al., 2013), should also increase the reading depth of buried tags.

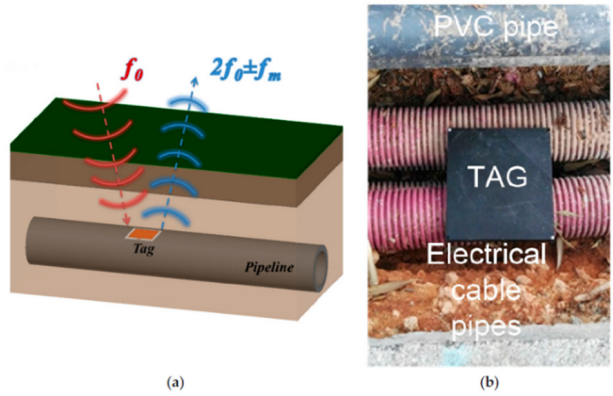

**Fig. 27: (a) Block diagram of the proposed harmonic system and (b) an image of a typical installation, buried down to 50 cm below ground. From (Abdelnour et al., 2018)**

### 6.2.3 Underwater tags

RFID, just like other radiofrequency technologies, is very sensitive to the high dielectric constant and ionic conductivity of water, which affects radiofrequency communication (Che et al., 2010). While water is not a natural conductor, the presence of dissolved impurities and salts transforms water into a partial conductor, which attenuates the propagated wave. From an electromagnetic point of view, two main parameters characterize water: its relative dielectric permittivity, typically 80, and its conductivity that depends on the impurities (salinity in particular). For pure water, which can be considered as an ideal lossless medium (conductivity of the order of $10^{-4}$ S/m), the propagation of electromagnetic signals is possible. However, there are three major sources of loss. First, the interface between air and water results in huge power reflection coefficients as high as 64%; it corresponds to a round trip air-water-air transmission penalty of 9 dB. Second, propagation of a distance equal to penetration depth does result roughly in a 9 dB attenuation. So, considering a link of length equal to the penetration depth, the penalty due to water attenuation and air-water interface reflection is approaching 20 dB. From Table 7 one can conclude that only LF and HF frequencies can be considered for the implementation of RFID in water environment (Y. Li et al., 2019). Third, antennas immersed under water will see a change in tuning frequency. Therefore underwater tags, particularly at UHF, would require an antenna that is specifically optimized for water (e.g., Sohrab et al., 2016).

The wave frequency is the main factor that impacts the link budget and the global performance. At Extremely Low Frequency (3–300 Hz), radio signals can travel in sea water over hundreds of meters. At Very Low Frequency (3–30 kHz), radio signals can penetrate sea water up to 20 meters. While these two frequency bands exhibit comfortable communication distances, they present severe technical limitations and practical challenges. In particular, their extremely long wavelengths require very bulky antennas of very large dimensions. Also, because of the narrow bandwidth, these frequencies can be used to transmit at very slow data rates only. Now considering the standardized RFID frequency bands, depending on quality of water, in particular its conductivity, the penetration depth of radio waves can be evaluated for different categories of waters and data are reported in Table 7.

**Table 7: Penetration depth (distance where the electrical and magnetic fields are reduced by a 1/e factor) of radio waves into water. From (Benelli and Pozzebon, 2013)**

|  | Low Frequency (125 kHz) | High Frequency (13.56 MHz) | Ultra High Frequency (900 MHz) | Microwaves (2.45 GHz) | Microwaves (5.8 GHz) |
|---|---|---|---|---|---|
| Fresh Water (0.003 S/m)<br>Fresh Water (0.2 S/m) | 26 000 cm<br>320 cm | 250 cm<br>30 cm | 26.5 cm<br>3 cm | 18.6 cm<br>2.3 cm | 12 cm<br>1.5 cm |

| Salt Water (4 S/m) | 71 cm | 6.8 cm | 0.7 cm | 0.5 cm | 0.3 cm |

Despite these limiting effects, some studies have demonstrated that in well-controlled situations, low-frequency RFID is usable even under water (Benelli and Pozzebon, 2013). Among the significant examples of underwater applications, RFID has been used for the monitoring of coastal dynamics by embedding tags into pebbles (see section 2). In addition to the river in applications, a few tests were made with HF (13.56 MHz) tags in salt water, but the read range was limited to a few cm (Benelli et al., 2009). The paper concluded that the results agree with the theoretical analysis *i.e.*, the achieved read range is lower than the penetration depth. In other contexts, low-frequency tags were applied to monitoring seaweed growth and sewers (Peres et al., 2020, 2021; Tatiparthi et al., 2021), and considered for supporting the navigation of autonomous underwater vehicles (Harasti et al., 2011).

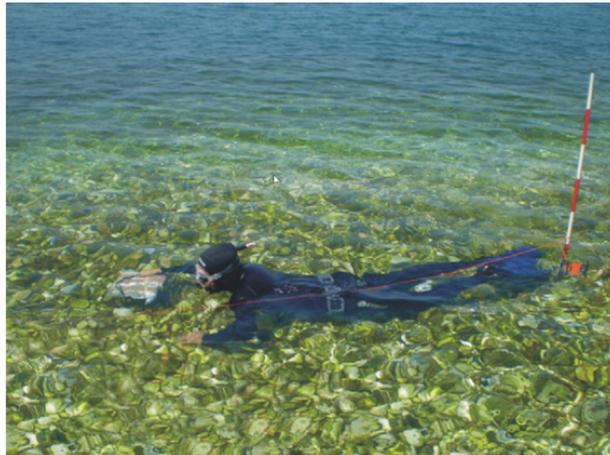

**Fig. 28: Example of underwater RFID survey. From (Benelli et al., 2009).**

### 6.2.4   Snowy and icy environments

Snow covers most of mountainous and cold regions in winter. It complicates the maintenance of wireless sensor networks supplied by wind turbines or solar panels. RFID tags have the advantage to work when covered by snow, without the need to be powered. However, very little literature deals with using RFID tags in snowy conditions. Nummela et al. (2008) showed that tags could continue to operate when covered with a few millimeters of snow. Le Breton (2019) has demonstrated that tags can be interrogated across several meters of snow. A snowpack on the ground affects the radiofrequency signal mostly through transmission in snow and at interfaces, interference from waves reflected at the snow surface, and proximity coupling with the antennas(Le Breton et al., in preparation). These influences depend strongly on the properties of the snowpack, particularly the snow depth, dry density, and liquid water content, which strongly vary in time and space. A transmitted wave can be slowed by a factor of 0–80%, and the signal loss attenuated by 0 – 12 dB per meter of transmission in snow. The influences on multipath interferences are of the order of ±2 rad (equivalent to ±5 cm of propagation at 866 MHz in the air) and ±12 dB on the amplitude. As a consequence, snowfall events have shown to induce a reversible and temporary error of tag localization outdoors of the order of ±8 cm (Le Breton et al., 2019). The reading range of tags under the snowpack or at <10 cm of its surface may also reduce drastically (Le Breton et al., in preparation)

The reading range with a tag under snow depends mostly on the snow liquid water content and on the angle of incidence with the snowpack surface. With vertical incidence and dry snow, the propagation is negligible (−1 to −2 dB). The tag antenna might be detuned by dry snow at its vicinity (the Confidex Survivor B rugged tags, with patch antennas, suffered no significant loss, yet other tags designs may experience a loss). With wet snow however (e.g., a liquid water content of 10% on the surface and 5% on average, with 500 kg/m³ density), the snowpack would attenuate the power by −3 dB/m (−6 dB/m both ways) and the tag's performance would fall (typically -10 dB to -20 dB based on the water's effect (Occhiuzzi et al., 2013). To provide orders of magnitude, a system with a 20m read range in the air would read a tag across 15-20m of dry snow at vertical incidence, only 2 m in wet snow, and <1 m just below the surface at an oblique angle. In practice, other influences on the reading range, such as the multipath interferences, the 3D heterogeneity of the liquid water content and the presence of impurities in the snowpack, may require to run a more detailed estimation.

The formation of ice layers on tags also influences the RFID signal, affecting both the phase and the signal strength, as observed by Le Breton et al. (2019) and quantified by (Wagih and Shi, 2021a). Two phenomena are identified. First, the ice layer forms progressively, typically during the night. Given the low permittivity of ice, it influences the phase and deteriorates the signal strength only moderately, and progressively. Then heat provided by solar radiation in the morning can rapidly melt this layer resulting in a strong loss of RSSI and increase of phase delay, temporarily until the water evaporates.

### 6.2.5 Vegetated environments

Radiofrequency methods have the advantage of transmitting across vegetation cover, as compared to remote optical sensors such as LIDAR, tacheometers or cameras. It could be useful to use RFID tags to collect sensor data across vegetation, for example on a vegetated landslide or an agriculture field. Le Breton (2019) shows that reading RFID tags across a high-grass vegetation can slow the radiofrequency wave by 0.1 – 1% and attenuate it by 0.2 – 1 dB/m. These numbers strongly depend on the density and on the moisture of the vegetal cover. The interrogation and the localization of RFID tags across a few meters of vegetal cover are therefore almost not disturbed. However, the communication could decrease the performances more significantly across tens of meters, both in terms of reading range and localization accuracy. This influence could potentially be exploited for sensing vegetation properties using RFID. Indeed, monitoring the slowness and attenuation of the radiofrequency wave might provide a rough estimate of the grass moisture content (Le Breton, 2019).

### *6.3 Read range, duration and cost of the tags*

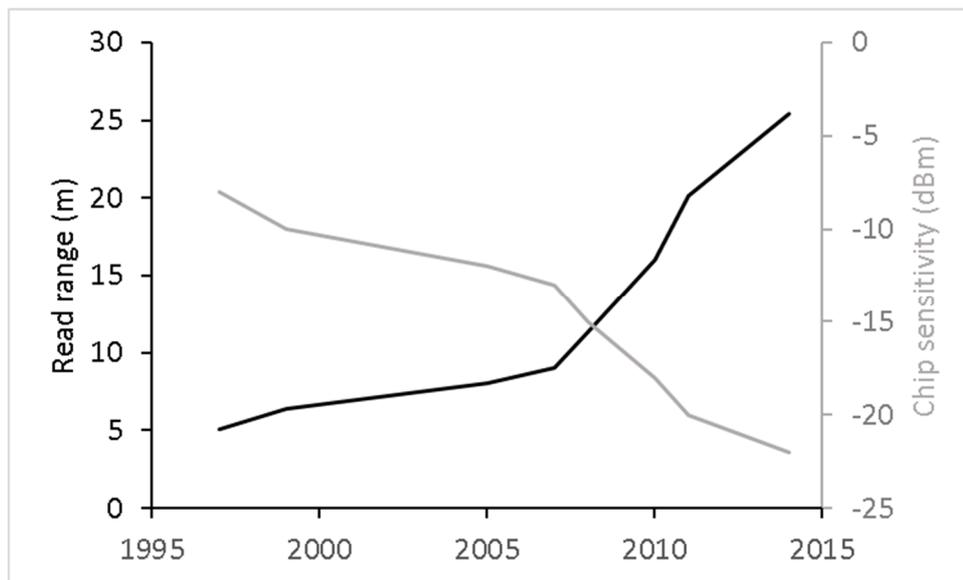

**Fig. 29: Evolution of purely passive RFID chip sensitivity (in dBm, right scale) and corresponding read range corresponding to the FCC (Federal Communication Commissions, USA) legislation using an ideal semi-dipole tag antenna (in meter, left scale). Adapted from (Nikitin et al., 2012; Durgin, 2016)**

Monitoring large areas, of several hundreds of meters long, would particularly benefit from a dense network of low-cost sensors. The reading range, of around 50 m in recent outdoor implementations (Le Breton et al., 2019), is increasing with the improvement in sensitivity of the RFID devices. Indeed, one of the essential properties of RFID chips is their sensitivity to radiofrequency signals which corresponds to the minimum power necessary to activate the RFID chip and therefore allow the communication with an RFID reader. Modern passive RFID chips require less than -22 dBm to be activated, resulting in a 25 m read range under the strict compliance with radiofrequency emission standards (European ETSI or North American FCC). The evolution of this sensitivity and the corresponding read range since the beginning of the 21st century is summarized in Fig. 29. Today, off-the-shelf readers with the same antenna for emission and reception (called monostatic) have typically sensitivities of −80 dBm to −95 dBm. The sensitivity of batteryless integrated circuits are typically −7 to −16 dBm with sensing capabilities that consume additional power, down to −23 dBm for standard passive tags, and down to − 35 dBm when adding a battery assistance. In this last case, the chip is no longer the limiting factor for the reading range, and the limit then comes from the reader. The sensitivity tends to decrease with time, both for the tags (Nikitin et al., 2012; Durgin, 2016) and for the readers, which will probably lead to increases in reading range in the future. Recently, research has reached reading ranges of several hundreds of meters using passive tags (Durgin, 2016; Amato et al., 2018; Qi et al., 2019), while remaining in the legislation limits, thanks to original microelectronic designs. Work still has to be performed to adapt these experimental tags to standard RFID interrogators. Another way to increase the reading range is to develop tags using directive antennas (e.g., Lin et al., 2016; Zuffanelli et al., 2016). Such tag antennas do not fit to the majority of the RFID use cases: they can be read only from specific directions, and they are larger and more expensive than with the typical small tags. But directive tag antennas make the monitoring of a large zone, such as a landslide, much more practical, by increasing the reading range.

The duration of sensing tags can be a challenge for monitoring over years to decades. The limit concerns the tags with advanced capabilities which need battery assistance, and the robustness of the tags over time. Battery assistance is often used today, to enable advanced capabilities of tags, such as permanent monitoring or longer reading range. However, batteries can sometimes be problematic in areas where maintenance is difficult. Energy harvesting could eliminate the need for battery assistance in most sensing tags, by converting small amounts of energy available in the environment into a usable electrical form (Ferdous et al., 2016). For example, energy harvesting was implemented on tags using photovoltaic cells (Abanob Abdelnour et al., 2019; Kantareddy et al., 2020, 2019, 2019; Valentine et al., 2015), electrochemical reactions (Kantareddy et al., 2018), vibrations (Lu et al., 2017), heat dissipation (Jauregi et al., 2017) or radiofrequency wave field (Allane et al., 2016; Deng et al., 2020; Zhu et al., 2021). Developments in energy harvesting could allow for monitoring earth processes with advanced sensing tags for decades. The second challenge related to duration concerns the long-term robustness of RFID tags outdoors. The longest deployment reported in the literature was six years under a bridge, and showed no apparent degradation (Watkins et al., 2007). Accelerated aging was also studied with cycles of high temperature (Taoufik, 2018) or changing humidity (Saarinen et al., 2014). Low-cost tags were rapidly damaged by this accelerated aging, but rugged industrial tags were designed to resist them. In our experience, rugged tags have worked well since their installation over at least four years (2017–2020), on a harsh mountainous environment.

The low cost per tag is a key advantage of the RFID technology, particularly on price-sensitive applications (e.g., soil moisture for agriculture). Yet, the cost per tag can increase significantly on applications that require a specific new tag. Indeed, developing a new sensing tag requires to design a tag and to set up a production (Węglarski and Jankowski-Mihułowicz, 2019). This initial investment could strongly increase the cost per tag with small volumes of production (e.g., thousands of tags). In this case, it is preferable to use a sensing approach that works with existing commercial tags. For example, localization and propagation-based sensing exploit physical measurement principles that are independent from the tag. Many antenna-based sensing can also use standard or self-tuning tags available on the market (e.g., J. Wang et al., 2020). Several industrial tags also embed temperature sensors (see Table 2). Finally, demonstration tags sold by producers of sensing chips (e.g., ASYGN, FARSENS) can sense a myriad of physical parameters. More sensing tags are also likely to be produced in the future as the RFID market continues its growth, for large outdoor applications such as structural health monitoring or precision agriculture (Zhang et al., 2017; Wasson et al., 2017). As for today, we suggest using commercial off-the-shelf tags when possible.

### 6.4   UAVs to access harsh fields

Unmanned Aerial Vehicles (UAVs) are now commonly used for earth surface sensing (Bemis et al., 2014). They can monitor long-term changes in topography with optical methods (Niethammer et al., 2012; Casagli et al., 2017; Peppa et al., 2017) or synthetic aperture array radar (Li and Ling, 2015). Passive targets are sometimes placed on the ground to improve the accuracy. Surveying the earth surface using UAVs proves particularly effective to monitor zones that are not accessible by land vehicles and difficult to access on foot, such as cliffs, landslides, steep hill slopes, non-wadable rivers, highly vegetated environments. They are also helpful when access is forbidden or dangerous, such as on volcanoes or minefields. A single UAV carrying the RFID reader antenna can replace a large series of interrogators fixed on the ground, thus reducing the operation cost (for time-lapse monitoring) and allowing deployment over very large areas. It may enable the exploitation of thousands of tags over wide areas. This section presents the latest research that uses UAVs to deploy, localize and collect data from sensor networks and passive RFID tags.

UAVs have been demonstrated to collect data from passive radiofrequency sensors on the deformation of a bridge, the humidity on a tree, and the soil moisture (Mascareñas et al., 2008; Wang et al., 2015) (Fig. 30). To ensure a reliable reading with standard antennas and transmit power, the UAV might need to fly a few meters away from the ground at a maximum velocity of a few m/s (Casati et al., 2017).

Localization of RFID tags could offer important advantages compared to widely used optical methods. Indeed, provided the UAV can fly with GPS-defined itinerary, it could work 24-24 h without sun under any meteorological conditions (except high-speed wind). Tags could also be read under a forest, under a dense vegetation, or under snow, allowing measurements of places where optical acquisition are difficult, if not impossible. The UAV-RFID method would also greatly reduce the size and processing time of the raw data.

Tags have first been localized from UAV with a metric accuracy using the signal amplitude measurement (RSSI) (PINC Inc., n.d.; Greco et al., 2015). Then, centimetric accuracy has been reached using synthetic aperture array radar with phase measurements (Buffi et al., 2017, 2019; Ma et al., 2017) which was presented in previous sections. Best accuracy was reached with increased synthetic aperture length, and with relative ranging of an array of tags. The bottleneck to reach higher accuracy is the localization of the UAV. Outdoors, Buffi et al. (2019) located the antenna carried by the UAV with a differential single-frequency GPS receiver, and estimated the error as +/- 2 cm. Many solutions—out of the scope of this review—exist to localize UAV with higher accuracy, such as dual-frequency GPS, optical positioning or ultra-wide band local positioning. Zhao et al. (2019) proposed a joint optimization algorithm for localizing tags using SAR methodology while compensating for aperture position error. The localization of the UAV itself from a grid of reference tags has also been investigated, using the signal amplitude (Longhi et al., 2018b), and suggested with using the phase (Ma et al., 2017; Savochkin et al., 2019). As a side usage, thanks to reciprocity, it might help to navigate UAVs in areas with bad GPS reception, such as on a cliff, steep mountainous slopes, river gorges, tunnels or caves.

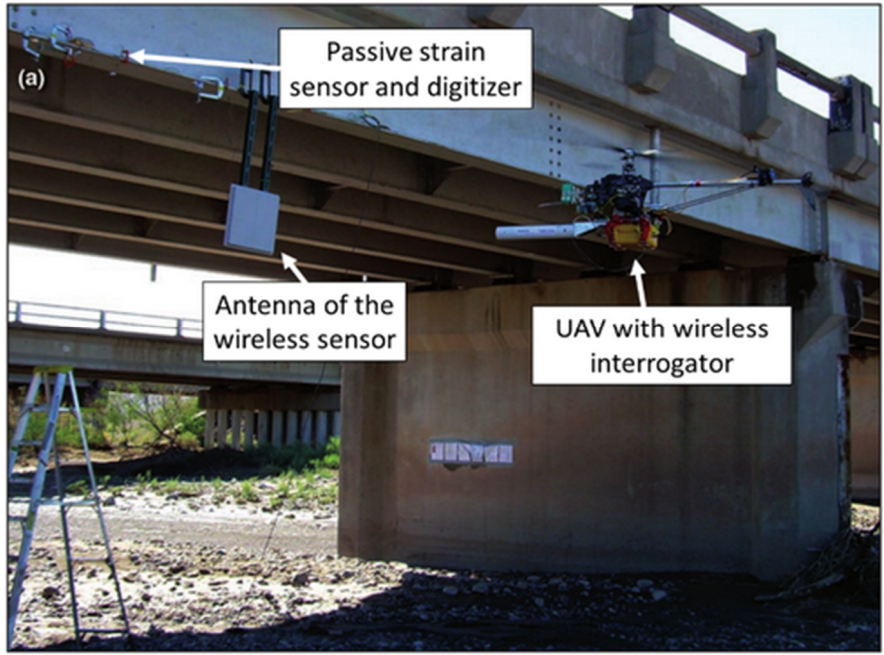

**Fig. 30: Bridge monitoring with an interrogator (reader antenna) attached to a UAV that powers wirelessly a passive displacement sensor. From Mascareñas et al. (2008)**

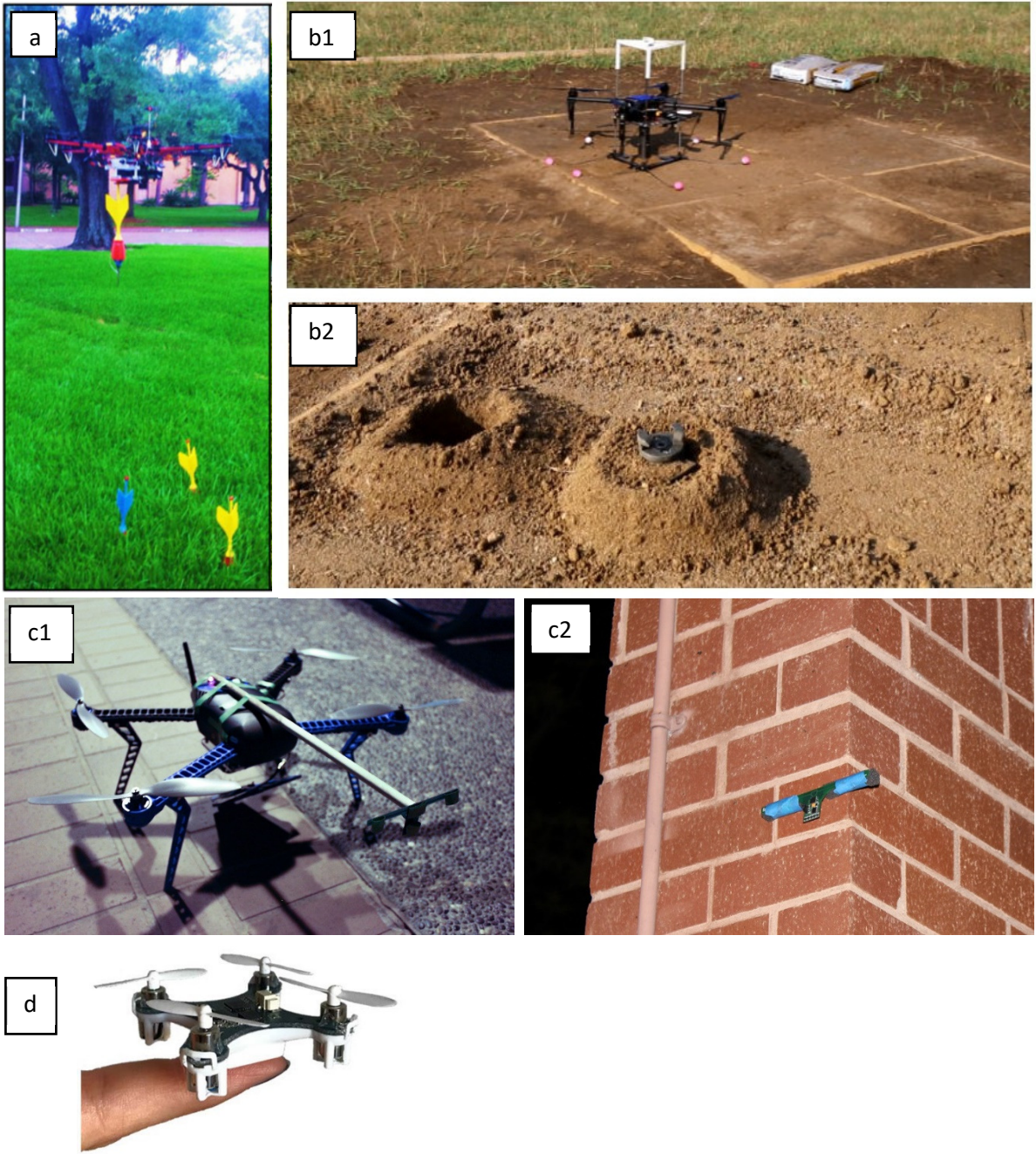

Fig. 31: UAV deployment of sensors that could be used for earth science: (a) seismic sensor dropped from a UAV and buried using their falling energy; (b1-2) soil moisture sensors buried after using a shallow-drilling-UAV; (c1-2) RFID tag installed high above ground by a UAV; (d) RFID tag attached to a micro-UAV deployed together on the measurement site. From (Sudarshan et al., 2017; Sun et al., 2018; Wang et al., 2015; Longhi and Marrocco, 2017).

Finally, deployment of sensors over low-accessibility terrains could be facilitated with the combination of UAV and RFID. Indeed, active wireless sensors were deployed from UAV on low accessibility terrains (e.g., Corke et al., 2004; Bernard et al., 2011), to measure air conditions (Martínez-de Dios et al., 2017; Melita et al., 2015), seismic vibrations (Sudarshan et al., 2017; Lys et al., 2018) and underground humidity (Sun et al., 2018) (Fig. 31a, b). However to recharging or replacing their battery would also require a UAV which is challenging (Bertoldo and Allegretti, 2015; Sheu et al., 2008). Deploying passive RFID tags instead would present many advantages: they are low-cost, light to carry, and batteryless (Luvisi et al., 2016). Recently, UAVs were tested to deploy sensing tags high above the ground or on a water body (Wang et al., 2015). Due to their very light weight, tags can be attached permanently on a microdrone and read from the ground (Fig. 31d). In this case, the microdrone moves to the zone of measurement, stays on this zone for the time of the monitoring, then comes back (Longhi et al., 2018b, 2018a; Longhi and Marrocco, 2017). The small size and low cost of microdrones make it less risky to deploy in areas with little space and high risk of a crash or collision.

Overall, UAVs could prove very useful to deploy RFID tags and collect data, over dangerous and low-accessibility terrains, such as cliffs, near-to-collapse unstable slopes, highly vegetated areas, minefields, highly contaminated sites, volcanoes, or water bodies. Multiple studies demonstrated the ability to collect sensing data and to localize tags with accuracy smaller than 10 cm, by carrying an RFID reader under a UAV. The ability for deployment is still challenging, but deploying light RFID tags should be easier than heavier and more costly active wireless sensors.

### 6.5   Sustainability and green RFID

The production and operation of large numbers of RFID sensors will generate electronic waste and consume energy. This challenge is tackled within the Green Internet of Things concept, which aims at reducing the waste and energy consumption of wireless sensing systems (Zhu et al., 2015). RFID and WSNs are considered among the most promising wireless systems to enable Green Internet of Things (Shaikh et al., 2017; Khan et al., 2021). RFID tags specifically are produced with little material and energy, and their ability for batteryless sensing is essential to increase sensor lifespan and reduces toxic wastes.

Yet, RFID tags generate electronic wastes (Condemi et al., 2019; Bukova et al., 2021) that are difficult to recycle (Aliaga et al., 2011). The waste is reduced with smaller tags that use less polluting material (Voipio et al., 2021) or even biodegradable materials, and by avoiding battery assistance. However, the tags deployed outdoors for years may oppose those guidelines. Indeed, the tags need a rugged protective casing to withstand harsh outdoor conditions. The tags may also require large antennas to increase the read range. Finally, they may need a battery assistance for sensing or for increasing the read range even further. Avoiding the battery assistance would both increase the tag's lifespan and reduce its toxic wastes: it appears as a priority to enable sustainable RFID applications. Removing batteries may require to reduce the energy consumption of the sensing tag and to increase the chip sensitivity, or otherwise to embed a complementary energy harvesting module (Ferdous et al., 2016). Finally, moving the reader close to the tag—for example using a UAV—also increases the energy received by the tag, allowing for smaller and batteryless tags.

In an RFID system, the reader consumes most of the energy and also generates electronic wastes. That is particularly true for permanent reading station installed on the field outdoors, that may need heavyweight batteries and powering through solar panels or wind turbines. In this case, reducing the reader power consumption strongly reduces the environmental cost of the infrastructure. The power can be reduced simply by alternating acquisition with sleeping time or by reducing the power emitted. In the future, more efficient communication protocols between the reader and tag should also reduce this energy consumption (Su et al., 2019). Finally, using a mobile reader would again solve this issue, by avoiding to install a permanent infrastructure on each monitored site.

Today, RFID is among the most promising technology to enable green Internet of Things: tags generate much less waste than other wireless sensors. RFID systems with mobile readers or mobile tags should allow for sustainable and low-cost monitoring installations. The low cost of tags and data collection should allow to obtain data from new places where standard instrumentation was too costly. The efficiency and saving allowed by this data, such as predictive maintenance or risk reduction, is likely to overcome the environmental cost of the RFID system itself (Bose and Yan, 2011; Nasiri et al., 2017; Gladysz et al., 2020).

# 7 Conclusions

RFID research develops the ability to use passive tags as sensors or location trackers. This opens many opportunities to monitor the dynamics of the earth surface with dense networks of low-cost and long-lasting tags.

As for now, the identification function of passive RFID has been used widely for the monitoring of sediment transport in rivers (more than sixty studies) and on coasts (fifteen studies). In these studies, low-frequency tags are inserted in sediments of various sizes (from centimeters to meters long) and localized with readers from distances below 1 m. In general, passive RFID offers higher efficiency in terms of cost and operation compared with other methods (e.g., visual, radioactive, magnetic). This enables longer studies (several years), with more tracers (up to thousands), that are tracked individually. The better information allows understanding more finely the natural processes observed.

The augmented RFID technologies—localization and sensing—are more recent; their usage is now emerging in earth science. The centimetric localization of ultra-high frequency tags allowed for continuous displacement monitoring on three landslides, during years. The detection of millimetric displacement thresholds have been deployed to monitor unstable rocks on a cliff. The monitoring of underground temperature with buried tags was used to monitor the efficiency of solarization processes used in agriculture. Monitoring the phase delay on tags on the ground allowed to estimate the increase of snow water equivalent caused by dry snowfall events during a whole winter. Finally, soil moisture sensors have been deployed both in industrial greenhouse to monitor the moisture of plant pots, and outdoors using buried tags.

The variety of deployment topologies across the studies confirms that passive RFID is enabling new possibilities for wireless sensing. Surveys for sediment tracers detection were firstly operated with handheld readers carried by a human operator. More efficient surveying methods have now been developed, using unmanned aerial vehicles or land vehicles. They provide more data (larger areas, better recovery rates, more frequent surveying) which again helps to better understand the monitored processes. As an alternative to surveying, several studies deployed fixed readers monitoring infrastructure. Continuous monitoring can improve the understanding of natural processes that are time sensitive (e.g., step-by-step sediment displacement or landslide reactivation). It also opens the door for operation surveillance applications (e.g., rockfall warning). However, the reader infrastructure is more expensive than the tags, it requires maintenance, and it has limitations in the monitored area (tag-reader distance below 100 m).

Many other RFID sensing techniques have been developed and tested in the laboratory, and could be adapted to a real field application, to measure soil indicators with high spatial density over several years. On landslides, tags may monitor soil moisture and tilt. On rocky structures, tags could monitor cracks, vibration modes and vibration intensity. Tags may also monitor changes in harsh areas such as the cryosphere, volcanoes, densely vegetated areas, or shallow underground.

The low-cost, low maintenance, small size and long duration characteristics of passive wireless tags should make them effective compared to active wireless sensors, in several scenarios: mostly, price-sensitive applications that use large amounts of sensors (e.g., soil moisture); monitored objects that need small, flexible or little-intrusive devices (e.g., small pebbles, or snow); monitoring during decades; environmental concerns that require minimal waste (e.g., to avoid leaving batteries in rivers); areas with difficult access for maintenance (e.g., underground); sensors that are likely to be lost (e.g., on a landslide, on a river, in the soil); measurement that exploits the phase delay (e.g., snowpack monitoring over its volume, or localization).

The drawbacks are mainly the reading range and the sensing accuracy. Reading range reaches an order of magnitude of 1 m with low-frequency tags, 10 m with ultra-high frequency batteryless sensing tags and up to 100 m with battery-assisted tags. Fortunately, the reading range is constantly growing with technological improvements. Furthermore, the tag localization or sensor data collection by a mobile reader (including unmanned aerial vehicles) should allow to cover larger areas. The accuracy of RFID sensing tags, often lower than conventional instruments, can also limit the potential applications. The accuracy is particularly challenging on batteryless tags with dedicated sensors (due to the little energy available to power the acquisition electronics) and on antenna-based sensors (due to the combination of many influences).

In perspective, many RFID sensing and localization techniques may be mature enough to deploy comprehensive "real-world" applications, reusing and assembling concepts previously demonstrated in the laboratory. That has recently led to new applications, for monitoring landslide displacements, soil moisture, soil temperature, and snow water equivalent. Furthermore, future deployments may grow from tens to thousands of RFID sensors, similarly to the earlier growth in the application of fluvial and coastal sediment tracking. This will fully exploit the low-cost advantage of passive RFID, as compared to wireless sensor networks or manual measurements. Finally, new concepts of propagation-based RFID sensing remain to be developed and demonstrated, with the advantage that they can use any commercially available tag. These concepts could particularly work outdoors where multipathing is less challenging than indoors. Overall, the development of RFID for geoscience instrumentation will require interdisciplinary collaborations between research laboratories and private companies, and between various scientific domains (from electronics and applied mathematics to bio/geosciences) and should create many opportunities for technological innovations.

**Table 8: Synthesis of the RFID monitoring systems reviewed in this article.**

| Measurand | Application | Measured values | Sensing approach | Maturity |
|---|---|---|---|---|
| Displacement | River & coastal sediments | 1–1000m+ | Localization - power | Field, >60 studies |
| | Landslides | 0.01–50m+ | Localization - phase | Field, 4 sites |
| | Rock fissure | >1 mm | Threshold extensometer | Field |
| Tilt | Boulder | >1° | MEMS or 1-bit | Laboratory |
| Vibrations | Shocks | 0.5–500 g | 1-bit switch | Laboratory |
| | Vibration control | 1 mg – 2 g | MEMS | To demonstrate |
| | Structure Resonance | 1–50 Hz | MEMS | Laboratory |
| Material characteristics | Soil moisture | 5 – 100% | Capacitive / resistive | Industry |
| | Snow water equivalent | 3-350 kg / m$^2$ | Phase delay | Field |
| | Soil temperature | 0 – 45 °C | Embedded sensor | Field |

## 8 Acknowledgment

We acknowledge funding and support from the project RISQID granted by the Région Auvergne-Rhône-Alpes region, from the French National Agency for research (ANR) through the LABCOM Geo3iLab project (ANR- 17- LCV2-0007-01), from the Labex OSUG@2020, and from the companies Géolithe and Géolithe Innov. We thank D. Jongmans, P. Benech, E. Rey, F. Guyoton and the engineers from Géolithe company for fruitful discussions.

## 9 References

3M EMS, n.d. Locating and Marking Full Portfolio Brochure.

Abdelnour, A., Fonseca, N., Rennane, A., Kaddour, D., Tedjini, S., 2019. Design of RFID Sensor Tag for Cheese Quality Monitoring, in: 2019 IEEE MTT-S International Microwave Symposium (IMS). Presented at the 2019 IEEE MTT-S International Microwave Symposium (IMS), pp. 290–292. https://doi.org/10.1109/MWSYM.2019.8700769

Abdelnour, Abanob, Hallet, A., Dkhil, S.B., Pierron, P., Kaddour, D., Tedjini, S., 2019. Energy Harvesting Based On Printed Organic Photovoltaic Cells for RFID Applications, in: 2019 IEEE International Conference on RFID Technology and Applications (RFID-TA). Presented at the 2019 IEEE International Conference on RFID Technology and Applications (RFID-TA), pp. 110–112. https://doi.org/10.1109/RFID-TA.2019.8892206

Abdelnour, A., Lazaro, A., Villarino, R., Kaddour, D., Tedjini, S., Girbau, D., Abdelnour, A., Lazaro, A., Villarino, R., Kaddour, D., Tedjini, S., Girbau, D., 2018. Passive Harmonic RFID System for Buried Assets Localization. Sensors 18, 3635. https://doi.org/10.3390/s18113635

Akbar, M.B., Taylor, D.G., Durgin, G.D., 2015. Hybrid Inertial Microwave Reflectometry for mm-Scale Tracking in RFID Systems. IEEE Transactions on Wireless Communications 14, 6805–6814. https://doi.org/10.1109/TWC.2015.2460250

Akyildiz, I.F., Stuntebeck, E.P., 2006. Wireless underground sensor networks: Research challenges. Ad Hoc Networks 4, 669–686. https://doi.org/10.1016/j.adhoc.2006.04.003

Akyildiz, I.F., Sun, Z., Vuran, M.C., 2009. Signal propagation techniques for wireless underground communication networks. Physical Communication 2, 167–183. https://doi.org/10.1016/j.phycom.2009.03.004

Aliaga, C., Ferreira, B., Hortal, M., Pancorbo, M.Á., López, J.M., Navas, F.J., 2011. Influence of RFID tags on recyclability of plastic packaging. Waste Management 31, 1133–1138. https://doi.org/10.1016/j.wasman.2010.12.015

Allan, J.C., Hart, R., Tranquili, J.V., 2006. The use of Passive Integrated Transponder (PIT) tags to trace cobble transport in a mixed sand-and-gravel beach on the high-energy Oregon coast, USA. Marine Geology 232, 63–86. https://doi.org/10.1016/j.margeo.2006.07.005

Allane, D., Vera, G.A., Duroc, Y., Touhami, R., Tedjini, S., 2016. Harmonic Power Harvesting System for Passive RFID Sensor Tags. IEEE Transactions on Microwave Theory and Techniques 64, 2347–2356. https://doi.org/10.1109/TMTT.2016.2574990

Alonso, D., Zhang, Q., Gao, Y., Valderas, D., 2017. UHF passive RFID-based sensor-less system to detect humidity for irrigation monitoring. Microwave and Optical Technology Letters 59, 1709–1715. https://doi.org/10.1002/mop.30611

Amato, F., Torun, H.M., Durgin, G.D., 2018. RFID Backscattering in Long-Range Scenarios. IEEE Transactions on Wireless Communications 17, 2718–2725. https://doi.org/10.1109/TWC.2018.2801803

Angeli, M.-G., Pasuto, A., Silvano, S., 2000. A critical review of landslide monitoring experiences. Engineering Geology 55, 133–147. https://doi.org/10.1016/S0013-7952(99)00122-2

Arnaud, F., Piégay, H., Béal, D., Collery, P., Vaudor, L., Rollet, A.-J., 2017. Monitoring gravel augmentation in a large regulated river and implications for process-based restoration. Earth Surface Processes and Landforms 42, 2147–2166. https://doi.org/10.1002/esp.4161

Arnaud, F., Piégay, H., Vaudor, L., Bultingaire, L., Fantino, G., 2015. Technical specifications of low-frequency radio identification bedload tracking from field experiments: Differences in antennas, tags and operators. Geomorphology 238, 37–46. https://doi.org/10.1016/j.geomorph.2015.02.029

Arnitz, D., Muehlmann, U., Witrisal, K., 2010. UWB ranging in passive UHF RFID: proof of concept. Electronics letters 46, 1401–1402. https://doi.org/10.1049/el.2010.1703

Aroca, R.V., Hernandes, A.C., Magalhães, D.V., Becker, M., Vaz, C.M.P., Calbo, A.G., 2018. Calibration of Passive UHF RFID Tags Using Neural Networks to Measure Soil Moisture. Journal of Sensors 2018, 1–12. https://doi.org/10.1155/2018/3436503

Aroca, R.V., Hernandes, A.C., Magalhães, D.V., Becker, M., Vaz, C.M.P., Calbo, A.G., 2016. Application of Standard EPC/GEN2 UHF RFID Tags as Soil Moisture Sensors. Proceedings 1, 10. https://doi.org/10.3390/ecsa-3-S5001

Arthaber, H., Faseth, T., Galler, F., 2015. Spread-Spectrum Based Ranging of Passive UHF EPC RFID Tags. IEEE Communications Letters 19, 1734–1737. https://doi.org/10.1109/LCOMM.2015.2469664

AS321X, n.d.

Atzori, L., Iera, A., Morabito, G., 2010. The Internet of Things: A survey. Computer Networks 54, 2787–2805. https://doi.org/10.1016/j.comnet.2010.05.010

Babar, A.A., Manzari, S., Sydanheimo, L., Elsherbeni, A.Z., Ukkonen, L., 2012. Passive UHF RFID Tag for Heat Sensing Applications. IEEE Transactions on Antennas and Propagation 60, 4056–4064. https://doi.org/10.1109/TAP.2012.2207045

Balaji, R.D., Malathi, R., Priya, M., Kannammal, K.E., 2020. A Comprehensive Nomenclature Of RFID Localization, in: 2020 International Conference on Computer Communication and Informatics (ICCCI). Presented at the 2020 International Conference on Computer Communication and Informatics (ICCCI), pp. 1–9. https://doi.org/10.1109/ICCCI48352.2020.9104083

Balanis, C.A., 2012. Advanced Engineering Electromagnetics, Second Edition. ed. John Wiley and Sons.

Bel, C., 2017. Analysis of debris-flow occurrence in active catchments of the French Alps using monitoring stations (PhD Thesis). Université Grenoble Alpes, Grenoble.

Bemis, S.P., Micklethwaite, S., Turner, D., James, M.R., Akciz, S., Thiele, S.T., Bangash, H.A., 2014. Ground-based and UAV-Based photogrammetry: A multi-scale, high-resolution mapping tool for structural geology and paleoseismology. Journal of Structural Geology 69, 163–178. https://doi.org/10.1016/j.jsg.2014.10.007

Benelli, G., Pozzebon, A., 2013. RFID Under Water: Technical Issues and Applications. Radio Frequency Identification from System to Applications. https://doi.org/10.5772/53934

Benelli, G., Pozzebon, A., Bertoni, D., Sarti, G., 2012. An RFID-Based Toolbox for the Study of Under- and Outside-Water Movement of Pebbles on Coarse-Grained Beaches. IEEE Journal of Selected Topics in Applied Earth Observations and Remote Sensing 5, 1474–1482. https://doi.org/10.1109/JSTARS.2012.2196499

Benelli, G., Pozzebon, A., Raguseo, G., Bertoni, D., Sarti, G., 2009. An RFID Based System for the Underwater Tracking of Pebbles on Artificial Coarse Beaches, in: 2009 Third International Conference on Sensor Technologies and Applications. Presented at the 2009 Third International Conference on Sensor Technologies and Applications (SENSORCOMM), IEEE, Athens, Greece, pp. 294–299. https://doi.org/10.1109/SENSORCOMM.2009.52

Benoit, L., Briole, P., Martin, O., Thom, C., Malet, J.-P., Ulrich, P., 2015. Monitoring landslide displacements with the Geocube wireless network of low-cost GPS. Engineering Geology 195, 111–121. https://doi.org/10.1016/j.enggeo.2015.05.020

Bernard, M., Kondak, K., Maza, I., Ollero, A., 2011. Autonomous transportation and deployment with aerial robots for search and rescue missions. J. Field Robotics 28, 914–931. https://doi.org/10.1002/rob.20401

Bernardini, F., Buffi, A., Fontanelli, D., Macii, D., Magnago, V., Marracci, M., Motroni, A., Nepa, P., Tellini, B., 2021. Robot-Based Indoor Positioning of UHF-RFID Tags: The SAR Method With Multiple Trajectories. IEEE Transactions on Instrumentation and Measurement 70, 1–15. https://doi.org/10.1109/TIM.2020.3033728

Bernardini, F., Buffi, A., Motroni, A., Nepa, P., Tellini, B., Tripicchio, P., Unetti, M., 2020. Particle Swarm Optimization in SAR-based Method enabling Real-Time 3D Positioning of UHF-RFID Tags. IEEE Journal of Radio Frequency Identification 1–1. https://doi.org/10.1109/JRFID.2020.3005351

Bertoldo, S., Allegretti, M., 2015. Recharging RFID Tags for Environmental Monitoring Using UAVs: A Feasibility Analysis. Wireless Sensor Network 7, 720–726. https://doi.org/10.4236/wsn.2015.72002

Bertoni, D., Grottoli, E., Ciavola, P., Sarti, G., Benelli, G., Pozzebon, A., 2013. On the displacement of marked pebbles on two coarse-clastic beaches during short fair-weather periods (Marina di Pisa and Portonovo, Italy). Geo-Mar Lett 33, 463–476. https://doi.org/10.1007/s00367-013-0341-3

Bertoni, D., Sarti, G., Benelli, G., Pozzebon, A., Raguseo, G., 2012. Transport trajectories of "smart" pebbles on an artificial coarse-grained beach at Marina di Pisa (Italy): Implications for beach morphodynamics. Marine Geology 291–294, 227–235. https://doi.org/10.1016/j.margeo.2011.08.004

Bertoni, D., Sarti, G., Benelli, G., Pozzebon, A., Raguseo, G., 2010. Radio Frequency Identification (RFID) technology applied to the definition of underwater and subaerial coarse sediment movement. Sedimentary Geology 228, 140–150. https://doi.org/10.1016/j.sedgeo.2010.04.007

Bhanushali, K., Zhao, W., Pitts, W.S., Franzon, P.D., 2021. A 125 $\mu$ m $\times$ 245 $\mu$ m Mainly Digital UHF EPC Gen2 Compatible RFID Tag in 55 nm CMOS Process. IEEE J. Radio Freq. Identif. 5, 317–323. https://doi.org/10.1109/JRFID.2021.3087448

Bhattacharyya, R., Floerkemeier, C., Sarma, S., 2010a. RFID tag antenna based sensing: Does your beverage glass need a refill?, in: 2010 IEEE International Conference on RFID (IEEE RFID 2010). Presented at the 2010 IEEE International Conference on RFID (IEEE RFID 2010), pp. 126–133. https://doi.org/10.1109/RFID.2010.5467235

Bhattacharyya, R., Floerkemeier, C., Sarma, S., 2009. Towards tag antenna based sensing - An RFID displacement sensor, in: 2009 IEEE International Conference on RFID. Presented at the 2009 IEEE International Conference on RFID, pp. 95–102. https://doi.org/10.1109/RFID.2009.4911195

Bhattacharyya, R., Floerkemeier, C., Sarma, S., Deavours, D., 2011. RFID tag antenna based temperature sensing in the frequency domain, in: 2011 IEEE International Conference on RFID. Presented at the 2011 IEEE International Conference on RFID, pp. 70–77. https://doi.org/10.1109/RFID.2011.5764639

Bhattacharyya, R., Leo, C.D., Floerkemeier, C., Sarma, S., Anand, L., 2010b. RFID tag antenna based temperature sensing using shape memory polymer actuation, in: 2010 IEEE Sensors. Presented at the 2010 IEEE Sensors, pp. 2363–2368. https://doi.org/10.1109/ICSENS.2010.5690951

Bianco, G.M., Occhiuzzi, C., Panunzio, N., Marrocco, G., 2021. A Survey on Radio Frequency Identification as a Scalable Technology to Face Pandemics. IEEE Journal of Radio Frequency Identification 1–1. https://doi.org/10.1109/JRFID.2021.3117764

Bibi, F., Guillaume, C., Gontard, N., Sorli, B., 2017. A review: RFID technology having sensing aptitudes for food industry and their contribution to tracking and monitoring of food products. Trends in Food Science & Technology 62, 91–103. https://doi.org/10.1016/j.tifs.2017.01.013

Biron, P.M., Carver, R.B., Carré, D.M., 2012. Sediment Transport and Flow Dynamics Around a Restored Pool in a Fish Habitat Rehabilitation Project: Field and 3d Numerical Modelling Experiments. River Research and Applications 28, 926–939. https://doi.org/10.1002/rra.1488

Boada, M., Lázaro, A., Villarino, R., Girbau, D., 2020. NFC Battery-Less Colour Sensor and its Applications, in: 2020 Global Congress on Electrical Engineering (GC-ElecEng). Presented at the 2020 Global Congress on Electrical Engineering (GC-ElecEng), pp. 46–50. https://doi.org/10.23919/GC-ElecEng48342.2020.9286288

Bogena, H.R., Huisman, J.A., Meier, H., Rosenbaum, U., Weuthen, A., 2009. Hybrid Wireless Underground Sensor Networks: Quantification of Signal Attenuation in Soil. Vadose Zone Journal 8, 755–761. https://doi.org/10.2136/vzj2008.0138

Bose, I., Yan, S., 2011. The Green Potential of RFID Projects: A Case-Based Analysis. IT Professional 13, 41–47. https://doi.org/10.1109/MITP.2011.15

Bottelin, P., Baillet, L., Larose, E., Jongmans, D., Hantz, D., Brenguier, O., Cadet, H., Helmstetter, A., 2017. Monitoring rock reinforcement works with ambient vibrations: La Bourne case study (Vercors, France). Engineering Geology 226, 136–145. https://doi.org/10.1016/j.enggeo.2017.06.002

Bottelin, P., Baillet, L., Mathy, A., Garnier, L., Cadet, H., Brenguier, O., 2020. Seismic study of soda straws exposed to nearby blasting vibrations. J Seismol. https://doi.org/10.1007/s10950-020-09922-7

Bottelin, P., Jongmans, D., Baillet, L., Lebourg, T., Hantz, D., Levy, C., Le Roux, O., Cadet, H., Lorier, L., Rouiller, J.-D., Turpin, J., Darras, L., 2013. Spectral Analysis of Prone-to-fall Rock Compartments using Ambient Vibrations. Journal of Environmental & Engineering Geophysics 18, 205–217. https://doi.org/10.2113/JEEG18.4.205

Bradford, J.H., Harper, J.T., Brown, J., 2009. Complex dielectric permittivity measurements from ground-penetrating radar data to estimate snow liquid water content in the pendular regime: MEASURING SWE WITH GPR. Water Resources Research 45. https://doi.org/10.1029/2008WR007341

Bradley, D.N., Tucker, G.E., 2012. Measuring gravel transport and dispersion in a mountain river using passive radio tracers. Earth Surface Processes and Landforms 37, 1034–1045. https://doi.org/10.1002/esp.3223

Brard, E., 1930. Process for radiotelegraphic or radiotelephonic communication. US1744036A.

Brousse, G., Arnaud-Fassetta, G., Liébault, F., Bertrand, M., Melun, G., Loire, R., Malavoi, J.-R., Fantino, G., Borgniet, L., 2020a. Channel response to sediment replenishment in a large gravel-bed river: The case of the Saint-Sauveur dam in the Buëch River (Southern Alps, France). River Research and Applications 36, 880–893. https://doi.org/10.1002/rra.3527

Brousse, G., Liébault, F., Arnaud-Fassetta, G., Breilh, B., Tacon, S., 2020b. Gravel replenishment and active-channel widening for braided-river restoration: The case of the Upper Drac River (France). Science of The Total Environment 142517. https://doi.org/10.1016/j.scitotenv.2020.142517

Brousse, G., Liébault, F., Arnaud-Fassetta, G., Vazquez-Tarrio, D., 2018. Experimental bed active-layer survey with active RFID scour chains: Example of two braided rivers (the Drac and the Vénéon) in the French Alps. Presented at the Proceedings of River Flow 2018.

Buchli, B., Sutton, F., Beutel, J., 2012. GPS-Equipped Wireless Sensor Network Node for High-Accuracy Positioning Applications, in: SpringerLink. Presented at the European Conference on Wireless Sensor Networks, Springer Berlin Heidelberg, pp. 179–195. https://doi.org/10.1007/978-3-642-28169-3_12

Buffi, A., Motroni, A., Nepa, P., Tellini, B., Cioni, R., 2019. A SAR-Based Measurement Method for Passive-Tag Positioning With a Flying UHF-RFID Reader. IEEE Transactions on Instrumentation and Measurement 68, 845–853. https://doi.org/10.1109/TIM.2018.2857045

Buffi, A., Nepa, P., Cioni, R., 2017. SARFID on drone: Drone-based UHF-RFID tag localization, in: 2017 IEEE International Conference on RFID Technology Application (RFID-TA). Presented at the 2017 IEEE International Conference on RFID Technology Application (RFID-TA), pp. 40–44. https://doi.org/10.1109/RFID-TA.2017.8098872

Bukova, B., Tengler, J., Brumercikova, E., 2021. A Model of the Environmental Burden of RFID Technology in the Slovak Republic. Sustainability 13, 3684. https://doi.org/10.3390/su13073684

Burland, J.B., Chapman, T., Institution of Civil Engineers (Great Britain) (Eds.), 2012. ICE manual of geotechnical engineering. ICE, London.

Caccami, M.C., Manzari, S., Marrocco, G., 2015. Phase-Oriented Sensing by Means of Loaded UHF RFID Tags. IEEE Transactions on Antennas and Propagation 63, 4512–4520. https://doi.org/10.1109/TAP.2015.2465891

Caccami, M.C., Marrocco, G., 2018. Electromagnetic Modeling of Self-Tuning RFID Sensor Antennas in Linear and Nonlinear Regimes. IEEE Transactions on Antennas and Propagation 66, 2779–2787. https://doi.org/10.1109/TAP.2018.2820322

Cain, A., MacVicar, B., 2020. Field tests of an improved sediment tracer including non-intrusive measurement of burial depth. Earth Surface Processes and Landforms 45, 3488–3495. https://doi.org/10.1002/esp.4980

Caizzone, S., DiGiampaolo, E., 2015. Wireless Passive RFID Crack Width Sensor for Structural Health Monitoring. IEEE Sensors Journal 15, 6767–6774. https://doi.org/10.1109/JSEN.2015.2457455

Camera, F., Marrocco, G., 2021. Electromagnetic-Based Correction of Bio-Integrated RFID Sensors for Reliable Skin Temperature Monitoring. IEEE Sensors Journal 21, 421–429. https://doi.org/10.1109/JSEN.2020.3014404

Capdevila, S., Jofre, L., Bolomey, J.C., Romeu, J., 2010. RFID Multiprobe Impedance-Based Sensors. IEEE Transactions on Instrumentation and Measurement 59, 3093–3101. https://doi.org/10.1109/TIM.2010.2063053

Capdevila, S., Jofre, L., Romeu, J., Bolomey, J.C., 2011. Passive RFID based sensing, in: 2011 IEEE International Conference on RFID-Technologies and Applications. Presented at the 2011 IEEE International Conference on RFID-Technologies and Applications, pp. 507–512. https://doi.org/10.1109/RFID-TA.2011.6068592

Cappelli, I., Fort, A., Mugnaini, M., Panzardi, E., Pozzebon, A., Tani, M., Vignoli, V., 2021. Battery-Less HF RFID Sensor Tag for Soil Moisture Measurements. IEEE Trans. Instrum. Meas. 70, 1–13. https://doi.org/10.1109/TIM.2020.3036061

Carré, D.M., Biron, P.M., Gaskin, S.J., 2007. Flow dynamics and bedload sediment transport around paired deflectors for fish habitat enhancement: a field study in the Nicolet River. Can. J. Civ. Eng. 34, 761–769. https://doi.org/10.1139/l06-083

Casagli, N., Frodella, W., Morelli, S., Tofani, V., Ciampalini, A., Intrieri, E., Raspini, F., Rossi, G., Tanteri, L., Lu, P., 2017. Spaceborne, UAV and ground-based remote sensing techniques for landslide mapping, monitoring and early warning. Geoenviron Disasters 4, 9. https://doi.org/10.1186/s40677-017-0073-1

Casati, G., Longhi, M., Latini, D., Carbone, F., Amendola, S., Frate, F.D., Schiavon, G., Marrocco, G., 2017. The Interrogation Footprint of RFID-UAV: Electromagnetic Modeling and Experimentations. IEEE Journal of Radio Frequency Identification 1, 155–162. https://doi.org/10.1109/JRFID.2017.2765619

Cassel, M., Dépret, T., Piégay, H., 2017a. Assessment of a new solution for tracking pebbles in rivers based on active RFID. Earth Surface Processes and Landforms 42, 1938–1951. https://doi.org/10.1002/esp.4152

Cassel, M., Piégay, H., Fantino, G., Lejot, J., Bultingaire, L., Michel, K., Perret, F., 2020. Comparison of ground-based and UAV a-UHF artificial tracer mobility monitoring methods on a braided river. Earth Surface Processes and Landforms 45, 1123–1140. https://doi.org/10.1002/esp.4777

Cassel, M., Piégay, H., Lavé, J., 2017b. Effects of transport and insertion of radio frequency identification (RFID) transponders on resistance and shape of natural and synthetic pebbles: applications for riverine and coastal bedload tracking. Earth Surface Processes and Landforms 42, 399–413. https://doi.org/10.1002/esp.3989

Casserly, C.M., Turner, J.N., O' Sullivan, J.J., Bruen, M., Magee, D., Coiléir, S.O., Kelly-Quinn, M., 2021. Coarse sediment dynamics and low-head dams: Monitoring instantaneous bedload transport using a stationary RFID antenna. Journal of Environmental Management 300, 113671. https://doi.org/10.1016/j.jenvman.2021.113671

Casserly, C.M., Turner, J.N., O'Sullivan, J.J., Bruen, M., Bullock, C., Atkinson, S., Kelly-Quinn, M., 2020. Impact of low-head dams on bedload transport rates in coarse-bedded streams. Science of The Total Environment 716, 136908. https://doi.org/10.1016/j.scitotenv.2020.136908

Catarinucci, L., Colella, R., Consalvo, S.I., Patrono, L., Rollo, C., Sergi, I., 2020. IoT-Aware Waste Management System Based on Cloud Services and Ultra-Low-Power RFID Sensor-Tags. IEEE Sensors Journal 20, 14873–14881. https://doi.org/10.1109/JSEN.2020.3010675

Chang, A.Y., Yu, C., Lin, S., Chang, Y., Ho, P., 2009. Search, Identification and Positioning of the Underground Manhole with RFID Ground Tag, in: 2009 Fifth International Joint

Conference on INC, IMS and IDC. Presented at the 2009 Fifth International Joint Conference on INC, IMS and IDC, pp. 1899–1903. https://doi.org/10.1109/NCM.2009.306

Chapuis, M., Bright, C.J., Hufnagel, J., MacVicar, B., 2014. Detection ranges and uncertainty of passive Radio Frequency Identification (RFID) transponders for sediment tracking in gravel rivers and coastal environments: DETECTION RANGES AND UNCERTAINTY OF PASSIVE RFID TRANSPONDERS. Earth Surface Processes and Landforms 39, 2109–2120. https://doi.org/10.1002/esp.3620

Chapuis, M., Dufour, S., Provansal, M., Couvert, B., de Linares, M., 2015. Coupling channel evolution monitoring and RFID tracking in a large, wandering, gravel-bed river: Insights into sediment routing on geomorphic continuity through a riffle–pool sequence. Geomorphology 231, 258–269. https://doi.org/10.1016/j.geomorph.2014.12.013

Charléty, A., Le Breton, M., Larose, É., Baillet, L., 2022a. Long-term Monitoring of Soil Surface Deformation with RFID. Presented at the 2022 IEEE 12th International Conference on RFID Technology and Applications (RFID-TA), Cagliary, Italy.

Charléty, A., Le Breton, M., Larose, E., Baillet, L., 2022b. 2D Phase-Based RFID Localization for On-Site Landslide Monitoring. Remote Sensing 14, 3577. https://doi.org/10.3390/rs14153577

Che, X., Wells, I., Dickers, G., Kear, P., Gong, X., 2010. Re-evaluation of RF electromagnetic communication in underwater sensor networks. IEEE Communications Magazine 48, 143–151. https://doi.org/10.1109/MCOM.2010.5673085

Colombero, C., Jongmans, D., Fiolleau, S., Valentin, J., Baillet, L., Bièvre, G., 2021. Seismic Noise Parameters as Indicators of Reversible Modifications in Slope Stability: A Review. Surv Geophys. https://doi.org/10.1007/s10712-021-09632-w

Condemi, A., Cucchiella, F., Schettini, D., 2019. Circular Economy and E-Waste: An Opportunity from RFID TAGs. Applied Sciences 9, 3422. https://doi.org/10.3390/app9163422

Corke, P., Hrabar, S., Peterson, R., Rus, D., Saripalli, S., Sukhatme, G., 2004. Autonomous deployment and repair of a sensor network using an unmanned aerial vehicle, in: IEEE International Conference on Robotics and Automation, 2004. Proceedings. ICRA '04. 2004. Presented at the IEEE International Conference on Robotics and Automation, 2004. Proceedings. ICRA '04. 2004, IEEE, New Orleans, LA, USA, pp. 3602-3608 Vol.4. https://doi.org/10.1109/ROBOT.2004.1308811

Costa, F., Genovesi, S., Borgese, M., Michel, A., Dicandia, F.A., Manara, G., 2021. A Review of RFID Sensors, the New Frontier of Internet of Things. Sensors 21, 3138. https://doi.org/10.3390/s21093138

Curtiss, G.M., Osborne, P.D., Horner-Devine, A.R., 2009. Seasonal patterns of coarse sediment transport on a mixed sand and gravel beach due to vessel wakes, wind waves, and tidal currents. Marine Geology 259, 73–85. https://doi.org/10.1016/j.margeo.2008.12.009

Dane, J.H., Topp, C.G., 2020. Methods of Soil Analysis, Part 4: Physical Methods. John Wiley & Sons.

De Donno, D., Catarinucci, L., Tarricone, L., 2014. RAMSES: RFID Augmented Module for Smart Environmental Sensing. IEEE Transactions on Instrumentation and Measurement 63, 1701–1708. https://doi.org/10.1109/TIM.2014.2298692

Del Gaudio, V., Muscillo, S., Wasowski, J., 2014. What we can learn about slope response to earthquakes from ambient noise analysis: An overview. Engineering Geology, Special Issue on The Long-Term Geologic Hazards in Areas Struck by Large-Magnitude Earthquakes 182, 182–200. https://doi.org/10.1016/j.enggeo.2014.05.010

Dell'Agnese, A., Brardinoni, F., Toro, M., Mao, L., Engel, M., Comiti, F., 2015. Bedload transport in a formerly glaciated mountain catchment constrained by particle tracking. Earth Surface Dynamics 3, 527–542. https://doi.org/10.5194/esurf-3-527-2015

Deng, F., Zuo, P., Wen, K., Wu, X., 2020. Novel soil environment monitoring system based on RFID sensor and LoRa. Computers and Electronics in Agriculture 169, 105169. https://doi.org/10.1016/j.compag.2019.105169

Denoth, A., 1994. An electronic device for long-term snow wetness recording. Annals of Glaciology 19, 104–106. https://doi.org/10.3189/S0260305500011058

Dey, S., Bhattacharyya, R., Karmakar, N., Sarma, S., 2019. A Folded Monopole Shaped Novel Soil Moisture and Salinity Sensor for Precision Agriculture Based Chipless RFID Applications, in: 2019 IEEE MTT-S International Microwave and RF Conference (IMARC). Presented at the 2019 IEEE MTT-S International Microwave and RF Conference (IMARC), IEEE, Mumbai, India, pp. 1–4. https://doi.org/10.1109/IMaRC45935.2019.9118618

Dey, S., Karmakar, N., Bhattacharyya, R., Sarma, S., 2016. Electromagnetic characterization of soil moisture and salinity for UHF RFID applications in precision agriculture, in: 2016 46th European Microwave Conference (EuMC). Presented at the 2016 46th European Microwave Conference (EuMC), IEEE, London, United Kingdom, pp. 616–619. https://doi.org/10.1109/EuMC.2016.7824418

Dickson, M.E., Kench, P.S., Kantor, M.S., 2011. Longshore transport of cobbles on a mixed sand and gravel beach, southern Hawke Bay, New Zealand. Marine Geology 287, 31–42. https://doi.org/10.1016/j.margeo.2011.06.009

DiGiampaolo, E., DiCarlofelice, A., Gregori, A., 2017. An RFID-Enabled Wireless Strain Gauge Sensor for Static and Dynamic Structural Monitoring. IEEE Sensors Journal 17, 286–294. https://doi.org/10.1109/JSEN.2016.2631259

DiGiampaolo, E., Martinelli, F., 2020. A Multiple Baseline Approach to Face Multipath. IEEE Journal of Radio Frequency Identification 4, 314–321. https://doi.org/10.1109/JRFID.2020.3022576

Dini, B., Bennett, G., Franco, A., Whitworth, M.R.Z., Senn, A., Cook, K., 2020. Monitoring boulder movement using the Internet of Things: towards a landslide early warning system (other). display. https://doi.org/10.5194/egusphere-egu2020-17392

Dobkin, D.M., 2008. The RF in RFID: passive UHF RFID in practice, Communications engineering series. Elsevier / Newnes, Amsterdam ; Boston.

Dolphin, T., Lee, J., Phillips, R., Taylor, C.J.L., Dyer, K.R., 2016. Velocity of RFID Tagged Gravel in a Non-uniform Longshore Transport System. Journal of Coastal Research 75, 363–367. https://doi.org/10.2112/SI75-073.1

Duan, K.-K., Cao, S.-Y., 2020. Emerging RFID technology in structural engineering – A review. Structures 28, 2404–2414. https://doi.org/10.1016/j.istruc.2020.10.036

Dunnicliff, J., Marr, W.A., Standing, J., 2012. Chapter 94 Principles of geotechnical monitoring, in: ICE Manual of Geotechnical Engineering: Volume II, ICE Manuals. Thomas Telford Ltd, pp. 1363–1377. https://doi.org/10.1680/moge.57098.1363

Durgin, G.D., 2016. RF thermoelectric generation for passive RFID, in: 2016 IEEE International Conference on RFID (RFID). Presented at the 2016 IEEE International Conference on RFID (RFID), pp. 1–8. https://doi.org/10.1109/RFID.2016.7488025

Duroc, Y., Tedjini, S., 2018. RFID: A key technology for Humanity. Comptes Rendus Physique, Radio science for Humanity / Radiosciences au service de l'humanité Journées scientifiques URSI-France 2017 – SophiaTech, Sophia Antipolis, France, 1–3 February 2017 / 1er–3 mars 2017 19, 64–71. https://doi.org/10.1016/j.crhy.2018.01.003

Dziadak, K., Kumar, B., Sommerville, J., 2009. Model for the 3D Location of Buried Assets Based on RFID Technology. J. Comput. Civ. Eng. 23, 148–159. https://doi.org/10.1061/(ASCE)0887-3801(2009)23:3(148)

Einstein, H.A., 1937. Bedload transport as a probability problem, in: Shen, H.W. (Ed.), Sedimentation. Colorado State University, Fort Collins, pp. C1–C105.

Elboushi, A., Telba, A., Sebak, A., Jamil, K., 2020. Electromagnetic Soil Characterization for Undergrounded RFID System Implementation. Electronics 9, 106. https://doi.org/10.3390/electronics9010106

Escobedo, P., Carvajal, M.A., Capitán-Vallvey, L.F., Fernández-Salmerón, J., Martínez-Olmos, A., Palma, A.J., 2016. Passive UHF RFID Tag for Multispectral Assessment. Sensors 16, 1085. https://doi.org/10.3390/s16071085

Evans, J.R., Allen, R.M., Chung, A.I., Cochran, E.S., Guy, R., Hellweg, M., Lawrence, J.F., 2014. Performance of Several Low-Cost Accelerometers. Seismological Research Letters 85, 147–158. https://doi.org/10.1785/0220130091

Fahmy, A., Altaf, H., Al Nabulsi, A., Al-Ali, A., Aburukba, R., 2019. Role of RFID Technology in Smart City Applications, in: 2019 International Conference on Communications, Signal Processing, and Their Applications (ICCSPA). Presented at the 2019 International Conference on Communications, Signal Processing, and their Applications (ICCSPA), pp. 1–6. https://doi.org/10.1109/ICCSPA.2019.8713622

Falco, A., Salmerón, J.F., Loghin, F.C., Lugli, P., Rivadeneyra, A., 2017. Fully Printed Flexible Single-Chip RFID Tag with Light Detection Capabilities. Sensors 17, 534. https://doi.org/10.3390/s17030534

Farsens, n.d. EVAL01-Kineo wireless, battery free orientation sensor tag. Farsens Wireless Sensors. URL http://www.farsens.com/en/products/eval01-kineo-rm/ (accessed 3.10.21).

FARSENS HYDRO H402, n.d.

Faseth, T., Winkler, M., Arthaber, H., Magerl, G., 2011. The influence of multipath propagation on phase-based narrowband positioning principles in UHF RFID, in: 2011 IEEE-APS Topical Conference on Antennas and Propagation in Wireless Communications (APWC). Presented at the 2011 IEEE-APS Topical Conference on Antennas and Propagation in Wireless Communications (APWC), pp. 1144–1147. https://doi.org/10.1109/APWC.2011.6046829

Ferdous, R.Md., Reza, A.W., Siddiqui, M.F., 2016. Renewable energy harvesting for wireless sensors using passive RFID tag technology: A review. Renewable and Sustainable Energy Reviews 58, 1114–1128. https://doi.org/10.1016/j.rser.2015.12.332

Fischbacher, R., Görtschacher, L., Amtmann, F., Priller, P., Bösch, W., Grosinger, J., 2020. Localization of UHF RFID Magnetic Field Sensor Tags, in: 2020 IEEE Radio and Wireless Symposium (RWS). Presented at the 2020 IEEE Radio and Wireless Symposium (RWS), pp. 169–172. https://doi.org/10.1109/RWS45077.2020.9050026

Fonseca, N., Freire, R., Fontgalland, G., Arruda, B., Tedjini, S., 2018. A Fully Passive UHF RFID Soil Moisture Time-Domain Transmissometry Based Sensor, in: 2018 3rd International Symposium on Instrumentation Systems, Circuits and Transducers (INSCIT). Presented at the 2018 3rd International Symposium on Instrumentation Systems, Circuits and Transducers (INSCIT), pp. 1–6. https://doi.org/10.1109/INSCIT.2018.8546707

Fonseca, N.S.S.M., Freire, R.C.S., Batista, A., Fontgalland, G., Tedjini, S., 2017. A passive capacitive soil moisture and environment temperature UHF RFID based sensor for low cost agricultural applications, in: 2017 SBMO/IEEE MTT-S International Microwave and Optoelectronics Conference (IMOC). Presented at the 2017 SBMO/IEEE MTT-S

International Microwave and Optoelectronics Conference (IMOC), IEEE, Aguas de Lindoia, pp. 1–4. https://doi.org/10.1109/IMOC.2017.8121099

Ford, M.R., 2014. The application of PIT tags to measure transport of detrital coral fragments on a fringing reef: Majuro Atoll, Marshall Islands. Coral Reefs 33, 375–379. https://doi.org/10.1007/s00338-014-1131-8

Galia, T., Škarpich, V., Ruman, S., 2021. Impact of check dam series on coarse sediment connectivity. Geomorphology 377, 107595. https://doi.org/10.1016/j.geomorph.2021.107595

Gareis, M., Fenske, P., Carlowitz, C., Vossiek, M., 2020. Particle Filter-Based SAR Approach and Trajectory Optimization for Real-Time 3D UHF-RFID Tag Localization, in: 2020 IEEE International Conference on RFID (RFID). Presented at the 2020 IEEE International Conference on RFID (RFID), pp. 1–8. https://doi.org/10.1109/RFID49298.2020.9244917

Gilet, L., Gob, F., Gautier, E., Houbrechts, G., Virmoux, C., Thommeret, N., 2020. Hydro-morphometric parameters controlling travel distance of pebbles and cobbles in three gravel bed streams. Geomorphology 358, 107117. https://doi.org/10.1016/j.geomorph.2020.107117

Gili, J.A., Corominas, J., Rius, J., 2000. Using Global Positioning System techniques in landslide monitoring. Engineering Geology 55, 167–192. https://doi.org/10.1016/S0013-7952(99)00127-1

Gladysz, B., Ejsmont, K., Kluczek, A., Corti, D., Marciniak, S., 2020. A Method for an Integrated Sustainability Assessment of RFID Technology. Resources 9, 107. https://doi.org/10.3390/resources9090107

Gómez-Pazo, A., Pérez-Alberti, A., Trenhaile, A., 2021. Tracking clast mobility using RFID sensors on a boulder beach in Galicia, NW Spain. Geomorphology 373, 107514. https://doi.org/10.1016/j.geomorph.2020.107514

Graff, K., Viel, V., Carlier, B., Lissak, C., Madelin, M., Arnaud-Fassetta, G., Fort, M., 2018. Traçage sédimentaire d'une lave torrentielle dans le bassin de la Peyronnelle (Queyras, Alpes françaises du Sud). Géomorphologie : relief, processus, environnement 24, 43–57. https://doi.org/10.4000/geomorphologie.11967

Greco, G., Lucianaz, C., Bertoldo, S., Allegretti, M., 2015. Localization of RFID tags for environmental monitoring using UAV, in: IEEE 1st Int. Forum Research and Technologies for Society and Industry. Presented at the 2015 IEEE 1st International Forum on Research and Technologies for Society and Industry (RTSI), Turin, Italy, pp. 480–483. https://doi.org/10.1109/RTSI.2015.7325144

Griffin, J.D., Durgin, G.D., 2009. Complete Link Budgets for Backscatter-Radio and RFID Systems. IEEE Antennas and Propagation Magazine 51, 11–25. https://doi.org/10.1109/MAP.2009.5162013

Gronz, O., Hiller, P.H., Wirtz, S., Becker, K., Iserloh, T., Seeger, M., Brings, C., Aberle, J., Casper, M.C., Ries, J.B., 2016. Smartstones: A small 9-axis sensor implanted in stones to track their movements. CATENA 142, 245–251. https://doi.org/10.1016/j.catena.2016.03.030

Gupta, G., Singh, B.P., Bal, A., Kedia, D., Harish, A.R., 2014. Orientation Detection Using Passive UHF RFID Technology [Education Column]. IEEE Antennas and Propagation Magazine 56, 221–237. https://doi.org/10.1109/MAP.2014.7011063

Habersack, H.M., 2001. Radio-tracking gravel particles in a large braided river in New Zealand: a field test of the stochastic theory of bed load transport proposed by Einstein. Hydrological Processes 15, 377–391.

Hamrita, T.K., Hoffacker, E.C., 2005. Development of a smart wireless soil monitoring sensor prototype using RFID technology. Applied Engineering in Agriculture 21, 139–143.

Han, M., Yang, D.Y., Yu, J., Kim, J.W., 2016. Typhoon Impact on a Pure Gravel Beach as Assessed through Gravel Movement and Topographic Change at Yeocha Beach, South Coast of Korea. Journal of Coastal Research. https://doi.org/10.2112/JCOASTRES-D-16-00104.1

Hansen, B.J., Carron, C.J., Jensen, B.D., Hawkins, A.R., Schultz, S.M., 2007. Plastic latching accelerometer based on bistable compliant mechanisms. Smart Mater. Struct. 16, 1967–1972. https://doi.org/10.1088/0964-1726/16/5/055

Harasti, T.J., Hertel, I.W.M., Howell, J.E., 2011. Underwater RFID Arrangement for Optimizing Underwater Operations. US2011095865A1.

Hasan, A., Bhattacharyya, R., Sarma, S., 2015. Towards pervasive soil moisture sensing using RFID tag antenna-based sensors, in: IEEE Int. Conf. RFID Technologies and Applications. Presented at the 2015 IEEE International Conference on RFID Technology and Applications (RFID-TA), Tokyo, Japan, pp. 165–170. https://doi.org/10.1109/RFID-TA.2015.7379812

Hasan, A., Bhattacharyya, R., Sarma, S., 2013. A monopole-coupled RFID sensor for pervasive soil moisture monitoring, in: 2013 IEEE Antennas and Propagation Soc. Int. Symp. Presented at the 2013 IEEE Antennas and Propagation Society International Symposium (APSURSI), pp. 2309–2310. https://doi.org/10.1109/APS.2013.6711813

Hasani, M., Vena, A., Sydänheimo, L., Ukkonen, L., Tentzeris, M.M., 2013. Implementation of a Dual-Interrogation-Mode Embroidered RFID-Enabled Strain Sensor. IEEE Antennas and Wireless Propagation Letters 12, 1272–1275. https://doi.org/10.1109/LAWP.2013.2283539

Haschenburger, J., Church, M., 1998. Bed material transport estimated from the virtual velocity of sediment. Earth Surface Processes and Landforms 23, 791–808.

Hassan, M.A., Ergenzinger, P., 2003. Use of tracers in fluvial geomorphology, in: Kondolf, G.M., Piégay, H. (Eds.), Tools in Fluvial Geomorphology. John Wiley and Sons, Chichester, pp. 397–423.

Hastewell, L., Inkpen, R., Bray, M., Schaefer, M., 2020. Quantification of contemporary storm-induced boulder transport on an intertidal shore platform using radio frequency identification technology. Earth Surface Processes and Landforms 45, 1601–1621. https://doi.org/10.1002/esp.4834

Hastewell, L.J., Schaefer, M., Bray, M., Inkpen, R., 2019. Intertidal boulder transport: A proposed methodology adopting Radio Frequency Identification (RFID) technology to quantify storm induced boulder mobility. Earth Surface Processes and Landforms 44, 681–698. https://doi.org/10.1002/esp.4523

Hautcoeur, J., Talbi, L., Nedil, M., 2013. High gain RFID tag antenna for the underground localization applications at 915 MHz band, in: 2013 IEEE Antennas and Propagation Society International Symposium (APSURSI). Presented at the 2013 IEEE Antennas and Propagation Society International Symposium (APSURSI), pp. 1488–1489. https://doi.org/10.1109/APS.2013.6711403

He, X., Zhu, J., Su, W., Tentzeris, M.M., 2020. RFID Based Non-Contact Human Activity Detection Exploiting Cross Polarization. IEEE Access 8, 46585–46595. https://doi.org/10.1109/ACCESS.2020.2979080

Herrera, G., Fernández-Merodo, J.A., Mulas, J., Pastor, M., Luzi, G., Monserrat, O., 2009. A landslide forecasting model using ground based SAR data: The Portalet case study. Engineering Geology 105, 220–230. https://doi.org/10.1016/j.enggeo.2009.02.009

Higgs-EC, n.d.

Hillier, A.J.R., Makarovaite, V., Gourlay, C.W., Holder, S.J., Batchelor, J.C., 2019. A Passive UHF RFID Dielectric Sensor for Aqueous Electrolytes. IEEE Sensors Journal 19, 5389–5395. https://doi.org/10.1109/JSEN.2019.2909353

Houbrechts, G., Campenhout, J.V., Levecq, Y., Hallot, E., Peeters, A., Petit, F., 2012. Comparison of methods for quantifying active layer dynamics and bedload discharge in armoured gravel-bed rivers. Earth Surface Processes and Landforms 37, 1501–1517. https://doi.org/10.1002/esp.3258

Houbrechts, G., Levecq, Y., Peeters, A., Hallot, E., Van Campenhout, J., Denis, A.-C., Petit, F., 2015. Evaluation of long-term bedload virtual velocity in gravel-bed rivers (Ardenne, Belgium). Geomorphology, Emerging geomorphic approaches to guide river management practices 251, 6–19. https://doi.org/10.1016/j.geomorph.2015.05.012

Hungr, O., Leroueil, S., Picarelli, L., 2014. The Varnes classification of landslide types, an update. Landslides 11, 167–194. https://doi.org/10.1007/s10346-013-0436-y

Hussain, Z., Sheng, Q.Z., Zhang, W.E., 2020. A review and categorization of techniques on device-free human activity recognition. Journal of Network and Computer Applications 167, 102738. https://doi.org/10.1016/j.jnca.2020.102738

Imhoff, K.S., Wilcox, A.C., 2016. Coarse bedload routing and dispersion through tributary confluences. Earth Surface Dynamics 4, 591–605. https://doi.org/10.5194/esurf-4-591-2016

Inserra, D., Hu, W., Li, Z., Li, G., Zhao, F., Yang, Z., Wen, G., 2020. Screw Relaxing Detection With UHF RFID Tag. IEEE Access 8, 78553–78564. https://doi.org/10.1109/ACCESS.2020.2986891

Intrieri, E., Carlà, T., Gigli, G., 2019. Forecasting the time of failure of landslides at slope-scale: A literature review. Earth-Science Reviews. https://doi.org/10.1016/j.earscirev.2019.03.019

Intrieri, E., Gigli, G., Gracchi, T., Nocentini, M., Lombardi, L., Mugnai, F., Frodella, W., Bertolini, G., Carnevale, E., Favalli, M., Fornaciai, A., Marturià Alavedra, J., Mucchi, L., Nannipieri, L., Rodriguez-Lloveras, X., Pizziolo, M., Schina, R., Trippi, F., Casagli, N., 2018. Application of an ultra-wide band sensor-free wireless network for ground monitoring. Engineering Geology 238, 1–14. https://doi.org/10.1016/j.enggeo.2018.02.017

Intrieri, E., Gigli, G., Mugnai, F., Fanti, R., Casagli, N., 2012. Design and implementation of a landslide early warning system. Engineering Geology 147–148, 124–136. https://doi.org/10.1016/j.enggeo.2012.07.017

ISO 4866, 2010. Mechanical vibration and shock — Vibration of fixed structures — Guidelines for the measurement of vibrations and evaluation of their effects on structures (No. ISO 4866:2010). International Organization for Standardization.

Ivanov, V., Radice, A., Papini, M., Longoni, L., 2020. Event-scale pebble mobility observed by RFID tracking in a pre-Alpine stream: a field laboratory. Earth Surface Processes and Landforms 45, 535–547. https://doi.org/10.1002/esp.4752

Jaboyedoff, M., Oppikofer, T., Abellán, A., Derron, M.-H., Loye, A., Metzger, R., Pedrazzini, A., 2012. Use of LIDAR in landslide investigations: a review. Nat Hazards 61, 5–28. https://doi.org/10.1007/s11069-010-9634-2

Jauregi, I., Solar, H., Beriain, A., Zalbide, I., Jimenez, A., Galarraga, I., Berenguer, R., 2017. UHF RFID Temperature Sensor Assisted With Body-Heat Dissipation Energy Harvesting. IEEE Sensors Journal 17, 1471–1478. https://doi.org/10.1109/JSEN.2016.2638473

Jayawardana, D., Kharkovsky, S., Liyanapathirana, R., Zhu, X., 2016. Measurement System With Accelerometer Integrated RFID Tag for Infrastructure Health Monitoring. IEEE Transactions on Instrumentation and Measurement 65, 1163–1171. https://doi.org/10.1109/TIM.2015.2507406

Jayawardana, D., Liyanapathirana, R., Zhu, X., 2019. RFID-Based Wireless Multi-Sensory System for Simultaneous Dynamic Acceleration and Strain Measurements of Civil

Infrastructure. IEEE Sensors Journal 19, 12389–12397. https://doi.org/10.1109/JSEN.2019.2937889

Jedermann, R., Ruiz-Garcia, L., Lang, W., 2009. Spatial temperature profiling by semi-passive RFID loggers for perishable food transportation. Computers and Electronics in Agriculture 65, 145–154. https://doi.org/10.1016/j.compag.2008.08.006

Kantareddy, S.N.R., Bhattacharyya, R., Sarma, S., 2018. UHF RFID tag IC power mode switching for wireless sensing of resistive and electrochemical transduction modalities, in: 2018 IEEE International Conference on RFID (RFID). Presented at the 2018 IEEE International Conference on RFID (RFID), IEEE, Orlando, FL, pp. 1–8. https://doi.org/10.1109/RFID.2018.8376201

Kantareddy, S.N.R., Mathews, I., Bhattacharyya, R., Peters, I.M., Buonassisi, T., Sarma, S.E., 2019. Long range battery-less PV-powered RFID tag sensors. IEEE Internet Things J. 6, 6989–6996. https://doi.org/10.1109/JIOT.2019.2913403

Kantareddy, S.N.R., Mathews, I., Sun, S., Layurova, M., Thapa, J., Correa-Baena, J.-P., Bhattacharyya, R., Buonassisi, T., Sarma, S.E., Peters, I.M., 2020. Perovskite PV-Powered RFID: Enabling Low-Cost Self-Powered IoT Sensors. IEEE Sensors Journal 20, 471–478. https://doi.org/10.1109/JSEN.2019.2939293

Karuppuswami, S., Mondal, S., Kumar, D., Chahal, P., 2020. RFID Coupled Passive Digital Ammonia Sensor for Quality Control of Packaged Food. IEEE Sensors Journal 20, 4679–4687. https://doi.org/10.1109/JSEN.2020.2964676

Kassal, P., Steinberg, I.M., Steinberg, M.D., 2013. Wireless smart tag with potentiometric input for ultra low-power chemical sensing. Sensors and Actuators B: Chemical 184, 254–259. https://doi.org/10.1016/j.snb.2013.04.049

Kenney, J.D., Poole, D.R., Willden, G.C., Abbott, B.A., Morris, A.P., McGinnis, R.N., Ferrill, D.A., 2009. Precise positioning with wireless sensor nodes: Monitoring natural hazards in all terrains, in: IEEE Int. Conf. Systems, Man and Cybernetics. Presented at the IEEE International Conference on Systems, Man and Cybernetics, San Antonio, TX, USA, pp. 722–727. https://doi.org/10.1109/ICSMC.2009.5346714

Khan, N., Sajak, A.A.B., Alam, M., Mazliham, M.S., 2021. Analysis of Green IoT. J. Phys.: Conf. Ser. 1874, 012012. https://doi.org/10.1088/1742-6596/1874/1/012012

Khokhlova, V., Delevoye, E., 2019. The Localization of Buried Objects in the Soil Using an RFID Tag: Protocol Description and Parameter Estimation With the Model of the Oscillating Magnetic Field in Media. IEEE Transactions on Magnetics 55, 1–16. https://doi.org/10.1109/TMAG.2019.2898840

Kim, S., Le, T., Tentzeris, M.M., Harrabi, A., Collado, A., Georgiadis, A., 2014. An RFID-enabled inkjet-printed soil moisture sensor on paper for "smart" agricultural applications, in: IEEE SENSORS Proc. Presented at the IEEE SENSORS 2014 Proceedings, Valencia, Spain, pp. 1507–1510. https://doi.org/10.1109/ICSENS.2014.6985301

Kim, Y., Zyl, J.J. van, 2009. A Time-Series Approach to Estimate Soil Moisture Using Polarimetric Radar Data. IEEE Transactions on Geoscience and Remote Sensing 47, 2519–2527. https://doi.org/10.1109/TGRS.2009.2014944

Kinar, N.J., Pomeroy, J.W., 2015. Measurement of the physical properties of the snowpack. Reviews of Geophysics 53, 481–544. https://doi.org/10.1002/2015RG000481

Kleinbrod, U., Burjánek, J., Fäh, D., 2019. Ambient vibration classification of unstable rock slopes: A systematic approach. Engineering Geology 249, 198–217. https://doi.org/10.1016/j.enggeo.2018.12.012

Koch, F., Henkel, P., Appel, F., Schmid, L., Bach, H., Lamm, M., Prasch, M., Schweizer, J., Mauser, W., 2019. Retrieval of Snow Water Equivalent, Liquid Water Content, and Snow Height of Dry and Wet Snow by Combining GPS Signal Attenuation and Time

Delay. Water Resources Research 55, 4465–4487. https://doi.org/10.1029/2018WR024431

Konstantakos, V., Kozalakis, K., Siozos, K., Siskos, S., Laopoulos, T., 2019. Earthquake instrumentation node with MEMS sensors, in: 2019 Panhellenic Conference on Electronics Telecommunications (PACET). pp. 1–6. https://doi.org/10.1109/PACET48583.2019.8956265

Korošak, Ž., Suhadolnik, N., Pleteršek, A., 2019. The Implementation of a Low Power Environmental Monitoring and Soil Moisture Measurement System Based on UHF RFID. Sensors 19, 5527. https://doi.org/10.3390/s19245527

Kumar, B., Sommerville, J., 2012. A model for RFID-based 3D location of buried assets. Automation in Construction 21, 121–131. https://doi.org/10.1016/j.autcon.2011.05.020

Lacroix, P., Handwerger, A.L., Bièvre, G., 2020. Life and death of slow-moving landslides. Nat Rev Earth Environ. https://doi.org/10.1038/s43017-020-0072-8

Lai, X., Cai, Z., Xie, Z., Zhu, H., 2018. A Novel Displacement and Tilt Detection Method Using Passive UHF RFID Technology. Sensors (Basel) 18. https://doi.org/10.3390/s18051644

Lamarre, H., MacVicar, B., Roy, A.G., 2005. Using Passive Integrated Transponder (PIT) Tags to Investigate Sediment Transport in Gravel-Bed Rivers. Journal of Sedimentary Research 75, 736–741. https://doi.org/10.2110/jsr.2005.059

Lamarre, H., Roy, A.G., 2008a. A field experiment on the development of sedimentary structures in a gravel-bed river. Earth Surface Processes and Landforms 33, 1064–1081. https://doi.org/10.1002/esp.1602

Lamarre, H., Roy, A.G., 2008b. The role of morphology on the displacement of particles in a step–pool river system. Geomorphology 99, 270–279. https://doi.org/10.1016/j.geomorph.2007.11.005

Lambert, S. (Ed.), 2011. Rockfall engineering. ISTE, London.

Larose, E., Carrière, S., Voisin, C., Bottelin, P., Baillet, L., Guéguen, P., Walter, F., Jongmans, D., Guillier, B., Garambois, S., Gimbert, F., Massey, C., 2015. Environmental seismology: What can we learn on earth surface processes with ambient noise? Journal of Applied Geophysics 116, 62–74. https://doi.org/10.1016/j.jappgeo.2015.02.001

Larson, K.M., 2016. GPS interferometric reflectometry: applications to surface soil moisture, snow depth, and vegetation water content in the western United States. Wiley Interdisciplinary Reviews: Water 3, 775–787. https://doi.org/10.1002/wat2.1167

Le Breton, M., 2019. Suivi temporel d'un glissement de terrain à l'aide d'étiquettes RFID passives, couplé à l'observation de pluviométrie et de bruit sismique ambiant (PhD Thesis). Université Grenoble Alpes, ISTerre, Grenoble, France.

Le Breton, M., Baillet, L., Larose, E., Rey, E., Benech, P., Jongmans, D., Guyoton, F., 2017. Outdoor UHF RFID: Phase Stabilization for Real-World Applications. IEEE Journal of Radio Frequency Identification 1, 279–290. https://doi.org/10.1109/JRFID.2017.2786745

Le Breton, M., Baillet, L., Larose, E., Rey, E., Benech, P., Jongmans, D., Guyoton, F., Jaboyedoff, M., 2019. Passive radio-frequency identification ranging, a dense and weather-robust technique for landslide displacement monitoring. Engineering Geology 250, 1–10. https://doi.org/10.1016/j.enggeo.2018.12.027

Le Breton, M., Baillet, L., Larose, É., Scheiblin, G., Lecci, M., Rey, É., in preparation. How RFID systems perform in the snow ?

Le Breton, M., Grunbaum, N., Baillet, L., Larose, É., 2021. Monitoring rock displacement threshold with 1-bit sensing passive RFID tag (No. EGU21-15305). Copernicus Meetings. https://doi.org/10.5194/egusphere-egu21-15305

Le Breton, M., Larose, É., Baillet, L., Lejeune, Y., van Herwijnen, A., 2022. Monitoring snowpack SWE and temperature using RFID tags as wireless sensors. https://doi.org/10.5194/egusphere-2022-761

Lekshmi, S.U.S., Singh, D.N., Shojaei Baghini, M., 2014. A critical review of soil moisture measurement. Measurement 54, 92–105. https://doi.org/10.1016/j.measurement.2014.04.007

Lévy, C., Jongmans, D., Baillet, L., 2011. Analysis of seismic signals recorded on a prone-to-fall rock column (Vercors massif, French Alps). Geophys J Int 186, 296–310. https://doi.org/10.1111/j.1365-246X.2011.05046.x

Li, C., Mo, L., Zhang, D., 2019. Review on UHF RFID Localization methods. IEEE Journal of Radio Frequency Identification 1–1. https://doi.org/10.1109/JRFID.2019.2924346

Li, C.-H., Lao, K.-W., Tam, K.-W., 2018. A Flooding Warning System based on RFID Tag Array for Energy Facility, in: 2018 IEEE International Conference on RFID Technology Application (RFID-TA). Presented at the 2018 IEEE International Conference on RFID Technology Application (RFID-TA), pp. 1–4. https://doi.org/10.1109/RFID-TA.2018.8552767

Li, C.J., Ling, H., 2015. Synthetic aperture radar imaging using a small consumer drone, in: 2015 IEEE International Symposium on Antennas and Propagation USNC/URSI National Radio Science Meeting. Presented at the 2015 IEEE International Symposium on Antennas and Propagation USNC/URSI National Radio Science Meeting, pp. 685–686. https://doi.org/10.1109/APS.2015.7304729

Li, P., An, Z., Yang, L., Yang, P., Lin, Q., 2021. RFID Harmonic for Vibration Sensing. IEEE Transactions on Mobile Computing 20, 1614–1626. https://doi.org/10.1109/TMC.2019.2963152

Li, Y., Wang, S., Jin, C., Zhang, Y., Jiang, T., 2019. A Survey of Underwater Magnetic Induction Communications: Fundamental Issues, Recent Advances, and Challenges. IEEE Communications Surveys Tutorials 21, 2466–2487. https://doi.org/10.1109/COMST.2019.2897610

Liébault, F., Bellot, H., Chapuis, M., Klotz, S., Deschâtres, M., 2012. Bedload tracing in a high-sediment-load mountain stream. Earth Surface Processes and Landforms 37, 385–399. https://doi.org/10.1002/esp.2245

Liébault, F., Laronne, J.B., 2008. Evaluation of bedload yield in gravel-bed rivers using scour chains and painted tracers: the case of the Esconavette Torrent (Southern French Prealps). Geodinamica Acta 21, 23–34. https://doi.org/10.3166/ga.21.23-34

Lin, C., Rommen, B., Floury, N., Schüttemeyer, D., Davidson, M.W.J., Kern, M., Kontu, A., Lemmetyinen, J., Pulliainen, J., Wiesmann, A., Werner, C.L., Mätzler, C., Schneebeli, M., Proksch, M., Nagler, T., 2016. Active Microwave Scattering Signature of Snowpack—Continuous Multiyear SnowScat Observation Experiments. IEEE Journal of Selected Topics in Applied Earth Observations and Remote Sensing 9, 3849–3869. https://doi.org/10.1109/JSTARS.2016.2560168

Lin, Y.F., Chang, M.J., Chen, H.M., Lai, B.Y., 2016. Gain Enhancement of Ground Radiation Antenna for RFID Tag Mounted on Metallic Plane. IEEE Transactions on Antennas and Propagation 64, 1193–1200. https://doi.org/10.1109/TAP.2016.2526047

Longhi, M., Marrocco, G., 2017. Ubiquitous Flying Sensor Antennas: Radiofrequency Identification Meets Micro Drones. IEEE J. Radio Freq. Identif. 1, 291–299. https://doi.org/10.1109/JRFID.2018.2801882

Longhi, M., Millane, A., Taylor, Z., Nieto, J., Siegwart, R., Marrocco, G., 2018a. An Integrated MAV-RFID System for Geo-referenced Monitoring of Harsh Environments, in: 2018 IEEE Conference on Antenna Measurements & Applications (CAMA). Presented at the 2018 IEEE Conference on Antenna Measurements & Applications (CAMA), IEEE, Västerås, pp. 1–4. https://doi.org/10.1109/CAMA.2018.8530596

Longhi, M., Taylor, Z., Popovic, M., Nieto, J., Marrocco, G., Siegwart, R., 2018b. RFID-Based Localization for Greenhouses Monitoring Using MAVs, in: 2018 IEEE-APS Topical Conference on Antennas and Propagation in Wireless Communications (APWC). Presented at the 2018 IEEE-APS Topical Conference on Antennas and Propagation in Wireless Communications (APWC), IEEE, Cartagena des Indias, pp. 905–908. https://doi.org/10.1109/APWC.2018.8503764

Lu, Y., Basset, P., Laheurte, J.-M., 2017. Performance Evaluation of a Long-Range RFID Tag Powered by a Vibration Energy Harvester. IEEE Antennas and Wireless Propagation Letters 16, 1832–1835. https://doi.org/10.1109/LAWP.2017.2682419

Łuczak, S., Grepl, R., Bodnicki, M., 2017. Selection of MEMS Accelerometers for Tilt Measurements. Journal of Sensors 2017, ID 9796146. https://doi.org/10.1155/2017/9796146

Luvisi, A., Panattoni, A., Materazzi, A., 2016. RFID temperature sensors for monitoring soil solarization with biodegradable films. Computers and Electronics in Agriculture 123, 135–141. https://doi.org/10.1016/j.compag.2016.02.023

Lys, P.-O., Elder, K., Archer, J., 2018. METIS, a disruptive R&D project to revolutionize land seismic acquisition, in: RDPETRO 2018: Research and Development Petroleum Conference and Exhibition, Abu Dhabi, UAE, 9-10 May 2018, SEG Global Meeting Abstracts. American Association of Petroleum Geologists, Society of Exploration Geophysicists, European Association of Geoscientists and Engineers, and Society of Petroleum Engineers, pp. 28–31. https://doi.org/10.1190/RDP2018-41752683.1

Ma, Y., Selby, N., Adib, F., 2017. Drone Relays for Battery-Free Networks, in: Proceedings of the Conference of the ACM Special Interest Group on Data Communication - SIGCOMM '17. Presented at the the Conference of the ACM Special Interest Group, ACM Press, Los Angeles, CA, USA, pp. 335–347. https://doi.org/10.1145/3098822.3098847

Ma, Y., Wang, B., Pei, S., Zhang, Y., Zhang, S., Yu, J., 2018. An Indoor Localization Method Based on AOA and PDOA Using Virtual Stations in Multipath and NLOS Environments for Passive UHF RFID. IEEE Access 6, 31772–31782. https://doi.org/10.1109/ACCESS.2018.2838590

MacVicar, B.J., Piégay, H., Henderson, A., Comiti, F., Oberlin, C., Pecorari, E., 2009. Quantifying the temporal dynamics of wood in large rivers: field trials of wood surveying, dating, tracking, and monitoring techniques. Earth Surface Processes and Landforms 34, 2031–2046. https://doi.org/10.1002/esp.1888

MacVicar, B.J., Roy, A.G., 2011. Sediment mobility in a forced riffle-pool. Geomorphology 125, 445–456. https://doi.org/DOI:10.1016/j.geomorph.2010.10.031

Magilligan, F.J., Roberts, M.O., Marti, M., Renshaw, C.E., 2021. The impact of run-of-river dams on sediment longitudinal connectivity and downstream channel equilibrium. Geomorphology 376, 107568. https://doi.org/10.1016/j.geomorph.2020.107568

Manzari, S., Caizzone, S., Rubini, C., Marrocco, G., 2014a. Feasibility of wireless temperature sensing by passive UHF-RFID tags in ground satellite test beds, in: 2014 IEEE International Conference on Wireless for Space and Extreme Environments (WiSEE). Presented at the 2014 IEEE International Conference on Wireless for Space and Extreme Environments (WiSEE), pp. 1–6. https://doi.org/10.1109/WiSEE.2014.6973074

Manzari, S., Catini, A., Pomarico, G., Natale, C.D., Marrocco, G., 2014b. Development of an UHF RFID Chemical Sensor Array for Battery-Less Ambient Sensing. IEEE Sensors Journal 14, 3616–3623. https://doi.org/10.1109/JSEN.2014.2329268

Mao, L., Dell'Agnese, A., Comiti, F., 2017. Sediment motion and velocity in a glacier-fed stream. Geomorphology, SEDIMENT DYNAMICS IN ALPINE BASINS 291, 69–79. https://doi.org/10.1016/j.geomorph.2016.09.008

Mao, L., Toro, M., Carrillo, R., Brardinoni, F., Fraccarollo, L., 2020. Controls over particle motion and resting times of coarse bed load transport in a glacier-fed mountain stream. Journal of Geophysical Research: Earth Surface 125, e2019JF005253. https://doi.org/10.1029/2019jf005253

Maroli, A., Narwane, V.S., Gardas, B.B., 2021. Applications of IoT for achieving sustainability in agricultural sector: A comprehensive review. Journal of Environmental Management 298, 113488. https://doi.org/10.1016/j.jenvman.2021.113488

Marrocco, G., 2010. Pervasive electromagnetics: sensing paradigms by passive RFID technology. IEEE Wireless Communications 17, 10–17. https://doi.org/10.1109/MWC.2010.5675773

Martínez-de Dios, J.R., De San Bernabé, A., Viguria, A., Torres-González, A., Ollero, A., 2017. Combining Unmanned Aerial Systems and Sensor Networks for Earth Observation. Remote Sensing 9, 336. https://doi.org/10.3390/rs9040336

Mascareñas, D., Flynn, E., Todd, M., Park, G., Farrar, C., 2008. Wireless Sensor Technologies for Monitoring Civil Structures 5.

McCoy, S.W., Coe, J.A., Kean, J.W., Tucker, G.E., Staley, D.M., Wasklewicz, T.A., 2011. Observations of debris flows at Chalk Cliffs, Colorado, USA: part 1, in situ measurements of flow dynamics, tracer particle movement and video imagery from the summer of 2009, in: Genevois, R., Hamilton, D.L., Prestininzi, A. (Eds.), . Presented at the 5th International Conference on Debris-Flow Hazards Mitigation: Mechanics, Prediction and Assessment, Italian Journal of Engineering Geology and Environment, pp. 715–726.

Mehner, H., Weise, C., Schwebke, S., Hampl, S., Hoffmann, M., 2015. A passive microsystem for detecting multiple acceleration events beyond a threshold. Microelectronic Engineering, Micro/Nano Devices and Systems 2014 An open focused special thematic issue of Microelectronic Engineering 145, 104–111. https://doi.org/10.1016/j.mee.2015.03.023

Melia-Segui, J., Vilajosana, X., 2019. Ubiquitous moisture sensing in automaker industry based on standard UHF RFID tags, in: 2019 IEEE International Conference on RFID (RFID). Presented at the 2019 IEEE International Conference on RFID (RFID), IEEE, Phoenix, AZ, USA, pp. 1–8. https://doi.org/10.1109/RFID.2019.8719092

Melita, C.D., Longo, D., Muscato, G., Giudice, G., 2015. Measurement and Exploration in Volcanic Environments, in: Valavanis, K.P., Vachtsevanos, G.J. (Eds.), Handbook of Unmanned Aerial Vehicles. Springer Netherlands, Dordrecht, pp. 2667–2692. https://doi.org/10.1007/978-90-481-9707-1_76

Merilampi, S., Björninen, T., Ukkonen, L., Ruuskanen, P., Sydänheimo, L., 2011. Embedded wireless strain sensors based on printed RFID tag. Sensor Review 31, 32–40. https://doi.org/10.1108/02602281111099062

Miesen, R., Ebelt, R., Kirsch, F., Schäfer, T., Li, G., Wang, H., Vossiek, M., 2011. Where is the Tag? IEEE Microwave Magazine 12, S49–S63. https://doi.org/10.1109/MMM.2011.942730

Miesen, R., Kirsch, F., Vossiek, M., 2013a. UHF RFID Localization Based on Synthetic Apertures. IEEE Transactions on Automation Science and Engineering 10, 807–815. https://doi.org/10.1109/TASE.2012.2224656

Miesen, R., Parr, A., Schleu, J., Vossiek, M., 2013b. 360° carrier phase measurement for UHF RFID local positioning, in: 2013 IEEE International Conference on RFID-Technologies and Applications (RFID-TA). Presented at the 2013 IEEE International Conference on

RFID-Technologies and Applications (RFID-TA), pp. 1–6. https://doi.org/10.1109/RFID-TA.2013.6694499

Milan, D.J., 2013. Sediment routing hypothesis for pool-riffle maintenance. Earth Surface Processes and Landforms 38, 1623–1641. https://doi.org/10.1002/esp.3395

Miller, I.M., Warrick, J.A., 2012. Measuring sediment transport and bed disturbance with tracers on a mixed beach. Marine Geology 299–302, 1–17. https://doi.org/10.1016/j.margeo.2012.01.002

Miller, I.M., Warrick, J.A., Morgan, C., 2011. Observations of coarse sediment movements on the mixed beach of the Elwha Delta, Washington. Marine Geology 282, 201–214. https://doi.org/10.1016/j.margeo.2011.02.012

Minasy, A.J., 1970. Method and apparatus for article theft detection. US3500373A.

Mondal, S., Kumar, D., Ghazali, Mohd.I., Chahal, P., Udpa, L., Deng, Y., 2018a. Monitoring and localization of buried plastic natural gas pipes using passive RF tags. AIP Conference Proceedings 1949, 020020. https://doi.org/10.1063/1.5031517

Mondal, S., Rothwell, E.J., Chahal, P., 2018b. A Wideband Antenna for Buried RFID Applications, in: 2018 IEEE International Symposium on Antennas and Propagation USNC/URSI National Radio Science Meeting. Presented at the 2018 IEEE International Symposium on Antennas and Propagation USNC/URSI National Radio Science Meeting, pp. 327–328. https://doi.org/10.1109/APUSNCURSINRSM.2018.8608414

Monza M700, n.d.

Moore, J., Geimer, P., Finnegan, R., Thorne, M., 2018. Use of Seismic Resonance Measurements to Determine the Elastic Modulus of Freestanding Rock Masses. Rock Mechanics and Rock Engineering 51. https://doi.org/10.1007/s00603-018-1554-6

Motroni, A., Nepa, P., Tripicchio, P., Unetti, M., 2018. A Multi-Antenna SAR-based method for UHF RFID Tag Localization via UGV, in: 2018 IEEE International Conference on RFID Technology Application (RFID-TA). Presented at the 2018 IEEE International Conference on RFID Technology Application (RFID-TA), pp. 1–6. https://doi.org/10.1109/RFID-TA.2018.8552780

Nanni, F., Nappi, S., Marrocco, G., 2022. Potentiometric Sensing by means of Self-tuning RFID ICs, in: 2022 IEEE International Conference on RFID (RFID). Presented at the 2022 IEEE International Conference on RFID (RFID), pp. 17–22. https://doi.org/10.1109/RFID54732.2022.9795959

Nappi, S., Marrocco, G., 2018. Inkjet-Printed RFID-Skins for the Detection of Surface Defects, in: 2018 2nd URSI Atlantic Radio Science Meeting (AT-RASC). Presented at the 2018 2nd URSI Atlantic Radio Science Meeting (AT-RASC), pp. 1–4. https://doi.org/10.23919/URSI-AT-RASC.2018.8471335

Nasiri, M., Tura, N., Ojanen, V., 2017. Developing Disruptive Innovations for Sustainability: A Review on Impact of Internet of Things (IOT), in: 2017 Portland International Conference on Management of Engineering and Technology (PICMET). Presented at the 2017 Portland International Conference on Management of Engineering and Technology (PICMET), IEEE, Portland, OR, pp. 1–10. https://doi.org/10.23919/PICMET.2017.8125369

Nguyen, S.D., Pham, T.T., Blanc, E.F., Le, N.N., Dang, C.M., Tedjini, S., 2013. Approach for quality detection of food by RFID-based wireless sensor tag. Electronics Letters 49, 1588–1589. https://doi.org/10.1049/el.2013.3328

Ni, L.M., Liu, Y., Lau, Y.C., Patil, A.P., 2003. LANDMARC: indoor location sensing using active RFID, in: Proceedings of the First IEEE International Conference on Pervasive Computing and Communications, 2003. (PerCom 2003). Presented at the Proceedings of the First IEEE International Conference on Pervasive Computing and

Communications, 2003. (PerCom 2003), pp. 407–415. https://doi.org/10.1109/PERCOM.2003.1192765

Nichols, M.H., 2004. A Radio Frequency Identification System for Monitoring Coarse Sediment Particle Displacement. Applied Engineering in Agriculture 20, 783–787. https://doi.org/10.13031/2013.17727

Niethammer, U., James, M.R., Rothmund, S., Travelletti, J., Joswig, M., 2012. UAV-based remote sensing of the Super-Sauze landslide: Evaluation and results. Engineering Geology, Integration of Technologies for Landslide Monitoring and Quantitative Hazard Assessment 128, 2–11. https://doi.org/10.1016/j.enggeo.2011.03.012

Nikitin, P., 2012. Leon Theremin (Lev Termen). IEEE Antennas and Propagation Magazine 54, 252–257. https://doi.org/10.1109/MAP.2012.6348173

Nikitin, P., Rao, K.V.S., Lam, S., 2012. UHF RFID tag characterization: overview and state-of-the-art, in: Antenna Measurement Techniques Association Symposium (AMTA).

Nikitin, P.V., Martinez, R., Ramamurthy, S., Leland, H., Spiess, G., Rao, K.V.S., 2010. Phase based spatial identification of UHF RFID tags, in: IEEE Int. Conf. RFID. Presented at the IEEE International Conference on RFID, IEEE, Orlando, FL, USA, pp. 102–109. https://doi.org/10.1109/RFID.2010.5467253

Nummela, J., Ukkonen, L., Sydänheimo, L., 2008. Passive UHF RFID tags in arctic environment. International Journal of Communications 2, 135–142.

Nunes-Silva, P., Hrncir, M., Guimarães, J.T.F., Arruda, H., Costa, L., Pessin, G., Siqueira, J.O., de Souza, P., Imperatriz-Fonseca, V.L., 2019. Applications of RFID technology on the study of bees. Insect. Soc. 66, 15–24. https://doi.org/10.1007/s00040-018-0660-5

Obe, R.T., 2003. Air Defence 1940–1970/Military IFF and Secondary Radar — A Historical Review. Measurement and Control 36, 246–251. https://doi.org/10.1177/002029400303600804

Occhiuzzi, C., Ajovalasit, A., Sabatino, M.A., Dispenza, C., Marrocco, G., 2015. RFID epidermal sensor including hydrogel membranes for wound monitoring and healing, in: 2015 IEEE International Conference on RFID (RFID). Presented at the 2015 IEEE International Conference on RFID (RFID), pp. 182–188. https://doi.org/10.1109/RFID.2015.7113090

Occhiuzzi, C., Amendola, S., Nappi, S., D'Uva, N., Marrocco, G., 2019. RFID Technology for Industry 4.0: Architectures and Challenges, in: 2019 IEEE International Conference on RFID Technology and Applications (RFID-TA). Presented at the 2019 IEEE International Conference on RFID Technology and Applications (RFID-TA), pp. 181–186. https://doi.org/10.1109/RFID-TA.2019.8892049

Occhiuzzi, C., Amendola, S., Nappi, S., D'Uva, N., Marrocco, G., 2018. Sensing-oriented RFID tag Response in High Temperature Conditions, in: 2018 3rd International Conference on Smart and Sustainable Technologies (SpliTech). Presented at the 2018 3rd International Conference on Smart and Sustainable Technologies (SpliTech), pp. 1–4.

Occhiuzzi, C., Caizzone, S., Marrocco, G., 2013. Passive UHF RFID antennas for sensing applications: Principles, methods, and classifcations. IEEE Antennas and Propagation Magazine 55, 14–34. https://doi.org/10.1109/MAP.2013.6781700

Occhiuzzi, C., Cippitelli, S., Marrocco, G., 2010. Modeling, Design and Experimentation of Wearable RFID Sensor Tag. IEEE Transactions on Antennas and Propagation 58, 2490–2498. https://doi.org/10.1109/TAP.2010.2050435

Occhiuzzi, C., Paggi, C., Marrocco, G., 2011a. Passive RFID Strain-Sensor Based on Meander-Line Antennas. IEEE Transactions on Antennas and Propagation 59, 4836–4840. https://doi.org/10.1109/TAP.2011.2165517

Occhiuzzi, C., Rida, A., Marrocco, G., Tentzeris, M., 2011b. RFID Passive Gas Sensor Integrating Carbon Nanotubes. IEEE Transactions on Microwave Theory and Techniques 59, 2674–2684. https://doi.org/10.1109/TMTT.2011.2163416

Olinde, L., Johnson, J.P.L., 2015. Using RFID and accelerometer-embedded tracers to measure probabilities of bed load transport, step lengths, and rest times in a mountain stream. Water Resour. Res. 51, 7572–7589. https://doi.org/10.1002/2014WR016120

Opasjumruskit, K., Thanthipwan, T., Sathusen, O., Sirinamarattana, P., Gadmanee, P., Pootarapan, E., Wongkomet, N., Thanachayanont, A., Thamsirianunt, M., 2006. Self-powered wireless temperature sensors exploit RFID technology. IEEE Pervasive Computing 5, 54–61. https://doi.org/10.1109/MPRV.2006.15

Osborne, P.D., Macdonald, N., Curtiss, G., 2011. Measurements and Modeling of Gravel Transport under Wind Waves, Vessel-Generated Waves, and Tidal Currents. Journal of Coastal Research 165–172. https://doi.org/10.2112/SI59-017.1

Papangelakis, E., MacVicar, B., 2020. Process-based assessment of success and failure in a constructed riffle-pool river restoration project. River Research and Applications 36, 1222–1241. https://doi.org/10.1002/rra.3636

Papangelakis, E., MacVicar, B., Ashmore, P., 2019a. Bedload sediment transport regimes of semi-alluvial rivers conditioned by urbanization and stormwater management. Water Resources Research 55, 10565–10587. https://doi.org/10.1029/2019WR025126

Papangelakis, E., Muirhead, C., Schneider, A., Macvicar, B., 2019b. Synthetic radio frequency identification tracer stones with weighted inner ball for burial depth estimation. Journal of Hydraulic Engineering 145. https://doi.org/10.1061/(ASCE)HY.1943-7900.0001650

Peeters, A., Houbrechts, G., Hallot, E., Van Campenhout, J., Gob, F., Petit, F., 2020. Can coarse bedload pass through weirs? Geomorphology 359, 107131. https://doi.org/10.1016/j.geomorph.2020.107131

Peplinski, N.R., Ulaby, F.T., Dobson, M.C., 1995. Dielectric properties of soils in the 0.3-1.3-GHz range. IEEE Transactions on Geoscience and Remote Sensing 33, 803–807. https://doi.org/10.1109/36.387598

Peppa, M.V., Mills, J.P., Moore, P., Miller, P.E., Chambers, J.E., 2017. Brief communication: landslide motion from cross correlation of UAV-derived morphological attributes. Natural Hazards and Earth System Sciences 17, 2143–2150. https://doi.org/Peppa, Maria V.; Mills, Jon P.; Moore, Phil; Miller, Pauline E.; Chambers, Jonathan E.. 2017 Brief communication: landslide motion from cross correlation of UAV-derived morphological attributes. Natural Hazards and Earth System Sciences, 17 (12). 2143-2150. https://doi.org/10.5194/nhess-17-2143-2017 <https://doi.org/10.5194/nhess-17-2143-2017>

Peres, C., Emam, M., Jafarzadeh, H., Belcastro, M., O'Flynn, B., 2021. Development of a Low-Power Underwater NFC-Enabled Sensor Device for Seaweed Monitoring. Sensors 21, 4649. https://doi.org/10.3390/s21144649

Peres, C., Pigeon, M., Rather, N., Gawade, D., Buckley, J., Jafarzadeh, H., O'Flynn, B., 2020. Theoretical models for underwater RFID and the impact of water salinity on the design of wireless systems. International Journal on Advances in Networks and Services 13, 45–59.

Peterson, J.R., 1993. Observations and modeling of seismic background noise (USGS Numbered Series No. 93–322), Open-File Report. U.S. Geological Survey.

Petit, F., Houbrechts, G., Peeters, A., Hallot, E., Van Campenhout, J., Denis, A.-C., 2015. Dimensionless critical shear stress in gravel-bed rivers. Geomorphology 250, 308–320. https://doi.org/10.1016/j.geomorph.2015.09.008

Philipose, M., Smith, J.R., Jiang, B., Mamishev, A., Roy, S., Sundara-Rajan, K., 2005. Battery-free wireless identification and sensing. IEEE Pervasive Computing 4, 37–45. https://doi.org/10.1109/MPRV.2005.7

Phillips, C.B., Jerolmack, D., 2014. Dynamics and Mechanics of Bed-Load Tracer Particles. Earth Surface Dynamics 513–530. https://doi.org/10.5194/esurf-2-513-2014

Phillips, C.B., Martin, R.L., Jerolmack, D.J., 2013. Impulse framework for unsteady flows reveals superdiffusive bed load transport. Geophysical Research Letters 40, 1328–1333.

Pichorim, S., Gomes, N., Batchelor, J., 2018. Two Solutions of Soil Moisture Sensing with RFID for Landslide Monitoring. Sensors 18, 452. https://doi.org/10.3390/s18020452

PINC Inc., n.d. Yard Management [WWW Document]. PINC. URL https://www.pinc.com/ (accessed 2.10.20).

Pursula, P., Marttila, I., Nummila, K., Seppa, H., 2013. High Frequency and Ultrahigh Frequency Radio Frequency Identification Passive Sensor Transponders for Humidity and Temperature Measurement Within Building Structures. IEEE Trans. Instrum. Meas. 62, 2559–2566. https://doi.org/10.1109/TIM.2013.2258763

Qi, C., Amato, F., Alhassoun, M., Durgin, G.D., 2019. Breaking the Range Limit of RFID Localization: Phase-based Positioning with Tunneling Tags, in: 2019 IEEE International Conference on RFID (RFID). Presented at the 2019 IEEE International Conference on RFID (RFID), IEEE, Phoenix, AZ, USA, pp. 1–8. https://doi.org/10.1109/RFID.2019.8719276

Qiao, Q., Zhang, L., Yang, F., Yue, Z., Elsherbeni, A.Z., 2013. Reconfigurable Sensing Antenna With Novel HDPE-BST Material for Temperature Monitoring. IEEE Antennas and Wireless Propagation Letters 12, 1420–1423. https://doi.org/10.1109/LAWP.2013.2286631

Rahmadya, B., Chen, X., Takeda, S., Kagoshima, K., Umehira, M., Kurosaki, W., 2020. Measurement of a UHF RFID-Based Battery-Less Vibration Frequency Sensitive Sensor Tag Using Tilt/Vibration Switches. IEEE Sensors Journal 20, 9901–9909. https://doi.org/10.1109/JSEN.2020.2992345

Rain RFID eBook, 2020. . Rain Alliance.

Rainato, R., Mao, L., Picco, L., 2018. Near-bankfull floods in an Alpine stream: Effects on the sediment mobility and bedload magnitude. International Journal of Sediment Research 33, 27–34. https://doi.org/10.1016/j.ijsrc.2017.03.006

Ravanel, L., Deline, P., Lambiel, C., Vincent, C., 2013. Instability of a high alpine rock ridge: the lower arête des cosmiques, mont blanc massif, france. Geografiska Annaler: Series A, Physical Geography 95, 51–66. https://doi.org/10.1111/geoa.12000

Ravazzolo, D., Mao, L., Picco, L., Lenzi, M.A., 2015. Tracking log displacement during floods in the Tagliamento River using RFID and GPS tracker devices. Geomorphology 228, 226–233. https://doi.org/10.1016/j.geomorph.2014.09.012

Raza, U., Salam, A., 2020a. Wireless Underground Communications in Sewer and Stormwater Overflow Monitoring: Radio Waves through Soil and Asphalt Medium. Information 11, 98. https://doi.org/10.3390/info11020098

Raza, U., Salam, A., 2020b. A Survey on Subsurface Signal Propagation. Smart Cities 3, 1513–1561. https://doi.org/10.3390/smartcities3040072

Reddy, R.R., Komeda, K., Okamoto, Y., Lebrasseur, E., Higo, A., Mita, Y., 2019. A zero-power sensing MEMS shock sensor with a latch-reset mechanism for multi-threshold events monitoring. Sensors and Actuators A: Physical 295, 1–10. https://doi.org/10.1016/j.sna.2019.05.036

Roberts, M.O., Renshaw, C.E., Magilligan, F.J., Brian Dade, W., 2020. Field measurement of the probability of coarse-grained sediment entrainment in natural rivers. Journal of Hydraulic Engineering 146. https://doi.org/10.1061/(ASCE)HY.1943-7900.0001694

Rose, D.P., Ratterman, M.E., Griffin, D.K., Hou, L., Kelley-Loughnane, N., Naik, R.R., Hagen, J.A., Papautsky, I., Heikenfeld, J.C., 2015. Adhesive RFID Sensor Patch for Monitoring of Sweat Electrolytes. IEEE Transactions on Biomedical Engineering 62, 1457–1465. https://doi.org/10.1109/TBME.2014.2369991

Ruiz-Garcia, L., Lunadei, L., 2011. The role of RFID in agriculture: Applications, limitations and challenges. Computers and Electronics in Agriculture 79, 42–50. https://doi.org/10.1016/j.compag.2011.08.010

Saarinen, K., Björninen, T., Ukkonen, L., Frisk, L., 2014. Reliability Analysis of RFID Tags in Changing Humid Environment. IEEE Transactions on Components, Packaging and Manufacturing Technology 4, 77–85. https://doi.org/10.1109/TCPMT.2013.2278182

Sadeghioon, A.M., Metje, N., Chapman, D.N., Anthony, C.J., 2014. SmartPipes: Smart Wireless Sensor Networks for Leak Detection in Water Pipelines. Journal of Sensor and Actuator Networks 3, 64–78. https://doi.org/10.3390/jsan3010064

Salam, A., Raza, U., 2020. Signals in the Soil: Developments in Internet of Underground Things. Springer International Publishing, Cham. https://doi.org/10.1007/978-3-030-50861-6

Salam, A., Vuran, M.C., Dong, X., Argyropoulos, C., Irmak, S., 2019. A Theoretical Model of Underground Dipole Antennas for Communications in Internet of Underground Things. IEEE Transactions on Antennas and Propagation 67, 3996–4009. https://doi.org/10.1109/TAP.2019.2902646

Salmerón, J.F., Molina-Lopez, F., Rivadeneyra, A., Quintero, A.V., Capitán-Vallvey, L.F., Rooij, N.F. de, Ozáez, J.B., Briand, D., Palma, A.J., 2014. Design and Development of Sensing RFID Tags on Flexible Foil Compatible With EPC Gen 2. IEEE Sensors Journal 14, 4361–4371. https://doi.org/10.1109/JSEN.2014.2335417

Savochkin, A., Timoshenko, A., Ivanov, M., 2019. Directed Antennas for RFID-Based Navigation for Agricultural Unmanned Vehicles, in: 2019 IEEE Conference of Russian Young Researchers in Electrical and Electronic Engineering (EIConRus). Presented at the 2019 IEEE Conference of Russian Young Researchers in Electrical and Electronic Engineering (EIConRus), pp. 2046–2050. https://doi.org/10.1109/EIConRus.2019.8657303

Scherhäufl, M., Pichler, M., Stelzer, A., 2015. UHF RFID Localization Based on Phase Evaluation of Passive Tag Arrays. IEEE Transactions on Instrumentation and Measurement 64, 913–922. https://doi.org/10.1109/TIM.2014.2363578

Schmugge, T.J., Jackson, T.J., McKim, H.L., 1980. Survey of methods for soil moisture determination. Water Resour. Res. 16, 961–979. https://doi.org/10.1029/WR016i006p00961

Schneider, J., Hegglin, R., Meier, S., Turowski, J.M., Nitsche, M., Rickenmann, D., 2010. Studying sediment transport in mountain rivers by mobile and stationary RFID antennas, in: Dittrich, K., Koll, A., Aberle, J., Geisenhainer, P. (Eds.), 5th International Conference on Fluvial Hydraulics (River Flow 2010). Bundesanstalt für Wasserbau, Braunschweig, Germany, pp. 1723–1730.

Schneider, J.M., Turowski, J.M., Rickenmann, D., Hegglin, R., Arrigo, S., Mao, L., Kirchner, J.W., 2014. Scaling relationships between bed load volumes, transport distances, and stream power in steep mountain channels. Journal of Geophysical Research: Earth Surface 119, 533–549. https://doi.org/10.1002/2013JF002874

Scudero, S., D'Alessandro, A., Greco, L., Vitale, G., 2018. MEMS technology in seismology: A short review, in: 2018 IEEE International Conference on Environmental Engineering (EE). pp. 1–5. https://doi.org/10.1109/EE1.2018.8385252

Segarra, P., Sanchidrián, J.A., Castedo, R., López, L.M., del Castillo, I., 2015. Performance of some coupling methods for blast vibration monitoring. Journal of Applied Geophysics 112, 129–135. https://doi.org/10.1016/j.jappgeo.2014.11.012

Shaikh, F.K., Zeadally, S., Exposito, E., 2017. Enabling Technologies for Green Internet of Things. IEEE Systems Journal 11, 983–994. https://doi.org/10.1109/JSYST.2015.2415194

Sheu, J.-P., Hsieh, K.-Y., Cheng, P.-W., 2008. Design and Implementation of Mobile Robot for Nodes Replacement in Wireless Sensor Networks. J. Inf. Sci. Eng. 24, 393–410.

Shi, X., Yang, F., Xu, S., Li, M., 2017. Design of A RFID patch antenna integrated with mercury switches for wireless tilt sensing, in: 2017 IEEE International Symposium on Antennas and Propagation USNC/URSI National Radio Science Meeting. Presented at the 2017 IEEE International Symposium on Antennas and Propagation USNC/URSI National Radio Science Meeting, pp. 2507–2508. https://doi.org/10.1109/APUSNCURSINRSM.2017.8073296

Siden, J., Zeng, X., Unander, T., Koptyug, A., Nilsson, H.E., 2007. Remote Moisture Sensing utilizing Ordinary RFID Tags, in: IEEE Sensors Conf. Presented at the 2007 IEEE Sensors, Atlanta, GA, USA, pp. 308–311. https://doi.org/10.1109/ICSENS.2007.4388398

Sohrab, A.P., Huang, Y., Hussein, M., Kod, M., Carter, P., 2016. A UHF RFID Tag With Improved Performance on Liquid Bottles. IEEE Antennas and Wireless Propagation Letters 15, 1673–1676. https://doi.org/10.1109/LAWP.2016.2521786

Squarzoni, C., Delacourt, C., Allemand, P., 2005. Differential single-frequency GPS monitoring of the La Valette landslide (French Alps). Engineering Geology 79, 215–229. https://doi.org/10.1016/j.enggeo.2005.01.015

Stähly, S., Franca, M.J., Robinson, C.T., Schleiss, A.J., 2020. Erosion, transport and deposition of a sediment replenishment under flood conditions. Earth Surface Processes and Landforms n/a. https://doi.org/10.1002/esp.4970

Stockman, H., 1948. Communication by Means of Reflected Power. Proceedings of the IRE 36, 1196–1204. https://doi.org/10.1109/JRPROC.1948.226245

Stoddard, B., Selker, Dr.J., Udell, Dr.C., 2019. Examining the Effectiveness of Commercial RFID Tags as Soil Moisture Sensors. https://doi.org/10.1002/essoar.10500837.1

Su, J., Sheng, Z., Leung, V.C.M., Chen, Y., 2019. Energy Efficient Tag Identification Algorithms For RFID: Survey, Motivation And New Design. IEEE Wireless Communications 26, 118–124. https://doi.org/10.1109/MWC.2019.1800249

Sudarshan, S.K.V., Montano, V., Nguyen, A., McClimans, M., Chang, L., Stewart, R.R., Becker, A.T., 2017. A Heterogeneous Robotics Team for Large-Scale Seismic Sensing. IEEE Robotics and Automation Letters 2, 1328–1335. https://doi.org/10.1109/LRA.2017.2666300

Sun, Y., Plowcha, A., Nail, M., Elbaum, S., Terry, B., Detweiler, C., 2018. Unmanned Aerial Auger for Underground Sensor Installation, in: 2018 IEEE/RSJ International Conference on Intelligent Robots and Systems (IROS). Presented at the 2018 IEEE/RSJ International Conference on Intelligent Robots and Systems (IROS), IEEE, Madrid, pp. 1374–1381. https://doi.org/10.1109/IROS.2018.8593824

Sun, Z., Wang, P., Vuran, M.C., Al-Rodhaan, M.A., Al-Dhelaan, A.M., Akyildiz, I.F., 2011. MISE-PIPE: Magnetic induction-based wireless sensor networks for underground pipeline monitoring. Ad Hoc Networks 9, 218–227. https://doi.org/10.1016/j.adhoc.2010.10.006

Swedberg, C., 2015. Smartrac's New Passive Sensor DogBone Transmits Moisture Levels. RFID Journal. URL https://www.rfidjournal.com/smartracs-new-passive-sensor-dogbone-transmits-moisture-levels (accessed 2.4.21).

Tan, J., Sathyamurthy, M., Rolapp, A., Gamez, J., Hennig, E., Schäfer, E., Sommer, R., 2019. A Fully Passive RFID Temperature Sensor SoC With an Accuracy of ±0.4 °C ($3\sigma$) From 0 °C to 125 °C. IEEE Journal of Radio Frequency Identification 3, 35–45. https://doi.org/10.1109/JRFID.2019.2896145

Taoufik, S., 2018. Fiabilité et analyse de défaillance des tags RFID UHF passifs sous contraintes environnementales sévères. (Ph.D. Thesis). Normandie Université.

Tatiparthi, S.R., De Costa, Y.G., Whittaker, C.N., Hu, S., Yuan, Z., Zhong, R.Y., Zhuang, W.-Q., 2021. Development of radio-frequency identification (RFID) sensors suitable for smart-monitoring applications in sewer systems. Water Research 198, 117107. https://doi.org/10.1016/j.watres.2021.117107

Tedesco, M., 2015. Electromagnetic properties of components of the cryosphere, in: Remote Sensing of the Cryosphere, The Cryosphere Science Series. Wiley.

Tedjini, S., Andia-Vera, G., Zurita, M., Freire, R.C.S., Duroc, Y., 2016. Augmented RFID Tags, in: 2016 IEEE Topical Conference on Wireless Sensors and Sensor Networks (WiSNet). Presented at the 2016 IEEE Topical Conference on Wireless Sensors and Sensor Networks (WiSNet), pp. 67–70. https://doi.org/10.1109/WISNET.2016.7444324

Todd, B., Phillips, M., Schultz, S.M., Hawkins, A.R., Jensen, B.D., 2009. Low-Cost RFID Threshold Shock Sensors. IEEE Sensors Journal 9, 464–469. https://doi.org/10.1109/JSEN.2009.2014410

Travelletti, J., Delacourt, C., Allemand, P., Malet, J.-P., Schmittbuhl, J., Toussaint, R., Bastard, M., 2012. Correlation of multi-temporal ground-based optical images for landslide monitoring: Application, potential and limitations. ISPRS Journal of Photogrammetry and Remote Sensing 70, 39–55. https://doi.org/10.1016/j.isprsjprs.2012.03.007

Tsakiris, A.G., Papanicolaou, A.N. (Thanos), Moustakidis, I.V. (Danny), Abban, B.K., 2015. Identification of the Burial Depth of Radio Frequency Identification Transponders in Riverine Applications. Journal of Hydraulic Engineering 141, 4015007–1. https://doi.org/10.1061/(ASCE)HY.1943-7900.0001001

Uchimura, T., Towhata, I., Lan Anh, T.T., Fukuda, J., Bautista, C.J.B., Wang, L., Seko, I., Uchida, T., Matsuoka, A., Ito, Y., Onda, Y., Iwagami, S., Kim, M.-S., Sakai, N., 2010. Simple monitoring method for precaution of landslides watching tilting and water contents on slopes surface. Landslides 7, 351–357. https://doi.org/10.1007/s10346-009-0178-z

UCODE 8, n.d.

Valentine, G., Vojtech, L., Neruda, M., 2015. Design of Solar Harvested Semi Active RFID Transponder with Supercapacitor Storage. Advances in Electrical and Electronic Engineering 13. https://doi.org/10.15598/aeee.v13i4.1485

Vaz, A., Ubarretxena, A., Zalbide, I., Pardo, D., Solar, H., Garcia-Alonso, A., Berenguer, R., 2010. Full Passive UHF Tag With a Temperature Sensor Suitable for Human Body Temperature Monitoring. IEEE Transactions on Circuits and Systems II: Express Briefs 57, 95–99. https://doi.org/10.1109/TCSII.2010.2040314

Vázquez-Tarrío, D., Recking, A., Liébault, F., Tal, M., Menéndez-Duarte, R., 2018. Particle transport in gravel-bed rivers: Revisiting passive tracer data. Earth Surface Processes and Landforms 44, 112–128. https://doi.org/10.1002/esp.4484

Vena, A., Sorli, B., Saggin, B., Garcia, R., Podlecki, J., 2019. Passive UHF RFID Sensor to Monitor Fragile Objects during Transportation, in: 2019 IEEE International Conference on RFID Technology and Applications (RFID-TA). Presented at the 2019 IEEE International Conference on RFID Technology and Applications (RFID-TA), pp. 415–420. https://doi.org/10.1109/RFID-TA.2019.8892033

Virtanen, J., Ukkonen, L., Bjorninen, T., Elsherbeni, A.Z., Sydänheimo, L., 2011. Inkjet-Printed Humidity Sensor for Passive UHF RFID Systems. IEEE Transactions on Instrumentation and Measurement 60, 2768–2777. https://doi.org/10.1109/TIM.2011.2130070

Voipio, V., Korpela, J., Elfvengren, K., 2021. Environmental RFID: measuring the relevance in the fashion industry. International Journal of Fashion Design, Technology and Education 14, 284–292. https://doi.org/10.1080/17543266.2021.1929510

Vossiek, M., Gulden, P., 2008. The Switched Injection-Locked Oscillator: A Novel Versatile Concept for Wireless Transponder and Localization Systems. IEEE Transactions on Microwave Theory and Techniques 56, 859–866. https://doi.org/10.1109/TMTT.2008.918158

Vyas, R., Tye, B., 2019. A Sequential RFID System for Robust Communication with Underground Carbon Steel Pipes in Oil and Gas Applications. Electronics 8, 1374. https://doi.org/10.3390/electronics8121374

Wagih, M., Shi, J., 2021a. Wireless Ice Detection and Monitoring Using Flexible UHF RFID Tags. IEEE Sensors Journal 21, 18715–18724. https://doi.org/10.1109/JSEN.2021.3087326

Wagih, M., Shi, J., 2021b. Towards the Optimal RF Sensing Strategy for IoT Applications: An Ice Sensing Case Study. https://doi.org/10.36227/techrxiv.14958174.v1

Wang, B., Law, M., Bermak, A., Luong, H.C., 2014. A Passive RFID Tag Embedded Temperature Sensor With Improved Process Spreads Immunity for a \$-\hbox 30^\circ\hbox C\$ to 60\$^\circ\hbox C\$ Sensing Range. IEEE Transactions on Circuits and Systems I: Regular Papers 61, 337–346. https://doi.org/10.1109/TCSI.2013.2278388

Wang, C., Chen, F., Wang, Y., Sadeghpour, S., Wang, Chenxi, Baijot, M., Esteves, R., Zhao, C., Bai, J., Liu, H., Kraft, M., 2020. Micromachined Accelerometers with Sub-$\mu$g/$\sqrt{Hz}$ Noise Floor: A Review. Sensors 20, 4054. https://doi.org/10.3390/s20144054

Wang, J., Chang, L., Aggarwal, S., Abari, O., Keshav, S., 2020. Soil moisture sensing with commodity RFID systems, in: Proceedings of the 18th International Conference on Mobile Systems, Applications, and Services. Presented at the MobiSys '20: The 18th Annual International Conference on Mobile Systems, Applications, and Services, ACM, Toronto Ontario Canada, pp. 273–285. https://doi.org/10.1145/3386901.3388940

Wang, J., Schluntz, E., Otis, B., Deyle, T., 2015. A New Vision for Smart Objects and the Internet of Things: Mobile Robots and Long-Range UHF RFID Sensor Tags. arXiv:1507.02373 [cs].

Wang, W., Owyeung, R., Sadeqi, A., Sonkusale, S., 2020a. Single Event Recording of Temperature and Tilt Using Liquid Metal With RFID Tags. IEEE Sensors Journal 20, 3249–3256. https://doi.org/10.1109/JSEN.2019.2956462

Wang, W., Sadeqi, A., Nejad, H.R., Sonkusale, S., 2020b. Cost-Effective Wireless Sensors for Detection of Package Opening and Tampering. IEEE Access 8, 117122–117132. https://doi.org/10.1109/ACCESS.2020.3004438

Wang, X., Zhang, J., Yu, Z., Mao, S., Periaswamy, S.C.G., Patton, J., 2019. On Remote Temperature Sensing Using Commercial UHF RFID Tags. IEEE Internet Things J. 6, 10715–10727. https://doi.org/10.1109/JIOT.2019.2941023

Wang, Z., Ye, N., Malekian, R., Xiao, F., Wang, R., 2016. TrackT: Accurate tracking of RFID tags with mm-level accuracy using first-order taylor series approximation. Ad Hoc Networks 53, 132–144. https://doi.org/10.1016/j.adhoc.2016.09.026

Want, R., 2004. Enabling ubiquitous sensing with RFID. Computer 37, 84–86. https://doi.org/10.1109/MC.2004.1297315

Wasson, T., Choudhury, T., Sharma, S., Kumar, P., 2017. Integration of RFID and sensor in agriculture using IOT, in: 2017 International Conference On Smart Technologies For Smart Nation (SmartTechCon). Presented at the 2017 International Conference On Smart Technologies For Smart Nation (SmartTechCon), pp. 217–222. https://doi.org/10.1109/SmartTechCon.2017.8358372

Watkins, S.E., Swift, T.M., Molander, M.J., 2007. RFID Instrumentation in a Field Application, in: IEEE Region 5 Technical Conf. Presented at the 2007 IEEE Region 5 Technical Conference, Fayetteville, AR, USA, pp. 400–403. https://doi.org/10.1109/TPSD.2007.4380343

Węglarski, M., Jankowski-Mihułowicz, P., 2019. Factors Affecting the Synthesis of Autonomous Sensors with RFID Interface. Sensors 19, 4392. https://doi.org/10.3390/s19204392

Wilcock, P.R., 1997. Entrainment, displacement and transport of tracer gravels. Earth Surface Processes and Landforms 22, 1125–1138.

Wixted, A., Kinnaird, P., Larijani, H., Tait, A., Ahmadinia, A., Strachan, N., 2017. Evaluation of LoRa and LoRaWAN for wireless sensor networks. IEEE SENSORS, 2016 1–3. https://doi.org/10.1109/ICSENS.2016.7808712

Wu, H., Tao, B., Gong, Z., Yin, Z., Ding, H., 2019. A Fast UHF RFID Localization Method Using Unwrapped Phase-Position Model. IEEE Transactions on Automation Science and Engineering 1–10. https://doi.org/10.1109/TASE.2019.2895104

Xu, L.D., He, W., Li, S., 2014. Internet of Things in Industries: A Survey. IEEE Transactions on Industrial Informatics 10, 2233–2243. https://doi.org/10.1109/TII.2014.2300753

Yan, P., Lu, W., Zhang, J., Zou, Y., Chen, M., 2017. Evaluation of human response to blasting vibration from excavation of a large scale rock slope: A case study. Earthq. Eng. Eng. Vib. 16, 435–446. https://doi.org/10.1007/s11803-017-0391-z

Yang, L., Li, Y., Lin, Q., Jia, H., Li, X.-Y., Liu, Y., 2017. Tagbeat: Sensing Mechanical Vibration Period With COTS RFID Systems. IEEE/ACM Trans. Networking 25, 3823–3835. https://doi.org/10.1109/TNET.2017.2769138

Yang, L., Zhang, R., Staiculescu, D., Wong, C.P., Tentzeris, M.M., 2009. A Novel Conformal RFID-Enabled Module Utilizing Inkjet-Printed Antennas and Carbon Nanotubes for Gas-Detection Applications. IEEE Antennas and Wireless Propagation Letters 8, 653–656. https://doi.org/10.1109/LAWP.2009.2024104

Yang, M.X., Hu, X., Akin, D., Poon, A., Wong, H.-S.P., 2021. Intracellular detection and communication of a wireless chip in cell. Sci Rep 11, 5967. https://doi.org/10.1038/s41598-021-85268-5

Yang, P., Feng, Y., Xiong, J., Chen, Z., Li, X.-Y., 2020. RF-Ear: Contactless Multi-device Vibration Sensing and Identification Using COTS RFID, in: IEEE INFOCOM 2020 - IEEE Conference on Computer Communications. Presented at the IEEE INFOCOM 2020 - IEEE Conference on Computer Communications, pp. 297–306. https://doi.org/10.1109/INFOCOM41043.2020.9155251

Yi, X., Cho, C., Cooper, J., Wang, Y., Tentzeris, M.M., Leon, R.T., 2013. Passive wireless antenna sensor for strain and crack sensing—electromagnetic modeling, simulation, and testing. Smart Mater. Struct. 22, 085009. https://doi.org/10.1088/0964-1726/22/8/085009

Yi, X., Wu, T., Wang, Y., Tentzeris, M.M., 2015. Sensitivity Modeling of an RFID-Based Strain-Sensing Antenna With Dielectric Constant Change. IEEE Sensors Journal 15, 6147–6155. https://doi.org/10.1109/JSEN.2015.2453947

Yin, J., Yi, J., Law, M.K., Ling, Y., Lee, M.C., Ng, K.P., Gao, B., Luong, H.C., Bermak, A., Chan, M., Ki, W.-H., Tsui, C.-Y., Yuen, M.M.-F., 2010. A system-on-chip EPC Gen-2 passive UHF RFID tag with embedded temperature sensor, in: 2010 IEEE International

Solid-State Circuits Conference - (ISSCC). Presented at the 2010 IEEE International Solid-State Circuits Conference - (ISSCC), pp. 308–309. https://doi.org/10.1109/ISSCC.2010.5433893

Yu, S.M., Feng, P., Wu, N.J., 2015. Passive and Semi-Passive Wireless Temperature and Humidity Sensors Based on EPC Generation-2 UHF Protocol. IEEE Sensors Journal 15, 2403–2411. https://doi.org/10.1109/JSEN.2014.2375180

Zannas, K., Matbouly, H.E., Duroc, Y., Tedjini, S., 2020. From Identification to Sensing: Augmented RFID Tags, in: Carvalho, N.B., Georgiadis, A. (Eds.), Wireless Power Transmission for Sustainable Electronics. Wiley, pp. 223–246. https://doi.org/10.1002/9781119578598.ch8

Zannas, K., Matbouly, H.E., Duroc, Y., Tedjini, S., 2018. Self-Tuning RFID Tag: A New Approach for Temperature Sensing. IEEE Transactions on Microwave Theory and Techniques 66, 5885–5893. https://doi.org/10.1109/TMTT.2018.2878568

Zarifi, M.H., Deif, S., Daneshmand, M., 2017. Wireless passive RFID sensor for pipeline integrity monitoring. Sensors and Actuators A: Physical 261, 24–29. https://doi.org/10.1016/j.sna.2017.04.006

Zhang, C., Fu, Y., Deng, F., Wei, B., Wu, X., 2018. Methane Gas Density Monitoring and Predicting Based on RFID Sensor Tag and CNN Algorithm. Electronics 7, 69. https://doi.org/10.3390/electronics7050069

Zhang, J., Tian, G.Y., Marindra, A.M.J., Sunny, A.I., Zhao, A.B., 2017. A Review of Passive RFID Tag Antenna-Based Sensors and Systems for Structural Health Monitoring Applications. Sensors 17, 265. https://doi.org/10.3390/s17020265

Zhao, R., Wang, D., Zhang, Q., Chen, H., Xu, H., 2019. PEC: Synthetic Aperture RFID Localization with Aperture Position Error Compensation, in: 2019 16th Annual IEEE International Conference on Sensing, Communication, and Networking (SECON). Presented at the 2019 16th Annual IEEE International Conference on Sensing, Communication, and Networking (SECON), pp. 1–9. https://doi.org/10.1109/SAHCN.2019.8824882

Zhou, C., Griffin, J.D., 2012. Accurate phase-based ranging measurements for backscatter RFID tags. IEEE Antennas Wirel. Propag. Lett 11, 152–155. https://doi.org/10.1109/LAWP.2012.2186110

Zhou, S., Deng, F., Yu, L., Li, B., Wu, X., Yin, B., 2016. A Novel Passive Wireless Sensor for Concrete Humidity Monitoring. Sensors 16, 1535. https://doi.org/10.3390/s16091535

Zhu, C., Leung, V.C.M., Shu, L., Ngai, E.C.-H., 2015. Green Internet of Things for Smart World. IEEE Access 3, 2151–2162. https://doi.org/10.1109/ACCESS.2015.2497312

Zhu, W., Zhang, Q., Matlin, M., Chen, Y., Wu, Y., Zhu, X., Zhao, H., Pollack, R., Xiao, H., 2021. Passive Digital Sensing Method and Its Implementation on Passive RFID Temperature Sensors. IEEE Sensors Journal 21, 4793–4800. https://doi.org/10.1109/JSEN.2020.3035756

Ziai, M.A., Batchelor, J.C., 2017. Supply chain integrity tilt sensing RFID tag, in: 2017 IEEE MTT-S International Microwave Workshop Series on Advanced Materials and Processes for RF and THz Applications (IMWS-AMP). Presented at the 2017 IEEE MTT-S International Microwave Workshop Series on Advanced Materials and Processes for RF and THz Applications (IMWS-AMP), pp. 1–3. https://doi.org/10.1109/IMWS-AMP.2017.8247357

Zuffanelli, S., Aguila, P., Zamora, G., Paredes, F., Martin, F., Bonache, J., 2016. A High-Gain Passive UHF-RFID Tag with Increased Read Range. Sensors 16, 1150. https://doi.org/10.3390/s16071150